\documentclass[useAMS,usenatbib,usecalrsfs,usegraphicx,onecolumn]{mn2e-astroph}

\usepackage{color}
\usepackage{amssymb}

\def\gapp{\ \lower 3pt\hbox{${\buildrel > \over \sim}$}\ }
\def\lapp{\ \lower 3pt\hbox{${\buildrel < \over \sim}$}\ }

\def\gl{\ \lower 3pt\hbox{${\buildrel > \over <}$}\ }

\newcommand{\ben}{\begin{enumerate}}
\newcommand{\een}{\end{enumerate}}
\newcommand{\rn}[1]{(\ref{#1})}

\newcommand{\be}{\begin{equation}}
\newcommand{\ee}{\end{equation}}
\newcommand{\bea}{\begin{eqnarray}}
\newcommand{\eea}{\end{eqnarray}}
\newcommand{\bean}{\begin{eqnarray*}}
\newcommand{\eean}{\end{eqnarray*}}
\newcommand{\ff}[2]{{\textstyle \frac{#1}{#2}}}

\newcommand{\next}{\nonumber\\}
\newcommand{\br}{{\bf r}}
\newcommand{\bR}{{\bf R}}
\newcommand{\hand}{\hspace{0.5cm}{\rm and}\hspace{0.5cm}}
\newcommand{\pp}[2]{\frac{\partial{#1}}{\partial{#2}}}

\title[General coplanar systems]
{New developments for modern celestial mechanics.\\
I. General coplanar three-body systems.\\ Application to exoplanets}
\author[Mardling]{Rosemary A. Mardling$^{1}$$^{2}$\thanks{E-mail:
rosemary.mardling@monash.edu} \\
$^{1}$School of Mathematical Sciences, Monash University, Victoria, 3800, Australia\\
$^{2}$Astronomy Department of the University of Geneva, Geneva Observatory,
51 Ch des Maillettes, CH1290 Versoix
}
\begin{document}

\date{Accepted ... Received ...; in original form ...}

\pagerange{\pageref{firstpage}--\pageref{lastpage}} \pubyear{2013}

\maketitle

\label{firstpage}

\begin{abstract}
Modern applications of celestial mechanics include the study of closely packed systems of exoplanets,
circumbinary planetary systems, binary-binary interactions in star clusters,
and the dynamics of stars near the galactic centre.
While developments have historically been guided by the architecture of the Solar System,
the need for more general formulations with as few restrictions on the parameters as possible is obvious.
Here we present clear and concise generalisations of two classic expansions of the 
three-body disturbing function,
simplifying considerably their original form and making them accessible to the non-specialist. 

Governing the interaction between the inner and outer orbits of a hierarchical triple, the
disturbing function in its general form
is the conduit for energy and angular momentum exchange and as such, governs the secular and resonant
evolution of the system and its stability characteristics.
Focusing here on coplanar systems,
the first expansion is one in the ratio of inner to outer semimajor axes and is valid for all
eccentricities, while the second is an expansion in eccentricity and is valid for all semimajor axis
ratios, except for systems in which the orbits cross (this restriction also applies to the first expansion).
Our generalizations make both formulations valid for arbitrary mass ratios.
The classic versions of these appropriate to the restricted three-body problem
are known as Kaula's expansion and the literal expansion respectively.
We demonstrate the equivalence of the new expansions, identifying the role of the spherical harmonic
order $m$ in both and its physical significance in the three-body problem,
and introducing the concept of {\it principal resonances}. 

Several examples of the accessibility of both expansions are given including resonance widths and the secular rates of change of the elements. Results in their final form are gathered together at the end of the paper for the reader mainly interested in their application, including a guide for the choice of expansion.

\end{abstract}

\begin{keywords}
celestial mechanics -- methods: analytical -- chaos --
planets and satellites: dynamical evolution and stability 

\end{keywords}

\section{Introduction}
The recent explosion in the study of exoplanets is bringing together specialists from diverse backgrounds.
On the observational side, spectroscopic and photometric techniques for binary star observations
have been refined to the point where radial velocity and eclipse contrast measurements at the $50\,{\rm cm\,s^{-1}}$
and $10^{-4}$ level respectively are possible \citep[with the HARPS spectrograph and the Kepler
satellite;][]{pepe,dumusque,borucki,batalha}, while the field is breathing new life into
observational techniques ranging from
direct imaging \citep{marois08,marois10,kalas,lagrange}, to 
infrared and even x-ray photometry and spectroscopy \citep{richardson,fortney,pillitteri}.
As a direct consequence of the ability to make high-precision estimates of the masses and radii of transiting planets,
as well as to observe their atmospheres directly \citep[][and references therein]{seager},
geophysical and atmospheric scientists are finding a place in this booming field, with new centres
for planet characterization and habitability springing up around the world.

On the theoretical side, the discovery of 51 Peg b in its 4.6 day orbit \citep{51peg}
immediately reinvigorated existing theories of planet formation,
especially the question of the role of planet migration which until then (and still) had specialists wondering why
the giant planets in the Solar System apparently migrated so little \citep{linpap,goldreichtremaine,linrichardson}.
The discovery of significantly eccentric exoplanets \citep{16cygb,naef}
and the detection of misaligned and even retrograde planetary orbits \citep{winn,triaud}
revived ideas about the secular evolution of inclined systems \citep{kozai,eggleton,fabrycky},
as well as dynamical interactions resulting in scattering during the formation process \citep{rasioscatter}.
The discovery of dynamically packed multiplanet systems \citep{lovis,kepler11}
reminds us of  
the age-old quest to understand the stability of the Solar System \citep{laskar96}, and indeed as Poincar\'e well
appreciated \citep{poincare}, the quest to understand stability in the ``simpler'' {\it three}-body problem
\citep{wisdom,robutel,cambody,stability}.

Large surveys have yielded a rich harvest of planets by now \citep{udry}, enabling a comparison 
with the results of Monte Carlo-type simulations of the planet-formation process \citep{mordasini}.
It is now becoming clear that the planet mass distribution continues to rise towards lower masses
with a hint of a deficit at $30 M_\oplus$ \citep{mayor}, and with almost no objects
in the range $25-45 M_J$ at the high-mass end of the distribution \citep{sahlmann}.
The latter is the so-called {\it brown dwarf desert}
and it supports the idea that at least two distinct mechanisms operate for the formation of planets and stars \citep{udry}.

Until very recently, the notion that planets might form in a circumbinary disk was purely 
theoretical,\footnote{Unless one considers the HD 202060 system to be in this category;
it is composed of an inner pair with masses
$1.13M_\odot$ and $0.017M_\odot$, and an outer planet with mass $2.44M_J$ \citep{correia05}.}
with no clear
concensus on whether or not the strong fluctuating gravitational field of the binary pair would prevent
their formation \citep{meschiari12a,meschiari12b,paardekooper,pz}.
The Kepler survey has revealed that Nature does, indeed, accomplish this feat 
\citep{kepler16,kepler34-35,kepler47,kepler38}, with planet-binary period ratios almost as low as they can
be stability-wise (the current range is 5.6 for Kepler 16 \citep{kepler16} to 10.4 for Kepler 34 \citep{kepler34-35}).
With our understanding of the growth of planetesimals in a relatively laminar environment still in its infancy 
\citep{meisner,okuzumi,takeuchi},
the very fact that planets exist in such
``hostile'' environments puts strong constraints on how and where planetesimals form.

At every step of the way, 
some knowledge of celestial mechanics and the dynamics of small-$N$ systems is essential.
A short list might include
using stability arguments to place limits on the masses in
a multi-planetary system observed spectrocopically \citep{mayor09}; 
including dynamical constraints in orbit-fitting algorithms \citep{lovis,laskar12};
modelling the planet-planet interaction in a resonant system to infer the presence 
of a low-mass planet \citep{rivera,correia};
using transit timing variations (TTVs) 
to infer the presence of unseen companions \citep{torres,ballard,nesvorny}  
or to estimate the masses of planets in a multi-transiting system \citep[where no spectroscopic data
is available:][]{kepler9,kepler11,steffen,fabrycky12}; 
understanding the origin of resonant and near resonant systems 
\citep[][Mardling \& Udry in preparation]{papres,wu,batyginmorbi};
deciphering the complex light curves of multi-transiting systems \citep{kepler11} and eclipsing binaries with circumbinary 
planets \citep{kepler47}; inferring the true orbit of a planet observed astrometrically \citep{mcarthur}; 
understanding the influence
of stellar and planetary tides on a system which includes a short-period planet \citep{wugoldreich,mardling07,hatp13};
inferring information about the internal structure of such a planet \citep{wugoldreich,batygin,hatp13}.
Most of these examples involve orbit-orbit interactions in systems with arbitrary mass ratios, eccentricities and inclinations,
and as such many researchers resort to expensive (time-wise)
direct integrations. 
Moreover, numerical studies offer shrouded insight into parameter dependence and physical processes.
The availability of a generalised 
and easy-to-use disturbing function therefore seems timely; this is the focus of the present paper.

\subsection{The disturbing function}\label{df}

As the interaction potential for a hierarchically arranged triple system of bodies, the disturbing function
({\it la fonction perturbatrice} of \citet{leverrier}) is developed as a Fourier series
expansion whose harmonic angles are linear combinations of the various orbital phase
and orientation angles in the
problem (except for the inclinations), 
and whose coefficients are functions of the masses, semimajor axes, eccentricities
and inclinations.
With the Solar System as the only planetary system known until 1992 \citep{wolszczan},
many expansions of the disturbing function from Laplace to \citet{leverrier} and on (see \citet{murray} for other references)
rely on the
smallness of eccentricities, inclinations and mass ratios, with ranges for
Solar System planets being $0.007-0.093$, $0^o-3.4^o$ and 
$3.2\times 10^{-7}-10^{-3}$ respectively (not including Mercury and Pluto).
Moreover, since the period ratios between adjacent pairs of planets are not very large, varying between 1.5
and 2.5 (including the mean period of the asteroids),
such expansions tend to place no restriction on the ratio of semimajor axes except
that the orbits should not cross. These include the {\it literal} expansions which are written in terms of 
{\it Laplace coefficients} \citep[][and Appendix~\ref{formulae} of this paper]{murray}.
They are often presented in somewhat long and unwieldy form (for example,
the paper by \citet{leverrier} is 84 pages long),
resulting in the reluctance of non-specialists to use them. Moreover, such presentations often obscure
the dependence of the disturbing function
on the various system parameters, making their use somewhat doubtful in
an era where Morse's law still holds.

Many elegant Hamiltonian formulations exist, for example, \citet{laskar95} which uses 
{\it Poincar\'e canonical heliocentric variables} to study the planetary three-body problem.
The Hamiltonian form is used, for example, when one wishes to exploit the integrability of the underlying
subsystems in the case that they are weakly interacting. The aim of \citet{laskar95} was to use this form
to study the stability of the planetary three-body problem \citep{robutel}
in the framework of the famous KAM theorem \citep{kolmogorov,arnold,moser}.
While the \citet{laskar95} formulation does not make any assumptions about the mass ratios, their dependence
is not explicit in the resulting expressions and it is hard to get a feel for the dependence 
on the eccentricities and inclinations. Moreover,
it has the drawback that except for the inner-most orbit,
ellipses described by the orbital elements are not tangential to the actual motion
(they are not ``osculating'') because the dominant body is used as the origin.
In contrast, Jacobi coordinates which refer the motion of a body to the centre of 
mass of the sub-system to its interior ensure that this is true. Jacobi coordinates are used here.

The study of systems with more substantial eccentricities and inclinations, for example comet orbits 
perturbed by the giant planets, or
stellar triples and quaduples etc., has largely relied on numerical integrations
and continues to do so since the discovery of significantly eccentric exoplanets. On the analytical side,
Legendre expansions in the ratio of semimajor axes 
generally use orbit averaging to produce a secular disturbing function which is suitable for the study of systems
in which resonance does not play a role. Examples include the formulations of \citet{innanen}, \citet{fordrasio},
\citet{boue} and \citet{naoz}. Each of these uses a Hamiltonian approach, each is
valid for arbitrary eccentricities and inclinations, and all but \citet{innanen} 
are valid for arbitrary mass ratios.
\citet{innanen} includes only quadrupole terms ($l=2$), 
\citet{fordrasio} and \citet{naoz} include quadrupole and octopole terms ($l=2,3$), while \citet{boue}
includes terms up to $l=5$ for inclined systems and $l=10$ for coplanar systems. 
\citet{naoz} demonstrate the importance of including the full mass dependence, especially
for systems for which significant angular momentum is transferred between the inner
and outer orbits.

While secular expansions give significant insight into the long-term evolution of arbitrary configurations,
care should be taken to determine that such an expansion is approriate for any given system.
The orbit averaging\label{secondorder}
technique effectively involves discarding all terms which depend on the fast-varying mean longitudes.
However, even if resonance doesn't play a role, that is, if no major harmonic angles are librating,
the influence of some non-secular harmonics on the secular evolution can be significant.
In effect, all harmonics in a Fourier expansion of the disturbing function force all other harmonics at some level (Mardling 2013;
hereafter Paper II in this series).
On the other hand, the idea behind the averaging technique is that this influence is only short term in nature,
and has no effect on the long-term evolution
(see Wisdom 1982 for an excellent historical and insightful discussion
of the use of the averaging principle).\footnote{The averaging principle relies on 
adiabatic invariance in the system; see \citet{landau} for an thorough discussion of this concept.}
In fact, as with all forced nonlinear oscillatory systems \citep[see, for example,][]{neyfehmook}, the characteristic frequencies
are modulated by forcing at some level, and for secularly varying triple configurations this becomes
more pronounced the closer to the stability boundary the system is.
A demonstration of this is given in \citet{giuppone} who study the secular variation of a test
particle orbiting the primary in a binary star system. In particular they demonstrate that predictions from the
usual ``first-order'' secular theory consistently underestimate the frequency and overestimate the
equilibrium value of the forced eccentricity, with relative errors as large as $80\%$ and $40\%$ respectively
when the ratio of binary to test particle semimajor axes is 10. Note that they take a binary mass ratio 
of 0.25 and a binary eccentricity of 0.36. Using a technique called Hori's averaging process \citep{hori},
they go on to calculate the ``second-order'' corrections to the secular frequency and amplitude, reducing
the errors to a few percent. While the authors do not identify the nature of the expansion parameter,
(and claim that their expressions are too complicated to write down),
in effect they are using the neglected harmonics to force the secular system, and in so doing obtain
corrected frequencies. 
In fact, Hori's averaging process is simply a version of the better-known {\it Lindstedt-Poincar\'e} method for 
correcting the frequencies of a forced nonlinear oscillator (see Neyfeh 1973 for some simple applications
including the Duffing equation). 

Here we present two new formulations
of the hierarchical three-body problem which are accessible to anyone interested in the short and long-term
evolution of small-$N$ systems, be they stellar or planetary systems or a mixture of both (for example,
circumbinary planets).
The two formulations are valid for arbitrary mass ratios, and are
distinguished by their choice of expansion parameter and hence
their range of validity in those parameters. For closely packed systems, our generalisation of 
the {\it literal} expansion\footnote{The origin of the use of the word ``literal''
in this context is unclear, but one should take it to indicate that the dependence on the ratio of semimajor axes is
via Laplace coefficients.} 
\citep[][and references therein]{leverrier,murray} with its lack of constraint on the
period ratio (except that the orbits should not cross) and its use of the eccentricities
as expansion parameters is appropriate, while more widely spaced systems are best studied with 
the {\it spherical harmonic} expansion, a generalisation of the work of \citet{kaula}
\citep[also see][]{murray},
which exploits the properties of spherical harmonics and which is valid for all 
eccentricities. Inclined systems will be studied in Paper III in this series.

Throughout the paper we refer to ``moderate mass ratio systems''. Our formal definition of such a system
is one whose stability characteristics are govered by the interaction of $N\!:\!1$ resonances 
(Paper II).
In practice, however, this corresponds roughly to systems for which {\it both} mass ratios $m_2/(m_1+m_2)$
and $m_3/(m_1+m_2+m_3)$ are greater than around 0.05, where $m_1\ge m_2$.

The paper is arranged as follows:

\vspace{2mm}

\noindent{\bf\ref{sh}. Spherical harmonic expansion \hfill \pageref{sh}}\\
\ref{derSH}. Derivation
\ref{dom}. Practical application: dominant terms.
\ref{secres}. Resonance widths and stability.
\ref{3.1.2}. Libration frequency.
\ref{sec1}. The secular disturbing function in the spherical harmonic expansion.

\vspace{2mm}

\noindent{\bf\ref{4}. Literal expansion \hfill \pageref{4}}\\
\ref{derL}. Derivation
\ref{4.1}. Eccentricity dependence.
\ref{4.1.1}. Power series representations of Hansen coefficients and the choice of expansion order.
\ref{4.2}. Dependence on the mass and semimajor axis ratios.
\ref{4.2.1}. Summary of leading terms in $\alpha$.
\ref{4.2.2}. Coefficients when $m_2/m_1\rightarrow 0$.
\ref{4.3}. The spherical harmonic order $m$ and principal resonances.
\ref{zee}. ``Zeeman splitting'' of resonances.
\ref{53}. A second-order resonance.
\ref{4.5}. First-order resonances.
\ref{4.6}. Resonance widths using the literal expansion.
\ref{libfreq} Libration frequency.
\ref{4.6.1}. Widths of first-order resonances.
\ref{sec2}. The secular disturbing function in the literal expansion.

\vspace{2mm}

\noindent{\bf\ref{acc}. Comparison of formulations to leading order in eccentricities \hfill \pageref{acc}}\\

\vspace{-2mm}

\noindent{\bf\ref{equiv}. Equivalence of formulations \hfill \pageref{equiv}}\\

\vspace{-2mm}

\noindent{\bf\ref{compclass}. Comparisons with classic expansions \hfill \pageref{compclass}}\\

\vspace{-2mm}

\noindent{\bf\ref{6}. Conclusion and highlights of new results \hfill \pageref{6}}\\

\vspace{-2mm}

\noindent{\bf \ref{quick}. Quick Reference \hfill \pageref{quick}}\\
\ref{2.1}. Harmonic coefficients for the semimajor axis expansion.
\ref{2.1.2}. Secular disturbing function to octopole order. 
\ref{domnon}. Dominant non-secular terms.
\ref{2.1.1}. Widths and libration frequencies of $[N\!:\!1](2)$ resonances. 
\ref{2.2}. Harmonic coefficients for the eccentricity expansion. 
\ref{sdf}. The secular disturbing function to second order in the eccentricities.
\ref{rw}. Widths and libration frequencies of $[n'\!:\!n](m)$ resonances.

\vspace{2mm}

\noindent{\bf{Appendices} \hfill \pageref{shapp}}\\
\ref{shapp}. Spherical harmonics.
\ref{hansen}. Hansen coefficients.
\ref{formulae}. Laplace coefficients.
\ref{lagrange}. Lagrange's planetary equations for the variation of the elements.
\ref{ml}. The mean longitude at epoch.
\ref{not}. Notation.

\section{Spherical harmonic expansion}\label{sh}

\subsection{Derivation}\label{derSH}
\begin{figure}
\centering
\includegraphics[width=120mm]{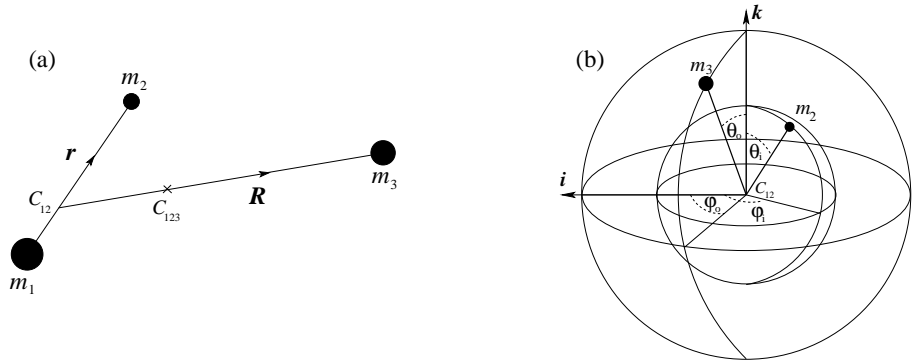}
\caption{(a): Jacobi coordinates $\br$ and $\bR$.  $C_{12}$ and $C_{123}$ refer to the centre of mass of bodies
1 and 2, and of the whole system respectively.
A hierarchical triple behaves like two weakly interacting binaries, with the inner binary
composed of bodies 1 and 2, and the outer binary composed of body 3 plus a body of mass $m_1+m_2$
situated at $C_{12}$. Bodies 1 and 2 are labelled such that $m_2\le m_1$.
(b): Spherical polar angles associated with $\br$ ($\theta_i,\varphi_i$) and
$\bR$ ($\theta_o,\varphi_o$). The origin corresponds to the centre of mass of
bodies 1 and 2.}
\label{hier}
\end{figure}

Both expansions presented in this paper make use of three-body {\it Jacobi} or {\it hierarchical} coordinates
and their associated osculating orbital elements
to describe the dynamics of the system \citep[eg.][]{murray}. Illustrated in Figure~\ref{hier}(a), %
these are used for
systems for which the motion of two of the bodies is predominantly Keplerian about their common centre of mass (the ``inner orbit''),
with the third body executing predominantly Keplerian motion about the centre of mass of the inner pair (the ``outer orbit'').
Note that the word ``hierarchical'' need not imply that
the orbits are necessarily well-spaced, but rather, that the orbits retain their identities for at least several outer periastron passages,
although the osculating orbital elements may vary dramatically from orbit to orbit if the system is unstable. With this broad definition,
even systems involving the exchange of the outer body with one of the inner pair can be considered as hierarchical.
Note also that one of the many advantages of using these coordinates is that the osculating semimajor axes are constant on average when
resonance does not play a role. When resonance does play a role, it is then easy to define the energy exchanged between the orbits.

Using Jacobi coordinates, the equations of motion for the inner and outer orbits are
\be
\mu_i\ddot{\bf r}+\frac{Gm_1m_2}{r^2}\hat{\bf r}=\pp{\cal R}{\bf r}
\label{inner}
\ee
and
\be
\mu_o\ddot{\bf R}+\frac{Gm_{12}m_3}{R^2}\hat{\bf R}=\pp{\cal R}{\bf R}
\label{outer}
\ee
where ${\bf r}$ is the position of body 2 relative to body 1, \label{p18}
${\bf R}$ is the position of body 3 relative to the centre of mass of 
the inner pair,
$m_1$ and $m_2$ are the masses of the bodies forming the inner orbit, 
$m_3$ is the mass of the outer body, $m_{12}=m_1+m_2$ \label{p1}, $\mu_i=m_1m_2/m_{12}$ \label{p5}
and $\mu_o=m_{12}m_3/m_{123}$ \label{p6}
are the inner and outer reduced masses with $m_{123}=m_1+m_2+m_3$, \label{p2}
$\hat{\bf r}$ and $\hat{\bf R}$ are unit vectors in the ${\bf r}$ and ${\bf R}$ directions respectively, 
$r=|{\bf r}|$, $R=|{\bf R}|$
and 
\be
{\cal R}=-\frac{Gm_{12}m_3}{R}+\frac{Gm_2m_3}{|{\bf R}-\beta_1{\bf r}|}
+\frac{Gm_1m_3}{|{\bf R}-\beta_2{\bf r}|}
\label{s1}
\ee
is the {\it disturbing function} or {\it interaction energy}, \label{p19}
with $\beta_1=m_1/m_{12}$ \label{p3}
and \label{p4}
$\beta_2=-m_2/m_{12}$.\footnote{Note that the use of Jacobi coordinates 
and the fact that the disturbing function has the
units of energy and not energy per unit mass means that it is only necessary to define a 
single disturbing function.
In constrast, in the classic expansions one distinguishes between an ``internal'' and ``external'' disturbing function, 
each composed of a ``direct'' component as well as individual ``indirect'' components
\citep{murray}.
Similarly, there is no distinction between ``internal'' and ``external'' resonances when the single general
disturbing function is used.}
In general we will use the subscripts $i$ and $o$ to represent quantities associated with the inner and
outer orbits respectively.
The notation $\partial/\partial{\bf r}$ refers to the {\it gradient} with respect to the spherical polar coordinates $(\beta_1 r,\theta_i,\varphi_i)$
associated with the position of body 2 relative to the centre of mass of bodies 1 and 2,
and similarly for $\partial/\partial{\bf R}$ for the position of body 3 with
spherical polar coordinates $(R,\theta_o,\varphi_o)$ relative to the same origin; see Figure~\ref{hier}(b).
The equations of motion \rn{inner} and \rn{outer} written in this form clearly demonstrate the perturbed
Keplerian nature of a hierarchical triple system when ${\cal R}$ and its gradients are 
small,\footnote{Relative to the binding energy of the individual orbits and the other terms in \rn{inner} and \rn{outer}
respectively.}
which is clearly the case when $r\ll R$. Note that the total energy is given by
\be
E=E_i+E_o-{\cal R},
\label{Etot}
\ee
where 
\be
E_i=\ff{1}{2}\mu_i\dot{\bf r}\cdot\dot{\bf r}-\frac{Gm_1m_2}{r}
\hand
E_o=\ff{1}{2}\mu_o\dot{\bf R}\cdot\dot{\bf R}-\frac{Gm_{12}m_3}{R}
\ee
are the instantaneous binding energies of the inner and outer orbits respectively.

A common way to expand terms of the form of the last two in \rn{s1} is in terms of 
Legendre polynomials. In this case we have for $s=1,2$
\be
\frac{1}{|{\bf R}-\beta_s{\bf r}|}=\frac{1}{R}\sum_{l=0}^\infty \left(\frac{\beta_s r}{R}\right)^l\,P_l(\cos\psi),
\label{s2}
\ee
where $\cos\psi=\hat{\bf r}\cdot\hat{\bf R}$ 
and $P_l$ is a Legendre polynomial.
A disadvantage of this is that the angles associated with each individual orbit do not appear explicitly.
The use of spherical harmonics overcomes this problem as follows.\footnote{The advantage of using spherical
harmonics becomes particularly apparent when inclined systems are considered; see Paper III of this series.}
Using the {\it addition theorem} for spherical harmonics \citep[eg.][]{jackson}
one can write
\be
P_l(\cos\psi)=\sum_{m=-l}^l\frac{4\pi}{2l+1}\,Y_{lm}(\theta_i,\phi_i)Y_{lm}^*(\theta_o,\varphi_o),
\label{s3}
\ee
where
$Y_{lm}$ is a spherical harmonic of {\it degree} $l$ and {\it order} $m$ \label{p27}
with $Y^*_{lm}$ its complex conjugate.
Using \rn{s2} and \rn{s3}, the disturbing function \rn{s1} becomes
\be
{\cal R}=G\mu_im_3\sum_{l=2}^\infty\sum_{m=-l}^l 
\left(\frac{4\pi}{2l+1}\right){\cal M}_l 
\left(\frac{r^l}{R^{l+1}}\right)Y_{lm}(\theta_i,\varphi_i)Y_{lm}^\ast(\theta_o,\varphi_o),
\ee
where the mass factor ${\cal M}_l$ \label{p4b}
is given by
\be
{\cal M}_l=\frac{m_1^{l-1}+(-1)^l m_2^{l-1}}{m_{12}^{l-1}}=\beta_1^{l-1}-\beta_2^{l-1}.
\label{Ml}
\ee
Note that ${\cal M}_2=1$ for any masses, while for equal masses 
${\cal M}_l=0$ when $l$ is odd.
In this paper we focus on {\it  coplanar} systems
for which we take $\theta_i=\theta_o=\pi/2$, $\varphi_i=f_i+\varpi_i$ and
$\varphi_o=f_o+\varpi_o$, where $f_i$ and $f_o$ \label{p7}
are the true anomolies of the inner and
outer orbits respectively with $\varpi_i$ and $\varpi_o$ \label{p10}
the corresponding longitudes of periastron. 
With the definition of the spherical harmonic used in
\citet{jackson}, 
\be
Y_{lm}(\theta,\varphi)=\sqrt{\frac{2l+1}{4\pi}\frac{(l-m)!}{(l+m)!}}
P_l^m(\cos\theta)\,e^{im\varphi},
\ee
where $P_l^m$ is an associated Legendre function,
the disturbing function becomes
\be  
{\cal R}=G\mu_i m_3
\sum_{l=2}^\infty\sum_{m=-l,2}^l
\ff{1}{2}c_{lm}^2{\cal M}_l\,
{\rm e}^{im(\varpi_i-\varpi_o)}
\left(r^l\,{\rm e}^{imf_i}\right)
\left(\frac{{\rm e}^{-imf_o}}{R^{l+1}}\right),
\label{calR2}
\ee
where \label{pclm}
\be
c_{lm}^2=\frac{8\pi}{2l+1}\left[Y_{lm}(\pi/2,0)\right]^2=c_{l\,-m}^2.
\label{clm}
\ee
A closed-form expression for these constants is given in Appendix~\ref{shapp} with specific values
given in Table~\ref{eccfuns}.
Note that the sum over $m$ is in steps of two because $Y_{lm}(\theta,\varphi)\propto\cos\theta$
when $l-m$ is odd so that $Y_{lm}(\pi/2,\varphi)=0$ in this case.

For stable systems including those near the stability boundary,
the expressions in the last two pairs of brackets in \rn{calR2} associated with the
inner and outer orbits are nearly periodic in their respective orbital
periods, and therefore can 
be expressed in terms of Fourier series
of the inner and outer mean anomalies, $M_i=\nu_i t+M_i(0)$ and $M_o=\nu_o t+M_o(0)$ respectively, \label{p8}
with $M_i(0)$ and $M_o(0)$ their values at $t=0$ and
$\nu_i$ and $\nu_o$ \label{p16}
the associated orbital frequencies (mean motions).
Thus
\be
r^l {\rm e}^{imf_i}=a_i^l\left(\frac{1-e_i^2}{1+e_i\cos f_i}\right)^l {\rm e}^{imf_i}=
a_i^l\sum_{n=-\infty}^\infty X_n^{l,m}(e_i){\rm e}^{inM_i}
\ee
and
\be
\frac{{\rm e}^{-imf_o}}{R^{l+1}}=a_o^{-(l+1)}\left(\frac{1-e_o^2}{1+e_o\cos f_o}\right)^{-(l+1)} {\rm e}^{-imf_o}=
a_o^{-(l+1)}
\sum_{n'=-\infty}^\infty X_{n'}^{-(l+1),m}(e_o){\rm e}^{-in'M_o},
\label{fourier}
\ee
where 
$a_i$ and $a_o$ are the inner and outer semimajor axes, $e_i$ and $e_o$ \label{p12}
are the corresponding eccentricities,
the Fourier coefficients \label{p28}
\be
X_{n}^{l,m}(e_i)=\frac{1}{2\pi}\int_0^{2\pi}(r/a_i)^l\,{\rm e}^{imf_i}
{\rm e}^{-inM_i}dM_i= {\cal O}(e_i^{|m-n|})
\label{s4}
\ee
and
\be
X_{n'}^{-(l+1),m}(e_o)=\frac{1}{2\pi}\int_0^{2\pi}\frac{{\rm e}^{-imf_o}}{(R/a_o)^{l+1}}{\rm e}^{in'M_o}dM_o
= {\cal O}(e_o^{|m-n'|})
\label{s5}
\ee
are called Hansen coefficients \citep{hughes} and we have indicated the order of the
leading terms (see Appendix~\ref{hansen} for graphical representations, closed-form expressions 
and approximations, {\it Mathematica} programs and 
in particular, Appendix~\ref{expan} for general expansions which demonstrate the form of the leading terms).
Note that since the real part of the integrands of \rn{s4} and \rn{s5} are even and the imaginary
parts are odd, the integrals are real so\label{pf7} that\footnote{Recall that even functions have the property that $\int_{-a}^af(x)dx=2\int_0^af(x)dx$
while odd functions have the property that $\int_{-a}^af(x)dx=0$.}
\be
\left[X_{n}^{l,m}\right]^*=X_{-n}^{l,-m}=X_n^{l,m}
\label{X1}
\ee
and
\be
\left[X_{n'}^{-(l+1),m}\right]^*=X_{-n'}^{-(l+1),-m}=X_{n'}^{-(l+1),m},
\label{X2}
\ee
justifying the notation used in \rn{s5}.
The disturbing function \rn{calR2} can then be expressed as
\bea
{\cal R}&=&\frac{G\mu_i m_3}{a_o}
\sum_{l=2}^\infty\sum_{m=-l,2}^l\sum_{n=-\infty}^\infty
\sum_{n'=-\infty}^\infty
\ff{1}{2}c_{lm}^2{\cal M}_l\,
\alpha^l\,
X_{n}^{l,m}(e_i)X_{n'}^{-(l+1),m}(e_o)\,
{\rm e}^{i\phi_{mnn'}}
\label{distR3a}\\
&=&\frac{G\mu_i m_3}{a_o}
\sum_{l=2}^\infty\sum_{m=m_{min},2}^l\sum_{n=-\infty}^\infty
\sum_{n'=-\infty}^\infty
\zeta_m c_{lm}^2\,
{\cal M}_l\,
\alpha^l\,
X_{n}^{l,m}(e_i)X_{n'}^{-(l+1),m}(e_o)\,
\cos\phi_{mnn'}
\label{distR3}
\eea
where $\alpha=a_i/a_o$, \label{p13}
\be
\phi_{mnn'}=nM_i-n'M_o+m(\varpi_i-\varpi_o)
\label{phi0}
\ee
is a {\it harmonic angle}, \label{p31}
\be
\zeta_m=\left\{\begin{array}{ll}
1/2, & m=0\\
1, &{\rm otherwise}
\end{array}
\right.
\hand
m_{min}=\left\{\begin{array}{ll}
0, & l\,\,{\rm even}\\
1, & l\,\,{\rm odd}.
\end{array}
\right.
\label{zetam}
\ee
In going from \rn{distR3a} to \rn{distR3} we have used the properties \rn{X1} and \rn{X2}
and have grouped together terms with the same value of $|m|$ (thus the
factor 1/2 in the definition of $\zeta_m$).
Writing the harmonic angle in terms of longitudes only (in anticipation of employing Lagrange's
planetary equations for the rates of change of the elements), \rn{phi0} becomes
\be
\phi_{mnn'}=n\lambda_i-n'\lambda_o+(m-n)\varpi_i-(m-n')\varpi_o,
\label{phi}
\ee
where $\lambda_i=M_i+\varpi_i$ and $\lambda_o=M_o+\varpi_o$ \label{p9}
are the inner and outer mean longitudes respectively.
Note that the harmonic angle should be invariant to a rotation of
the coordinate axes. Since such a rotation changes all longitudes by
the same amount, their coefficients should add up to zero thereby
satisfying the {\it d'Alembert relation} \citep{murray}, which indeed \rn{phi} does.

Expression \rn{distR3} for the disturbing function 
may be compared with that derived by \citet{kaula}
for the case where $m_2$ is a test particle
(see also \citet{murray}, p232). Note that the Kaula expression is valid for arbitrary inclinations;
this case will be considered Paper III in this series.

Defining the coefficient of $\cos\phi_{mnn'}$ as ${\cal R}_{mnn'}$, it is desirable to change
the order of summation of $l$ and $m$ so that $m$ no longer depends on $l$ (ie., it becomes a
free index independent of any other index).
The simplest way to see how this works is to write
out the first few terms, grouping them appropriately. Thus
\bea
\sum_{l=2}^\infty\sum_{m=m_{min},2}^l T_{lm}
&=&[T_{20}+T_{22}]+[T_{31}+T_{33}]+[T_{40}+T_{42}+T_{44}]+[T_{51}+T_{53}+T_{55}]+\ldots\next
&=&
[T_{20}+T_{40}+\ldots]
+[T_{31}+T_{51}+\ldots]
+[T_{22}+T_{42}+\ldots]
+[T_{33}+T_{53}+\ldots]+\ldots\next
&=&
\sum_{m=0}^\infty\sum_{l=l_{min},2}^\infty T_{lm},
\label{32}
\eea
where \label{p32}
\be
l_{min}=\left\{\begin{array}{ll}
2, & m=0\\
3, & m=1\\
m, &m\ge 2
\end{array}
\right.
\label{s7b}
\ee
so that the disturbing function \rn{distR3} becomes
\be
{\cal R}=\sum_{m=0}^\infty \sum_{n=-\infty}^\infty \sum_{n'=-\infty}^\infty\,{\cal R}_{mnn'}\,\cos\phi_{mnn'}
\label{s6}
\ee
with \label{p21}
\bea
{\cal R}_{mnn'}&=&\frac{G\mu_i m_3}{a_o}\sum_{l=l_{min},2}^\infty
\zeta_m c_{lm}^2\,
{\cal M}_l\,
\alpha^l\,
X_{n}^{l,m}(e_i)X_{n'}^{-(l+1),m}(e_o)\label{s7}\\
&=&\frac{G\mu_i m_3}{R_p}\sum_{l=l_{min},2}^\infty
\zeta_m c_{lm}^2\,
{\cal M}_l\,
\rho^l\,
X_{n}^{l,m}(e_i)Z_{n'}^{-(l+1),m}(e_o).
\label{rhoform}
\eea
Here $\rho=a_i/R_p$ with $R_p=a_o(1-e_o)$ the outer periastron distance, and we will refer to \label{p29}
\be
Z_{n'}^{-(l+1),m}(e_o)=(1-e_o)^{l+1}X_{n'}^{-(l+1),m}(e_o)
\ee
as a {\it modified} Hansen coefficient.
The form \rn{rhoform} is especially useful for systems with high outer eccentricity since $X_{n'}^{-(l+1),m}(e_o)$
is singular at $e_o=1$ while $Z_{n'}^{-(l+1),m}(e_o)$ is not.
Note that the summation over $l$ in \rn{s7} is in steps of 2. Moreover, note that there are only 
{\it three} independent indices associated with each harmonic for a coplanar system, although usually
an additional one is included erroneously (see discussion in Section~\ref{compclass}).
This makes sense because there are three independent frequencies in the problem, that is,
the two orbital frequencies and the rate of change of the relative orientation of the orbits.
We will refer to the quantity ${\cal R}_{mnn'}$ as the {\it harmonic coefficient} associated
with the harmonic angle $\phi_{mnn'}$.
Moreover, the classical 
nomenclature for terms associated with $l=2$ and $l=3$ is ``quadrupole'' and ``octopole'' (or
``octupole'') respectively.

\subsection{Practical application: dominant terms}\label{dom}

\begin{figure}
\centering
\includegraphics[width=150mm]{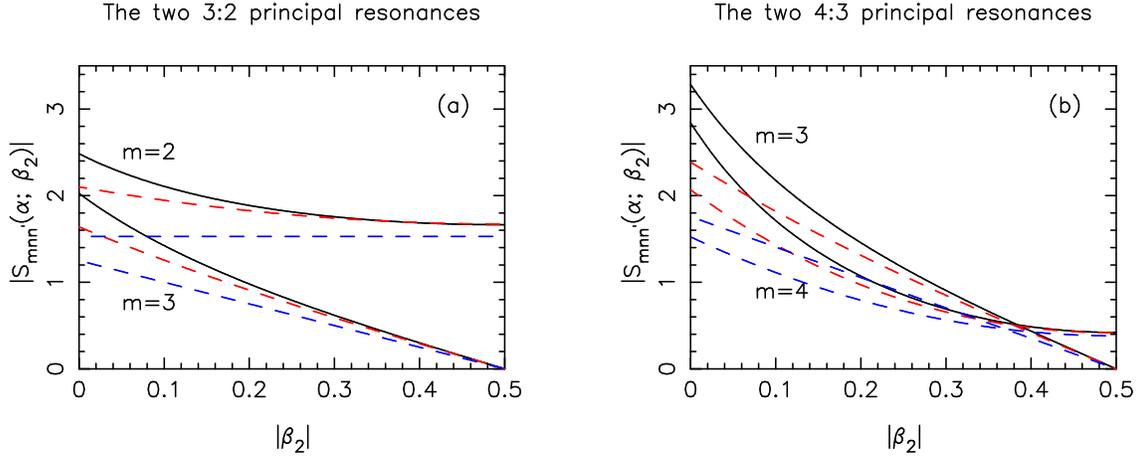}
\caption{Similar to Figure~\ref{PR32} for two first-order harmonics. Two values of $l$ are inadequate
for such close systems when $|\beta_2|=m_2/m_{12}$ is small. Here (a): $\alpha=0.76$ and (b): $\alpha=0.83$.
}
\label{PR32}
\end{figure}

The spherical harmonic expansion is significantly simpler to use than the literal expansion
when the accuracy required can be achieved with only one or two values of $l$,
and hence is recommended for use in preference to the latter except for the very closest systems
(period-ratio-wise).
In Section~\ref{acc} we compare the two expansions to leading order in the eccentricities with the aim of determining
the minimum period ratio for which the spherical harmonic expansion is acceptably accurate when only
the two lowest values of $l$ are included. Figures~\ref{PR21} and \ref{PR32} suggest that
this minimum is around 2 (panel (a) of Figure~\ref{PR21}), although the expansion is still reasonably accurate for 
a period ratio as low as 1.5 (panel (a) of Figure~\ref{PR32}).

The question then arises: which harmonics in the triple-infinite series \rn{s6} should one include
for a given application? How does one know whether or not resonant harmonics play a role?
If they don't, is it only necessary to include the {\it secular} terms in \rn{s6}, that is, terms 
which do not depend on the mean longitudes (those with
$n=n'=0$; see Section~\ref{sec1})? What about non-resonant, non-secular harmonics?
A few general comments can be offered here, however, in general the answers depend on the questions
being asked, on the timescales of interest and of course on the configuration itself.

Timescales on which point-mass three-body systems which are coplanar and non-orbit crossing vary
can generally be arranged according to the following hierarchy:
\be
P_i<T_p\le P_o<P_{lib}<P_{sec}\gtrless \tau_{stab}
\ee
where $P_i$ and $P_o$ \label{p15}
are the inner and outer orbital periods, $T_p\equiv (1-e_o)^{3/2}P_o$ is the ``time of periastron passage''
of the outer body, a timescale of interest when the outer orbit is significantly eccentric and the orbit-orbit interaction is
effective only around outer periastron,
$P_{lib}=2\pi/\omega_{lib}$ is the libration period in the case that the system is in (or near) resonance, with $\omega_{lib}$
given by \rn{libw}, $P_{sec}$ is the period on which
the eccentricities vary secularly, and $\tau_{stab}$ is the dynamical stability timescale
in the case that the system is unstable to the escape of one of the bodies (Lagrange instability).
If one is interested in studying short-period variations on timescales up to a few times $P_o$, non-secular harmonics whose coefficients are
zeroth and/or first order in $e_i$ are generally included, independent of the value of the inner eccentricity.
However, the selection from amongst such harmonics depends on the value of the outer eccentricity, and 
these are not necessarily those which are low-order in $e_o$ except when $e_o$ is small.
Inspection of Figure~\ref{flmfig} shows that for significant values of $e_o$, harmonics spanning a wide range
of values of $n'$  have similar amplitudes so that in principle, many harmonics should be included in 
such cases. One can avoid this by using {\it overlap integrals}, one for each value of $n$ (and $l$ and $m$);
this technique will be discussed in a future paper in this series.

While the eccentricities and semimajor axes of
resonant or near resonant systems vary on the timescale of the orbital periods, these variations tend
to accumulate on the libration timescale and it is the resonant harmonics which govern the behaviour. For stable systems
with significant outer eccentricities
it is usually adequate to include only one term in the analysis, that term being the $[N\!:\!1](2)$ harmonic with 
$N\simeq\sigma\equiv\nu_i/\nu_o$ \label{p17}
as discussed in Section~\ref{secres} and Paper II. Here and later the notation $[n'\!:\!n](m)$ refers to the harmonic term
associated with the angle $\phi_{mnn'}$ and coefficient ${\cal R}_{mnn'}$ (see Section~\ref{4.3}).

Unstable coplanar systems are also governed by resonant harmonics, but it is their interaction with ``neighbouring''
non-resonant terms which result in the chaotic behviour of the system. In this case it is usually sufficient to include only
the resonant harmonic $[N\!:\!1](2)$ and its neighbour $[N+1\!:\!1](2)$ (Section~\ref{secres}), although one often needs to take into account
the forced and secular variation of the eccentricities (Paper II).

For stable systems one is often interested in the long-term secular variation of the elements, in which case it is 
usually sufficient only to include the secular $[0\!:\!0](m)$ harmonics, that is, those which do not depend on 
the mean longitudes. In addition, it is normally only necessary to include the first two of these, that is, $m=0$
and $m=1$. However, for stable systems which are relatively close period-ratio-wise, it may be necessary to
include the forcing effect of some non-secular harmonics; this is discussed in Section~\ref{df}.

On the subject of the secular variation of the elements, it is worth mentioning here that while the purely secular
coplanar three-body system governed by the single angle $\phi_{100}=\varpi_i-\varpi_o$ is integrable and therefore not
admitting of chaotic solutions, {\it non-coplanar} secular three-body systems as well as coplanar (and non-coplanar)
higher-order (four-body etc) systems are governed by two or more independent angles and hence
do admit chaotic solutions (see, for example, Laskar 1988 for a numerical study of the long-term secular evolution of the Solar System).

\subsection{Resonance widths and stability}\label{secres}
Amongst systems with moderate mass ratios, stable systems tend to have significant period ratios because
strong mutual interactions between the bodies tend to destabilize closer systems.
The stability of a system can be studied using the heuristic
resonance overlap stability criterion which involves calculating the widths of resonances 
(Chirikov 1979; Wisdom 1980; Mardling 2008; Paper II).
For systems with moderate mass ratios, it is the $[n'\!:\!1](2)$ resonances which
govern the exchange of energy between the orbits and hence it is these which are responsible 
for the stability of the system. 
In this section we summarize the derivation of a simple expression for the widths of these resonances,
the full derivation of which can be found in Paper II where a thorough 
study of resonance and stability in hierarchical systems with moderate mass ratios
is presented.

The general harmonic angle has the form
$\phi_{mnn'}=n\lambda_i-n'\lambda_o+(m-n)\varpi_i-(m-n')\varpi_o$,
and unless $n'/n$ is sufficiently close to the period ratio $\nu_i/\nu_o$, this angle will
{\it circulate}, that is, it will pass through
all values $[0,2\pi]$ because there is no commensurability between the rates of change of the individual
angles making up $\phi_{mnn'}$. 
In fact, were there no (nonlinear) coupling between the inner and outer orbits, $\phi_{mnn'}$ would
circulate no matter how close $n'/n$ was to $\nu_i/\nu_o$ (except if $\dot\phi_{mnn'}=0$
precisely). But because the orbits are able to exchange
energy, the period ratio changes slightly each outer orbit allowing for the possibility of {\it libration} of $\phi_{mnn'}$
when the energy is coherently transferred (ie, when conjunction occurs at almost the same place
in the orbit; see, for example, \citet{peale} for a general discussion).
In that case, $\phi_{mnn'}$ will
oscillate between two values such that 
$\oint \cos\phi_{mnn'}d\phi_{mnn'}\ne 0$, where the integral is taken over one libration cycle.
Then we refer to the harmonic angle in question as a {\it resonance angle} and say that the system is
{\it in resonance}.\footnote{Note \label{dellask} that it is possible for the harmonic angle to librate and for the system
to be not in resonance; formally the latter requires the existence of a {\it hyperbolic point} in the phase
space $(\dot\phi_{mnn'},\phi_{mnn'})$ and this may not be the case when the eccentricities are very 
small \citep{delisle} which is never the case when at least one of the mass ratios is significant (except 
when the period ratio is very large).}
The period of libration may be tens to hundreds or even thousands 
of outer orbital periods, depending on the system parameters;
an expression for this is given below in terms
of ``distance'' from exact commensurability in dimensionless units of period ratio. 

In order to study resonant behaviour, we use the {\it pendulum model for resonance} \citep[see, for example,][]{chirikov,wisdom,murray} which involves deriving a pendulum-like differential equation for $\phi_{mnn'}$.\footnote{Another model used to study resonance is 
{\it the second fundamental model of resonance} of \citet{henrard}. The associated Hamiltonian was designed specifically 
for the study of {\it resonance capture}, although it is possible to study this phenomenon 
without the Hamiltonian formalism using the pendulum model (Mardling \& Udry in preparation).}
To do this, we need to take into account the 
the dependence of the orbital frequency on time. Following \citet{brouwer} p285,
the mean longitude is defined in terms of the orbital frequency such that
\be
\lambda_i=M_i+\varpi_i=\int_{T_0}^t \nu_i(t')\,dt'+M_i(T_0)+\varpi_i=\int_{T_0}^t \nu_i(t')\,dt'+\epsilon_i,
\ee
where $\epsilon_i$ \label{p11}
is the mean longitude at epoch $t=T_0$.\footnote{See Appendix~\ref{lagrange}
for a discussion of the this orbital element.}
With a similar expression for $\lambda_o$,
the rate of change of a harmonic angle is then
\be
\dot\phi_{mnn'}=n\nu_i-n'\nu_o+n\dot\epsilon_i-n'\dot\epsilon_o+
(m-n)\dot\varpi_i-(m-n')\dot\varpi_o.
\ee
Except for systems with very small eccentricities, we have in general that
$\dot\varpi_i\ll\nu_o$ and $\dot\varpi_o\ll\nu_o$ (see Section~\ref{zee} for an example which
illustrates this). Moreover, for all systems, $\dot\epsilon_i\ll\nu_o$ and $\dot\epsilon_o\ll\nu_o$.
Since some eccentricity is always induced\footnote{An expression for the induced eccentricity is given in Paper II.}
and this is only small when the mass ratios are very small,
for systems with moderate mass ratios it is a reasonable approximation to take
\be
\dot\phi_{mnn'}\simeq n\nu_i-n'\nu_o.
\ee
Now consider the $[n'\!:\!n](m)=[N\!:\!1](2)$ harmonic,
where $N$ is an integer close to the
period ratio $\nu_i/\nu_o$.
Libration of the angle $\phi_{21N}\equiv\phi_N$ will occur when \label{p24}
\be
\dot\phi_{N}=\nu_i-N\nu_o\simeq 0.
\label{dotphi}
\ee
Thus one can ask: how close to exact commensurability should the system be for this angle to librate?
This is equivalent to asking for the {\it width} of the resonance. We can get a good 
answer to this question by showing that $\phi_{N}$ satisfies approximately a pendulum equation of the form
\be
\ddot\phi_{N}=-\omega_N^2\sin\phi_{N},
\ee
where $\omega_N^2$ depends on the parameters of the system. \label{p26}
Note that the $[N\!:\!1](2)$ resonance librates about $\phi_N=0$ as we show below.
Once $\omega_N^2$ is known, 
the range of values of $\dot\phi_N$ for which $\phi_N$ librates is determined from the equation
for the pendulum separatrix, that is,
\be
\dot\phi_N=\pm2\,\omega_N\cos\left(\frac{\phi_N}{2}\right),
\ee
so that libration occurs if $\dot\phi_N<2\,\omega_N$ when $\phi_N=0$. In order to determine $\omega_N$,
we start by writing
\be
\ddot\phi_N=\dot\nu_i-N\dot\nu_o
=\nu_o\left(\frac{\nu_i}{\nu_o}\frac{\dot\nu_i}{\nu_i}-N\frac{\dot\nu_o}{\nu_o}\right)
=-\frac{3}{2}\nu_o\left(\sigma\frac{\dot a_i}{a_i}-N\frac{\dot a_o}{a_o}\right),
\label{phidd}
\ee
where $\sigma=\nu_i/\nu_o$ is the period ratio, and we have used Kepler's
third law to replace $\dot\nu_i/\nu_i$ by $-\ff{3}{2}\dot a_i/a_i$ and similarly for the outer orbit.
The rates of change of the semimajor axes are given by Lagrange's planetary equation \rn{dadt}.
For the latter, consider
a reduced disturbing function which contains only the $[n'\!:\!1](2)$ harmonics
truncated at $l=2$;
one might call this the quadrupole contribution to the disturbing function, although it does not contain
terms with $m=0$ or $n\ne 1$. Referring to this as ${\cal R}_q$, we have from \rn{s6} and \rn{s7} that
\be
{\cal R}_q=
\frac{3}{4}\frac{G\mu_i m_3}{a_o}
\alpha^2X_{1}^{2,2}(e_i)
\sum_{n'=-\infty}^\infty
X_{n'}^{-3,2}(e_o)\cos\phi_{n'},
\label{Rq}
\ee
where 
\be
\phi_{n'}=\lambda_i-n'\lambda_o+\varpi_i+(n'-2)\varpi_o
\label{phin}
\ee
and we have put $c_{22}=3/4$ and ${\cal M}_2=1$.
Now suppose that the harmonic angle with $n'=N$ librates and for now, assume that
it is unaffected by all the harmonics contributing to ${\cal R}_q$ except itself.
Retaining only the $[N:1](2)$ harmonic in \rn{Rq}, 
\rn{phidd} together with Lagrange's planetary equation \rn{dadt} gives
\be
\ddot\phi_N=-\omega_N^2\sin\phi_N=
\frac{9}{4}\nu_o^2\left[\left(\frac{m_3}{m_{123}}\right)+N^{2/3}
\left(\frac{m_{12}}{m_{123}}\right)^{2/3}\left(\frac{m_1m_2}{m_{12}^2}\right)\right]
X_{1}^{2,2}(e_i)X_{N}^{-3,2}(e_o)\sin\phi_{N},
\label{w}
\ee
where we have replaced $\sigma$ by $N$ and used Kepler's third law
to replace $\alpha$ by $(m_{12}/m_{123})^{1/3}N^{-2/3}$.
If we further replace the Hansen coefficients by the approximations given in Table~\ref{hansentab} and \rn{asymX},
we obtain
\be
\ddot\phi_N=-\nu_o^2\left\{\frac{9{\cal H}_{22}}{\sqrt{2\pi}}\left[\left(\frac{m_3}{m_{123}}\right)+N^{2/3}
\left(\frac{m_{12}}{m_{123}}\right)^{2/3}\left(\frac{m_1m_2}{m_{12}^2}\right)\right]
\left(\frac{e_i}{e_o^2}\right)(1-\ff{13}{24}e_i^2)(1-e_o^2)^{3/4}{N}^{3/2}\,{\rm e}^{-N\xi(e_o)}\right\}\,\sin\phi_{N},
\ee
where $\xi(e_o)={\rm Cosh}^{-1}(1/e_o)-\sqrt{1-e_o^2}$ and
from Table~\ref{hansentab} ${\cal H}_{22}=0.71$,
giving us an expression for $\omega_N$ and hence the range of values of $\dot\phi_N$ for which 
$\phi_N$ librates. Moreover, we see that $\phi_N$ does indeed librate around $\phi_N=0$ due
to the fact that $X_1^{2,2}(e_i)< 0$ for all $0<e_i\le 1$.

A more practical definition of the resonance width is in terms of the ``distance''
from exact commensurability in dimensionless units of period ratio. 
Rewriting \rn{dotphi} and incorporating the libration condition, we have
that libration occurs when
\be
\dot\phi_N=\nu_o(\sigma-N)<2\,\omega_N
\ee
so that the width of the $[N\!:\!1](2)$ resonance is approximately \label{p23}
\be
\Delta\sigma_N=2\,\omega_N/\nu_o
=\frac{6{\cal H}_{22}^{1/2}}{(2\pi)^{1/4}}\left[\left(\frac{m_3}{m_{123}}\right)+N^{2/3}
\left(\frac{m_{12}}{m_{123}}\right)^{2/3}\left(\frac{m_1m_2}{m_{12}^2}\right)\right]^{1/2}
\left(\frac{e_i^{1/2}}{e_o}\right)(1-\ff{13}{24}e_i^2)^{1/2}(1-e_o^2)^{3/8}{N}^{3/4}\,{\rm e}^{-N\xi(e_o)/2},
\label{res}
\ee
where 
$\lim_{e_o\rightarrow 0}\Delta\sigma_N$ is infinite for $N=1$, finite for $N=2$ and zero for $N\ge 3$.
Notice the steep dependence on the quantity $N\xi(e_o)$; since $\xi(e_o)$
is a monotonically decreasing function of $e_o$, the widths of high-$N$ 
resonances are only signficant when $e_o$ is also high.
Notice also that $\Delta\sigma_N=0$ when $e_i=0$; this implies that systems with circular
inner orbits are always stable which is most certainly not the case. In fact one needs to know how
much eccentricity is induced dynamically to calculate the true resonance width, and moreover one
needs to know the maximum inner eccentricity the system acquires during a secular cycle
to study its stability. This is thoroughly addressed in Paper II in which
{\it stability maps} are plotted which clearly demonstrate the success of \rn{res} as a predictor of instability
using the concept of resonance overlap.
Simple algorithms are also provided for determining the stability of any moderate-mass ratio hierarchical triple.

Note that our definition of the resonance width does not involve the usual concept of ``internal'' and ``external''
resonance \citep{murray}, in the same way that the present formulation
does not involve separate internal and external disturbing functions.

\subsubsection{Libration frequency}\label{3.1.2}
While the libration frequency of a pendulum depends on the amplitude, for small amplitudes it is
independent of amplitude and is given approximately by $\omega_N$.
Thus the libration frequency of the angle $\phi_N$ is
\be
\omega_N=\nu_o\Delta\sigma_N/2.
\label{libw}
\ee
For example, for an equal mass system with $e_i=0.1$, $e_o=0.5$ and $\sigma=20$,
the libration period is 1000 outer orbital periods, while increasing $e_o$ to 0.6 decreases
this to only 66 outer orbital periods (with the same factor increase in the resonance width).

\subsection{The secular disturbing function in the spherical harmonic expansion}\label{sec1}

Keeping in mind the caveats discussed in the Introduction, one can use the averaging principle 
to eliminate fast-varying terms from 
the disturbing function \rn{s6}, a process involving integrating over the two mean longitudes individually
(as if they were independent) for an orbital period of each.
In practice this is achieved simply by retaining only the $n=n'=0$ terms in \rn{s6}. 
Using the
notation $\tilde{\cal R}$ \label{p20}
for the averaged disturbing function, we obtain
\be
\tilde{\cal R}=
\sum_{m=0}^\infty
\tilde{\cal R}_m\,\cos\left[m(\varpi_i-\varpi_o)\right],
\label{Rsecsh}
\ee
where
\be
\tilde{\cal R}_m=\frac{G\mu_i m_3}{a_o}\sum_{l=l_{min},2}^\infty
\zeta_m c_{lm}^2\,
{\cal M}_l\,
\alpha^l\,
X_{0}^{l,m}(e_i)X_{0}^{-(l+1),m}(e_o).
\label{s7sec}
\ee
Closed-form expressions exist for $X_n^{l,m}(e_i)$ and $X_{n'}^{-(l+1),m}(e_o)$ when
$n=n'=0$; these are given in Appendix~\ref{hansen}, with some explicit forms given in 
Table~\ref{eccfuns}.
Expanding to octopole order, the disturbing function \rn{Rsecsh} becomes
\be
\tilde{\cal R}=\frac{G\mu_i m_3}{a_o}
\left[\frac{1}{4}\,\left(\frac{a_i}{a_o}\right)^2\,\frac{1+\ff{3}{2}e_i^2}{(1-e_o^2)^{3/2}}\,
-\frac{15}{16}\,\left(\frac{a_i}{a_o}\right)^3\,\left(\frac{m_1-m_2}{m_{12}}\right)\,
\frac{e_i e_o(1+\ff{3}{4}e_i^2)}{(1-e_o^2)^{5/2}}\,\cos(\varpi_i-\varpi_o)
\right].
\label{Rsecoct}
\ee
For coplanar secular systems, only the rates of change of the eccentricities and longitudes
of the periastra are of interest. From Lagrange's planetary equations (Appendix~\ref{lagrange}), these are
\be
\frac{de_i}{dt}=-\nu_i\frac{15}{16}\left(\frac{m_3}{m_{12}}\right)
\left(\frac{m_1-m_2}{m_{12}}\right)\left(\frac{a_i}{a_o}\right)^4
\frac{e_o(1+\ff{3}{4}e_i^2)\left(1-e_i^2\right)^{1/2}}{\left(1-e_o^2\right)^{5/2}}
\sin(\varpi_i-\varpi_o),
\label{ei}
\ee
\be
\frac{d\varpi_i}{dt}=\nu_i\left(\frac{m_3}{m_{12}}\right)
\left[
\frac{3}{4}\left(\frac{a_i}{a_o}\right)^3
\frac{\left(1-e_i^2\right)^{1/2}}{\left(1-e_o^2\right)^{3/2}}
-\frac{15}{16}\left(\frac{m_1-m_2}{m_{12}}\right)\left(\frac{a_i}{a_o}\right)^4
\frac{e_o(1+\ff{9}{4}e_i^2)\left(1-e_i^2\right)^{1/2}}{e_i\left(1-e_o^2\right)^{5/2}}
\cos(\varpi_i-\varpi_o)\right],
\label{varpii}
\ee
\be
\frac{de_o}{dt}=\nu_o\frac{15}{16}\left(\frac{m_1m_2}{m_{12}^2}\right)
\left(\frac{m_1-m_2}{m_{12}}\right)\left(\frac{a_i}{a_o}\right)^3
\frac{e_i(1+\ff{3}{4}e_i^2)}{\left(1-e_o^2\right)^2}
\sin(\varpi_i-\varpi_o),
\ee
\be
\frac{d\varpi_o}{dt}=\nu_o\left(\frac{m_1m_2}{m_{12}^2}\right)
\left[
\frac{3}{4}\left(\frac{a_i}{a_o}\right)^2
\frac{\left(1+\ff{3}{2}e_i^2\right)}{\left(1-e_o^2\right)^2}
-\frac{15}{16}\left(\frac{m_1-m_2}{m_{12}}\right)\left(\frac{a_i}{a_o}\right)^3
(1+\ff{3}{4}e_i^2)\frac{e_i}{e_o}\frac{\left(1+4e_o^2\right)}{\left(1-e_o^2\right)^3}
\cos(\varpi_i-\varpi_o)\right].
\label{wo}
\ee
Thus, for example, it is clear that for systems with $m_1=m_2$, there is no
secular variation in the eccentricities at this level of approximation and consequently, the inner
and outer rates of apsidal motion are constant. 

\section{Literal expansion}\label{4}

\subsection{Derivation}\label{derL}

The original literal expansions \citep[for example,][]{leverrier} were especially devised to study planetary orbits in the Solar System,
and in particular, to take advantage of the small planet-to-star mass ratios, small eccentricities
and inclinations, while putting essentially no restrictions on the ratio of semimajor axes
except that they should not cross. Our aims here are to generalise the formulation so that no 
assumptions about the mass ratios are made, and to present the formulation in a clear and
concise way which makes it easy to use to any order in the eccentricities and for
any appropriate application. Again, for this paper we consider coplanar configurations only,
and for clarity of presentation, we will repeat the definition of some quantities already defined in Section~\ref{sh}.

We start by writing down the disturbing function in a form which is useful for the coming analysis:
\bea
{\cal R}&=&-\frac{Gm_{12}m_3}{R}+\frac{Gm_2m_3}{|{\bf R}-\beta_1{\bf r}|}
+\frac{Gm_1m_3}{|{\bf R}-\beta_2{\bf r}|}\next
&=&
\frac{G\mu_i m_3}{a_o}\left[
\beta_1^{-1}\beta_2^{-1}\left(\frac{a_o}{R}\right)+\beta_1^{-1}\frac{a_o}{|{\bf R}-\beta_1{\bf r}|}
-\beta_2^{-1}\frac{a_o}{|{\bf R}-\beta_2{\bf r}|}\right],
\label{distlit}
\eea
where again,
$\beta_1=m_1/m_{12}$ and $\beta_2=-m_2/m_{12}$. 
Now consider the second and third term in \rn{distlit}
and write
\be
\frac{a_o}{|{\bf R}-\beta_s{\bf r}|}
=
\frac{a_o}{\sqrt{R^2-2\beta_s{\bf r}\cdot{\bf R}+(\beta_sr)^2}}
=
\frac{a_o}{R}\frac{1}{\sqrt{1-2x_s\cos\psi+x_s^2}},
\label{e0}
\ee
with $s$ being 1 or 2,
$x_s=\beta_sr/R$ and $\cos\psi=\hat{\bf r}\cdot\hat{\bf R}$ so that for coplanar systems,
\be
\psi=f_i+\varpi_i-f_o-\varpi_o,
\label{psi}
\ee
where again, $f_i$ and $f_o$, and $\varpi_i$ and $\varpi_o$ are the inner and outer true anomalies and 
longitudes of periastron respectively.

The literal expansion involves (1): a Taylor series expansion about the circular state $x_s=\alpha_s\equiv \beta_s\alpha$ \label{p14}
with small parameter $\epsilon$ related to the eccentricities, followed by (2): a Fourier expansion
in the angle $\psi$; this step introduces the {\it Laplace coefficients}, 
then (3): binomial expansions of powers of  $\epsilon$; (4): Fourier series in the mean anomalies,
and finally (5): combining the three contributions to the disturbing function in a simple expression.
This procedure is set out in \citet{murray} for the case of the restricted problem; here we present a significantly more
compact formulation for the general problem. Now let 
\be
g(x_s,\psi)=\left[1-2x_s\cos\psi+x_s^2\right]^{-1/2}.
\ee
Writing
\bea
x_s&=&\frac{\beta_s r}{R}=\beta_s\left(\frac{a_i}{a_o}\right)\frac{(r/a_i)}{(R/a_o)}\next
&=&\alpha_s\left(\frac{1-e_i^2}{1+e_i\cos f_i}\right)\left(\frac{1+e_o\cos f_o}{1-e_o^2}\right)\next
&\equiv& \alpha_s(1+\epsilon)
\eea
where 
\be
\epsilon=(r/a_i)/(R/a_o)-1
\label{epsi}
\ee
is first-order in the eccentricities, the first two steps described above are
\bea
\left[1-2x_s\cos\psi+x_s^2\right]^{-1/2}
&=&g(\alpha_s+\alpha_s\epsilon,\psi)\next
&=&\sum_{j=0}^\infty \frac{(\alpha_s\epsilon)^j}{j!}\left.\frac{\partial^j g}{\partial x_s^j}\right|_{x_s=\alpha_s} 
\hspace{8.7cm} \ldots{\it Step\ 1}\next
&=&\sum_{j=0}^\infty\frac{(\alpha_s\epsilon)^j}{j!}\frac{\partial^j}{\partial\alpha_s^j}\left[g(\alpha_s,\psi)\right]\next
&=&\sum_{j=0}^\infty \frac{(\alpha_s\epsilon)^j}{j!}\frac{\partial^j}{\partial\alpha_s^j}
\left[\sum_{m=-\infty}^\infty \ff{1}{2}b_{1/2}^{(m)}(\alpha_s){\rm e}^{im\psi}\right] 
\hspace{6.3cm} \ldots{\it Step\ 2}\next
&=&\sum_{j=0}^\infty \sum_{m=-\infty}^\infty {\rm e}^{im(\varpi_i-\varpi_o)}\,
\ff{1}{2}{\cal B}_{1/2}^{(j,m)}(\alpha_s)\,\epsilon^j\,{\rm e}^{im(f_i-f_o)},
\label{laplace1}
\eea
where we have used \rn{psi} for $\psi$ in the last step.
The Fourier coefficient in {\it Step 2} is a Laplace coefficient defined by \label{p30}
\be
\ff{1}{2}b^{(m)}_{1/2}(\alpha_s)=\frac{1}{2\pi}\int_0^{2\pi}\frac{{\rm e}^{-im\psi}}{\sqrt{1-2\alpha_s\cos\psi+\alpha_s^2}}d\psi
\label{e1b}
\ee
\citep{murray},
with the factor 1/2 (as opposed to the subscript 1/2) introduced to obtain the standard
definition of $b_{1/2}^{(m)}$. Note 
that since the real part of the integrand is even and the imaginary
part is odd, the integral is real\footnote{See footnote on page \pageref{pf7}.}
so that
\be
b^{(-m)}_{1/2}(\alpha_s)=\left[b^{(m)}_{1/2}(\alpha_s)\right]^*=b^{(m)}_{1/2}(\alpha_s).
\ee
The function in \rn{laplace1} involving the $jth$ derivative of the Laplace coefficient is
\be
{\cal B}_{1/2}^{(j,m)}(\alpha_s)=\frac{\alpha_s^j}{j!}\frac{d^j}{d\alpha_s^j}b^{(m)}_{1/2}(\alpha_s).
\label{bigB}
\ee
General properties of Laplace coefficients and their derivatives are given in Appendix~\ref{formulae}.
Note also that we have used $m$ for the Fourier summation index because
in fact it corresponds to the spherical harmonic order $m$ (see Section~\ref{equiv} where 
the equivalence of the two formulations is demonstrated). 

Referring to \rn{e0} and introducing the factor $a_o/R$, the next step in the procedure is a binomial expansion of $\epsilon^j$,
so that
\bea
(a_o/R)\left[1-2x_s\cos\psi+x_s^2\right]^{-1/2}
&=&\sum_{j=0}^\infty \sum_{m=-\infty}^\infty {\rm e}^{im(\varpi_i-\varpi_o)}\,
\ff{1}{2}{\cal B}_{1/2}^{(j,m)}(\alpha_s)\left[\frac{(r/a_i)}{(R/a_o)}-1\right]^j \left(\frac{a_o}{R}\right) {\rm e}^{im(f_i-f_o)}
\label{eps}\\
&=&
\sum_{j=0}^\infty \sum_{m=-\infty}^\infty{\rm e}^{im(\varpi_i-\varpi_o)}\ff{1}{2}{\cal B}_{1/2}^{(j,m)}(\alpha_s)\sum_{k=0}^j
{j\choose k}(-1)^{j-k}\left[(r/a_i)^k{\rm e}^{imf_i}\right]\left[\frac{{\rm e}^{-imf_o}}{(R/a_o)^{k+1}}\right],
\label{e3}\\
&&\hspace{10.8cm}\ldots{\it Step\ 3}\nonumber
\eea
where we have gathered together in the square brackets 
quantities associated with the inner and outer orbits in preparation for the next step.
As functions of the eccentricities and the sine and cosine of the true anomalies,
these terms can be expanded in Fourier series with period $2\pi$ such that
\be
(r/a_i)^k{\rm e}^{imf_i}=\left[\frac{1-e_i^2}{1+e_i\cos f_i}\right]^k{\rm e}^{imf_i}
=\sum_{n=-\infty}^\infty X_{n}^{k,m}(e_i) \,{\rm e}^{inM_i}
\label{ra}
\ee
and
\be
\frac{{\rm e}^{-imf_o}}{(R/a_o)^{k+1}}=\left[\frac{1+e_o\cos f_o}{1-e_o^2}\right]^{k+1}{\rm e}^{-imf_o}
=\sum_{n=-\infty}^\infty X_{n'}^{-(k+1),m}(e_o)\, {\rm e}^{-in'M_o},
\label{Ra}
\ee
where $X_{n}^{k,m}(e_i)$ and $X_{n'}^{-(k+1),m}(e_o)$ are again Hansen coefficients given by
\rn{s4} and \rn{s5} (recall that $X_{-n'}^{-(k+1),-m}=X_{n'}^{-(k+1),m}$).
Substituting \rn{ra} and \rn{Ra} into \rn{e3} and
gathering together the angles, we obtain for $s=1,2$
\bea
\frac{a_o}{|{\bf R}-\beta_s{\bf r}|}&=&
\sum_{m=-\infty}^\infty
\sum_{n=-\infty}^\infty\sum_{n'=-\infty}^\infty
\sum_{j=0}^\infty
\ff{1}{2}{\cal B}_{1/2}^{(j,m)}(\alpha_s)\,
\left[\sum_{k=0}^j 
(-1)^{j-k}{j\choose k}
X_{n}^{k,m}(e_i)\,X_{n'}^{-(k+1),m}(e_o)\right]
\,{\rm e}^{i\phi_{mnn'}}
\hspace{0.7cm}\ldots{\it Step\ 4}
\label{e50}\\
&=&
\sum_{m=0}^\infty
\sum_{n=-\infty}^\infty\sum_{n'=-\infty}^\infty
\left[\sum_{j=0}^\infty
\zeta_m\,{\cal B}_{1/2}^{(j,m)}(\alpha_s)\,
F_{mnn'}^{(j)}(e_i,e_o)\right]\,\cos\phi_{mnn'},
\label{e5}
%\\&&\hspace{10.8cm}\ldots{\it Step\ 4}\nonumber
\eea
where \label{pFmnn}
\be
F_{mnn'}^{(j)}(e_i,e_o)=\sum_{k=0}^j 
(-1)^{j-k}{j\choose k}
X_{n}^{k,m}(e_i)\,X_{n'}^{-(k+1),m}(e_o),
\label{Fmnn}
\ee
\bea
\phi_{mnn'}&=&
nM_i-n'M_o+m(\varpi_i-\varpi_o)\next
&=&
n\lambda_i-n'\lambda_o+(m-n)\varpi_i-(m-n')\varpi_o
\label{resang2}
\eea
is again a harmonic angle, \label{p13b}
and $\zeta_m$ is 1/2 when $m=0$ and 1 otherwise.
As with the spherical harmonic formulation, in going from \rn{e50} to \rn{e5}
we have paired together terms with positive and negative values of $m$ to 
make the expression manifestly real. 

The final step involves writing down the full literal expansion for the disturbing function.
Substituting \rn{e5} into \rn{distlit}, one obtains
\bea
{\cal R}&=&\frac{G\mu_i m_3}{a_o}
\sum_{m=0}^\infty\sum_{n=-\infty}^\infty
\sum_{n'=-\infty}^\infty
\sum_{j=0}^\infty
{\cal A}_{jm}(\alpha;\beta_2)\,
F_{mnn'}^{(j)}(e_i,e_o)
\,\cos\phi_{mnn'}
\hspace{6.0cm} \ldots{\it Step\ 5}\next
&=&
\sum_{m=0}^\infty\sum_{n=-\infty}^\infty
\sum_{n'=-\infty}^\infty
{\cal R}_{mnn'}\,\cos\phi_{mnn'},
\label{R}
\eea
where the harmonic coefficient associated with the angle $\phi_{mnn'}$ is \label{p22}
\be
{\cal R}_{mnn'}=\frac{G\mu_i m_3}{a_o}\sum_{j=0}^\infty
{\cal A}_{jm}(\alpha;\beta_2)\,F_{mnn'}^{(j)}(e_i,e_o),
\label{Rmnn}
\ee
with \label{pA}
\be
{\cal A}_{jm}(\alpha;\beta_2)
=\zeta_m\,\left[\beta_1^{-1}{\cal B}_{1/2}^{(j,m)}(\alpha_1)-\beta_2^{-1}{\cal B}_{1/2}^{(j,m)}(\alpha_2)\right]
\label{Alm}
\ee
for all $j$, $m$ except when $j=m=0$ in which case
\be
{\cal A}_{00}(\alpha;\beta_2)
=\ff{1}{2}\,\left[\beta_1^{-1}b_{1/2}^{(0)}(\alpha_1)-\beta_2^{-1}b_{1/2}^{(0)}(\alpha_2)\right]
+\beta_1^{-1}\beta_2^{-1}.
\label{A00}
\ee
Recall here that $-\beta_2=m_2/m_{12}=1-\beta_1$,
$\alpha=a_1/a_2$ and $\alpha_s=\beta_s\alpha$, $s=1,2$.
Note that the order of the expansion is given by the number of terms included in the summation in \rn{Rmnn},
that is, it is given by the maximum value of $j$.
The equivalence of the the literal and spherical harmonic formulations is demonstrated
in Section~\ref{equiv}. 

In contrast to the classical literal expansion which is valid for $m_2/m_1\ll 1$ (see Section~\ref{4.2.2})
and involves Laplace coefficients as functions of the ratio of semimajor axes $\alpha$, 
\rn{R} is valid for any mass ratios and expresses the disturbing function in terms of Laplace coefficients whose
arguments are $\alpha_s=\beta_s\alpha$, that is, the ratio of semimajor axes scaled by the mass ratios
$m_s/m_{12}$, $s=1,2$. 

By calculating the harmonic coefficients for a second-order resonance
as well as those for general first-order resonances and for the secular harmonics,
and also by calculating resonance widths,
we demonstrate in Sections~\ref{53}, \ref{4.5}, \ref{4.6} and \ref{sec2},
the ease with which this form of the literal expansion can be used to any required
order in eccentricity.

\subsection{Eccentricity dependence}\label{4.1}
The dependence of the disturbing function on the eccentricity is via
$F_{mnn'}^{(j)}(e_i,e_o)$, defined in \rn{Fmnn}. Although each term in the finite summation over $k$
is ${\cal O}(e_i^{|m-n|}e_o^{|m-n'|})$, the leading order of $F_{mnn'}^{(j)}(e_i,e_o)$ will be either $j$ or $j+1$
when $j>|m-n|+|m-n'|\equiv\eta$. 
This results from the fact that from \rn{eps},
$F_{mnn'}^{(j)}(e_i,e_o)\propto \epsilon^j$ with $\epsilon={\cal O}[{\rm max}(e_i,e_o)]$, and also that
the Hansen coefficients which make up $F_{mnn'}^{(j)}(e_i,e_o)$ are either odd or even
functions of the eccentricity (so their power series expansions have only odd or even powers).
In particular,
\be
F_{mnn'}^{(j)}(e_i,e_o)=\left\{\begin{array}{lll}
{\cal O}(e_i^{|m-n|}\,e_o^{|m-n'|}), & j\le \eta &\\
{\cal O}(e_i^{|m-n|}\,e_o^{j-|m-n|}),\ {\cal O}(e_i^{|m-n|+2}\,e_o^{j-2-|m-n|}),
\ldots,{\cal O}(e_i^{j-|m-n'|}\,e_o^{|m-n'|}), & j> \eta, & j-\eta\,\,\,{\rm even}\\
{\cal O}(e_i^{|m-n|}\,e_o^{j+1-|m-n|}),\ {\cal O}(e_i^{|m-n|+2}\,e_o^{j-1-|m-n|}),
\ldots,{\cal O}(e_i^{j+1-|m-n'|}\,e_o^{|m-n'|}), & j> \eta, & j-\eta\,\,\,{\rm odd}\\
\end{array}
\right.
\ee
Thus for example, 
the leading-order term in an expansion of $F_{423}^{(0)}(e_i,e_o)$ is 
${\cal O}(e_i^2e_o)$ as it is for $F_{423}^{(1)}(e_i,e_o)$, $F_{423}^{(2)}(e_i,e_o)$ and $F_{423}^{(3)}(e_i,e_o)$,
while the leading-order terms of both of 
$F_{423}^{(4)}(e_i,e_o)$ and $F_{423}^{(5)}(e_i,e_o)$ are
${\cal O}(e_i^2e_o^3)$ and ${\cal O}(e_i^4e_o)$.
As a consequence, 
\label{discuss} if one requires an expansion of the disturbing function which is correct to order
$j_{max}$ in the eccentricities, then one should include terms up to and including $j=j_{max}$ in \rn{Rmnn}, and
moreover, expand each of $F_{mnn'}^{(j)}(e_i,e_o)$, $j\le j_{max}$, to order $j_{max}$ in the eccentricities.
Similarly, one should only include harmonics which are such that $|m-n|+|m-n'|\le j_{max}$.
Conversely, if one is particularly interested in a term whose harmonic angle is $\phi_{mnn'}$,
one should include terms {\it at least} up to $j=|m-n|+|m-n'|$ in order to obtain a non-zero coefficient for 
such a term.

For practical applications, it is usually most efficient 
to evaluate the functions $F_{mnn'}^{(j)}(e_i,e_o)$ using {\it Mathematica} (Section~\ref{math})
or similar for the particular range of values of $m,n,n',j$ of interest, summing over the
various Hansen coefficients which contribute. However, it is of considerable interest
to examine the functional dependence of the power-series representations of the Hansen coefficients,
not only on the eccentricity, but also on the associated indices $n$, $j$ and $m$.

\subsubsection{Power series representations of Hansen coefficients and the choice of expansion order}\label{4.1.1}
Power series expansions for $X_n^{j,m}(e)$ are given in Appendix~\ref{expan} for
arbitrary $j$ and $n$, and for $m=n\pm p$, $p=0,1,2,3,4$, with the choice of values for $m$ being
guided by the use of Hansen coefficients in the study of resonance, since the {\it order} of a resonance
is $|m-n|+|m-n'|$ (see Section \ref{53}).
Two features in particular emerge from these general series expansions. 
First one sees that their leading terms are indeed proportional to $e^{|m-n|}$, 
and secondly that the {\it coefficients} of the leading terms contain contributions
which are ${\cal O}(j^p)$ and ${\cal O}(n^p)$. 
This means that $X_n^{j,m}(e)$ is in fact ${\cal O}\{{\rm max}[(ne)^{|m-n|},(je)^{|m-n|}]\}$, which
in turn has implications for the radius of convergence of the series (the range of values of each
of the eccentricities for which the series converges), and the order at which one should truncate the series
for given eccentricities. This should be kept in mind when using these expansions, especially
for systems with period ratio close to one. 
The effect is evident when one compares, for example, 
the first-order Hansen coefficients in
Figures~\rn{hansencomp21} and \rn{hansencomp76}. 
In both of these figures, numerically evaluated integrals (solid curves) are compared with 
their fourth-order correct series approximations (dashed curves). 
For example, the series representation of 
$X_1^{0,2}(e_i)={\cal O}(e_i)$ in panel (a) of Figure~\rn{hansencomp21} approximates
the actual function well for $0\le e_i\lapp 0.8$,
while that of 
$X_6^{0,7}(e_i)={\cal O}(6e_i)$ in panel (a) of Figure~\rn{hansencomp76} is only 
accurate for $0\le e_i\lapp 0.1\simeq 0.8/6=0.13$. 

While it may be tempting to avoid the series
expansions of Hansen coefficients with high values of $n$, one should remember that any particular harmonic
coefficient is only accurate to order $j_{max}\ge|m-n|+|m-n'|$, even if individual Hansen 
coefficients (and hence $F_{mnn'}^{(j)}(e_i,e_o)$) are evaluated accurately.
On the other hand, if computational efficiency is required, it is best to calculate individual series
expansions of $F_{mnn'}^{(j)}(e_i,e_o)$ to adequate order in the eccentricities.
A {\it Mathematica} program is provided in Appendix~\ref{math} for this purpose.

While we do not attempt a formal convergence analysis here,
the most straightforward way to gain confidence in any particular expansion is to compare its 
predictions with direct numerical integration of the equations of motion. 

\subsection{Dependence on the mass and semimajor axis ratios}\label{4.2}
The coefficient of any particular harmonic term in the Fourier expanded
disturbing function \rn{R} is given by \rn{Rmnn}, and
this itself may be expressed as an infinite series of terms in increasing orders of eccentricity.
Each term contributing to a coefficient depends on the mass ratio $m_2/m_1$ and the
semimajor axis ratio $\alpha$ though
the factor ${\cal A}_{jm}(\alpha;\beta_2)$.
To obtain an idea of this dependence, we can use the expansion \rn{e11}
for ${\cal B}_{1/2}^{(j,m)}(\alpha_i)$ in \rn{Alm}.
Noting that the leading term in this expansion depends on whether $j\le m$ or $j> m$,
we have in the first case that
\bea
{\cal A}_{jm}(\alpha;\beta_2)
&=&\zeta_m\sum_{p=0}^\infty
E_p^{(j,m)}\left[\beta_1^{m+2p-1}-\beta_2^{m+2p-1}\right]\alpha^{m+2p}\next
&=&\zeta_m \, E_0^{(j,m)}\left[\frac{m_1^{m-1}+(-1)^m m_2^{m-1}}{m_{12}^{m-1}}\right]\,\alpha^m+\ldots\next
&=&\zeta_m \, E_0^{(j,m)}\left[1+(m-1)(m_2/m_1)+\ldots+(-1)^m(m_2/m_1)^{m-1}+\ldots\right]\alpha^m+\ldots,
\hspace{0.25cm} j\leq m,
\hspace{0.25cm} m\ge 2,
\label{mass}
\eea
where $E_0^{(j,m)}$ is given by \rn{E0}.
However, the leading term is zero when $m=1$ in which case
\bea
{\cal A}_{j1}(\alpha;\beta_2)
&=&E_1^{(j,1)}\left[\frac{m_1-m_2}{m_{12}}\right]\alpha^3+\ldots\next
&=&
E_1^{(j,1)}\left[1-2(m_2/m_1)+\ldots\right]\alpha^3+\ldots,
\hspace{0.25cm} j=0,1,
\label{566}
\eea
while for $m=j=0$ we have from \rn{A00} that
\bea
{\cal A}_{00}(\alpha;\beta_2)
&=&\ff{1}{2}E_1^{(0,0)}\alpha^2+\ff{1}{2}E_2^{(0,0)}\left[\frac{m_1^3+m_2^3}{m_{12}^3}\right]\alpha^4
+\ldots\next
&=&
\ff{1}{4}\alpha^2+\ff{9}{64}\left(1-3(m_2/m_1)+\ldots\right)\alpha^4+\ldots 
\label{A00s}
\eea
The fact that there are no monopole or dipole terms (ie., no power of $\alpha$ less than 2) is consistent
with the spherical harmonic expansion \rn{s7}.
When $j >m$,
\bea
{\cal A}_{jm}(\alpha;\beta_2)
&=&\zeta_m\sum_{p=p_*}^\infty
E_p^{(j,m)}\left[\beta_1^{m+2p-1}-\beta_2^{m+2p-1}\right]\alpha^{m+2p}\next
&=&\left\{\begin{array}{ll}
\displaystyle{\zeta_m \, E_{p_*}^{(j,m)}\left[\frac{m_1^{j-1}+(-1)^j m_2^{j-1}}{m_{12}^{j-1}}\right]\,\alpha^j+\ldots,}
& j>m,\,\,\,j-m\,\,\,{\rm even},\\
\displaystyle{\zeta_m \, E_{p_*}^{(j+1,m)}\left[\frac{m_1^j+(-1)^{j+1} m_2^j}{m_{12}^j}\right]\,\alpha^{j+1}+\ldots,}
& j>m,\,\,\,j-m\,\,\,{\rm odd}.
\end{array}\right.
\label{massj}
\eea
where $p_*=\lfloor(j-m+1)/2\rfloor$ with $\lfloor\,\,\rfloor$ denoting the nearest lowest integer and
$E_{p_*}^{(j,m)}$ is given by \rn{E0b}. 
Note that since the leading terms of the harmonic coefficient ${\cal R}_{mnn'}$ are of order $|m-n|+|m-n'|$ in eccentricity, they will
be such that $0\le j\le m$ (see example in the next Section). 

\subsubsection{Summary of leading terms in $\alpha$}\label{4.2.1}

We can summarize the above as follows. For ${\cal A}_{jm}$ we have
\be
{\cal A}_{jm}(\alpha;\beta_2)=\left\{\begin{array}{lll}
{\cal O}(\alpha^m), & m\ge 2, &  j\le m\\
{\cal O}(\alpha^j), & j>m, & j-m \,\,\, {\rm even}\\
{\cal O}(\alpha^{j+1}), & j>m, & j-m\,\,\,{\rm odd}\\
{\cal O}(\alpha^2), & m=0, &  j=0\\
{\cal O}(\alpha^3), & m=1, &  j=0,1,\\
\end{array}\right.
\ee
so that the harmonic coefficients are such that
\be
{\cal R}_{mnn'}=\left\{\begin{array}{ll}
{\cal O}(\alpha^2), & m=0\\
{\cal O}(\alpha^3), & m=1\\
{\cal O}(\alpha^m), & m\ge 2.\\
\end{array}\right.
\ee

\subsubsection{Coefficients when $m_2/m_1\rightarrow 0$}\label{4.2.2}

The standard literal expansion is derived assuming that one or other of the mass ratios $m_2/m_1$ and $m_3/m_1$ is 
zero \citep{murray}. Putting $\beta_1=1$ and $\beta_2=0$ in \rn{Alm} and \rn{A00}, we have
\be
\lim_{\beta_2\rightarrow 0}A_{jm}=\zeta_m {\cal B}_{1/2}^{(j,m)}(\alpha)
\ee
for all $j$, $m$ except when $m=1$, $j=0,1$ and when $j=m=0$. In these cases, from \rn{e1bb} and \rn{e11} we have
\be
\lim_{\beta_2\rightarrow 0}A_{01}=b_{1/2}^{(1)}(\alpha)-\alpha,
\label{01}
\ee
\be
\lim_{\beta_2\rightarrow 0}A_{11}=\alpha\frac{db_{1/2}^{(1)}(\alpha)}{d\alpha}-\alpha,
\label{11}
\ee
and
\be
\lim_{\beta_2\rightarrow 0}A_{00}=\ff{1}{2}b_{1/2}^{(0)}(\alpha)-1
\ee
(recall that $b_{1/2}^{(0)}(0)=2$; see \rn{e1bb} and \rn{e1bbb}).
Using these approximations 
makes the disturbing function zeroth-order correct in the mass ratio $m_2/m_1$ 
(times the factor $m_2\,m_3$ when it has dimensions of energy as in the formulations presented here),
so that the rates of change of the elements are first-order in $m_3/m_1$ for the inner elements,
and first-order in $m_2/m_1$ for the outer elements (see Sections~\ref{sec1} and \ref{sec2}).

\subsection{The spherical harmonic order $m$ and principal resonances}\label{4.3}

For coplanar systems there are three labels,
$m$, $n$ and $n'$, associated with each harmonic.
In turn, each label is associated with an independent frequency of the system: $n$ and $n'$ are
associated with the inner and outer orbital frequencies respectively, while $m$ is assocated
with the difference in the rates of apsidal advance 
$\dot\varpi_i-\dot\varpi_o$; see the definition of the harmonic angle \rn{resang2}.
In Section \ref{equiv} we demonstrate the equivalence of the spherical harmonic and literal expansions,
where the index $m$ in the literal expansion
is shown to correspond to the spherical harmonic order $m$.
This plays an important role in many physical systems, and the
three-body problem is no exception. For example, we will show in a future paper in this series that one can
define the concept of ``modes of oscillation of a binary'' which are excited in the presence
of a triple companion, in analogy 
with the modes of oscillation of a star which are excited in the presence of a binary companion.
The spherical harmonic order $m$ acts as an azimuthal mode number in the formalism,
with the analogy between the two physical problems revealing a rich vein of exploration.

First note that $m$ distinguishes resonant states with the same values
of $n$ and $n'$. 
In general, a harmonic coefficient is ${\cal O}(e_i^{|m-n|}e_o^{|m-n'|})$, and since
\be
|m-n|+|m-n'|=\left\{\begin{array}{ll}
n'-n, & n\le m\le n',\\
|2m-n-n'|, & {\rm otherwise},
\end{array}\right.
\label{mnn}
\ee
the order in eccentricity is minimized at $n'-n$ when $n\le m\le n'$.
Using the notation $[n'\!:\!n](m)$ introduced in Section~\ref{dom} to emphasise 
the association of the harmonic angle $\phi_{mnn'}$ with 
the $n'\!:\!n$ resonance of spherical harmonic order $m$,
we refer to the 
$[n'\!:\!n](m)$ resonances, $n\le m\le n'$, as the 
{\it principal resonances} or
{\it principal harmonics} of the $n'\!:\!n$ resonance.
For example, the two principal $2\!:\!1$ resonances
are $[2\!:\!1](1)$ and $[2\!:\!1](2)$, with harmonic angles 
$\phi_{112}=\lambda_i-2\lambda_o+\varpi_o$ and $\phi_{212}=\lambda_i-2\lambda_o+\varpi_i$ respectively
and with harmonic coefficients
${\cal R}_{112}={\cal O}(e_o)$ and ${\cal R}_{212}={\cal O}(e_i)$.
Similarly, there are four principal $5\!:\!2$ resonances, each third-order in eccentricity, 
namely $[5\!:\!2](2)$, $[5\!:\!2](3)$, $[5\!:\!2](4)$ and $[5\!:\!2](5)$,
with resonance angles 
$\phi_{225}=2\lambda_i-5\lambda_o+3\varpi_o$,
$\phi_{325}=2\lambda_i-5\lambda_o+\varpi_i+2\varpi_o$,
$\phi_{425}=2\lambda_i-5\lambda_o+2\varpi_i+\varpi_o$,
and
$\phi_{525}=2\lambda_i-5\lambda_o+3\varpi_i$,
and harmonic coefficients
${\cal R}_{225}={\cal O}(e_o^3)$, ${\cal R}_{325}={\cal O}(e_ie_o^2)$, ${\cal R}_{425}={\cal O}(e_i^2e_o)$
and ${\cal R}_{525}={\cal O}(e_i^3)$. In general there are $n'-n+1$ principle harmonics associated with the $n'\!:\!n$ resonance.

The $2\!:\!1$ resonance is referred to as a {\it first-order resonance} because the 
minimum order in eccentricity of either of the principal harmonic coefficients is first order.
Using the nomenclature introduced here, we can be more definite and 
say that in general, a resonance is $p$th-order if the 
{\it principal resonances} are $p$th-order in eccentricity.

It is sometimes desirable to express the ``largeness'' or otherwise of the values of $n$ and $n'$, especially
for first-order resonances. The author is aware that the term
{\it resonance degree} is occasionally used for this purpose, however, in the context of 
spherical harmonics the words ``degree'' and `` order'' are associated with the indices $l$ and $m$
respectively. In hindsight this is unfortunate because $m$ could have been
used for this purpose had the word ``order'' not already refered to the value of $n'-n$.
Moreover, one correctly refers to a polynomial's {\it degree} rather than {\it order} 
when describing its highest power (although the latter is often used), and had degree been adopted
for describing the order in eccentricity of a resonance, all would be consistent. 
But history takes precedence for words in common use, and the $7\!:\!6$ resonance
continues to be a seventh-degree first-order resonance.

Finally recall from Section~\ref{4.2.1} that ${\cal R}_{mnn'}={\cal O}(\alpha^m)$, $m\ge 2$,
while ${\cal R}_{0nn'}={\cal O}(\alpha^2)$ and ${\cal R}_{1nn'}={\cal O}(\alpha^3)$.
This implies that in general, unless $e_o\ll e_i$, it is the principal resonance with $m=n$ which tends to make
the largest contribution to the disturbing function,
except when $n=1$ in which case the $m=2$ harmonic tends to make the largest contribution.

\subsubsection{``Zeeman splitting'' of resonances}\label{zee}
Just as a magnetic field introduces fine stucture to atomic energy levels (Zeeman splitting), 
apsidal advance of the inner and outer orbits introduces fine structure in the positions of the 
centres of resonances relative to exact commensurability. In both cases it is the spherical harmonic
order $m$ which labels the associated frequencies
and moreover physically, it is the introduction of one or more
distinguished directions (magnetic field or third body)
which breaks the otherwise symmetric state of the system. The slow rotation of the system
about these directions introduces new (generally low) frequencies, splitting the
otherwise degenerate state.
To get an idea of the magnitude of this effect,
consider a two-planet system near the $2\!:\!1$ resonance
with stellar and planetary masses $m_*$, $m_i$ and $m_o$,
with $m_i$ the mass of the inner planet and $m_i,m_o\ll m_*$.
The rate of apsidal advance is given by Lagrange's planetary equation \rn{D2} which involves 
a partial derivative with respect to the eccentricity.
To obtain a quick estimate of the rates for both orbits which is 
correct to first-order in the eccentricities, it is simplest to use the leading terms in 
the spherical harmonic expression \rn{s7} for the harmonic coefficients,
including in the disturbing function ${\cal R}_{000}$ and ${\cal R}_{100}$ for the secular contributions
(all others are more than second-order in the eccentricities),
and ${\cal R}_{212}$ and ${\cal R}_{112}$ for the quadrupole and octopole resonant contributions
corresponding to the two principal resonant angles $\phi_{212}$ and $\phi_{112}$.
The error incurred in using only the leading term in the sum over $l$ is discussed in Section~\ref{acc}.
To second-order in the eccentricities and to zeroth-order in $m_i/m_*$ these are
\be
{\cal R}_{000}=\frac{1}{4}\frac{Gm_im_o}{a_o}\alpha^2(1+\ff{3}{2}e_i^2)(1+\ff{3}{2}e_o^2),
\hspace{0.5cm}
{\cal R}_{100}=-\frac{15}{16}\frac{Gm_im_o}{a_o}\alpha^3\,e_i\,e_o,
\label{90}
\ee
\be
{\cal R}_{212}=-\frac{9}{4}\frac{Gm_im_o}{a_o}\alpha^2 e_i
\hand
{\cal R}_{112}=\frac{9}{8}\frac{Gm_im_o}{a_o}\alpha^3 e_o,
\ee
where we have used the expansions for the Hansen coefficents in Section~\ref{expan}.
If all of $\varpi_i-\varpi_o$, $\phi_{212}$ and $\phi_{112}$ librate,
the rates of apsidal advance are then
\be
\dot\varpi_i=\frac{3}{4}\nu_i\left(\frac{m_o}{m_*}\right)\,
\alpha^3\left[1-\frac{5}{4}\alpha\left(\frac{e_o}{e_i}\right)\cos(\varpi_i-\varpi_o)
-\left(\frac{3}{e_i}\right)\cos(\lambda_i-2\lambda_o+\varpi_i)\right]
\label{wdoti}
\ee
and
\be
\dot\varpi_o=\frac{3}{4}\nu_o\left(\frac{m_i}{m_*}\right)\,
\alpha^2\left[1-\frac{5}{4}\alpha\left(\frac{e_i}{e_o}\right)\cos(\varpi_i-\varpi_o)
+\left(\frac{3}{2e_o}\right)\cos(\lambda_i-2\lambda_o+\varpi_o)\right].
\label{wdoto}
\ee
Whether or not a particular angle contributes
on average to $\dot\varpi_i$ and $\dot\varpi_o$ depends on whether it librates or not,
that is, whether or not the average value of its cosine is non-zero.
If it does librate, the sign of its contribution will depend on whether it does so around zero or $\pi$
(or some other angle in some cases).
Using \rn{phiddl}, one can show that for small eccentricities, when $\phi_{212}$ 
and $\phi_{112}$ librate, they do so around zero and $\pi$ respectively.
One may then ask whether it is possible for both angles to librate at the same time.
It is possible to show that if the harmonic angle $\phi_{21N}$ librates,
then all other angles of the form $\phi_{21n'}$, $n'\ne N$, must circulate. This is not necessarily
true for a set of principal resonances because for any two angles from the set, labeled,
say, by $m_1$ and $m_2$ (not to be confused with the masses),
\be
\phi_{m_2nn'}=\phi_{m_1nn'}+(m_2-m_1)(\varpi_i-\varpi_o).
\ee
Thus if one resonance angle librates {\it and in addition},
$\varpi_i-\varpi_o$ librates, then all other associated principal resonance angles will librate.\footnote{In fact,
{\it all} $n'\!:\!n$ resonance angles will librate, not just the principal angles.}
Now, the angle $\varpi_i-\varpi_o$ will librate 
if the eccentricities are small enough (see, for example, \citet{mardling07}
for a study of the libration and circulation of this angle in the case of secular evolution). 
If this occurs, then since $\phi_{212}-\phi_{112}=\varpi_i-\varpi_o$, 
then $\varpi_i-\varpi_o$ must librate around $-\pi$
(because $\phi_{212}$ librates around zero and $\phi_{122}$ librates around $\pi$) so that the average
value of $\cos(\varpi_i-\varpi_o)$ is $-1$, while the average values of $\cos\phi_{212}$
and $\cos\phi_{112}$ are 1 and $-1$ respectively (at exact resonance).
From \rn{wdoti} and \rn{wdoto}, the average rates of apsidal advance in this case are therefore
\be
\dot\varpi_i=\frac{3}{4}\nu_i\left(\frac{m_o}{m_*}\right)\,
\alpha^3\left[1-\left(\frac{3-\ff{5}{4}\alpha e_o}{e_i}\right)
\right]\simeq -\ff{9}{4}\nu_o(m_o/m_*)\sigma^{-1}e_i^{-1}
\ee
and
\be
\dot\varpi_o=\frac{3}{4}\nu_o\left(\frac{m_i}{m_*}\right)\,
\alpha^2\left[1-\frac{1}{4}\left(\frac{6-5\alpha e_o}{e_i}\right)
\right]\simeq -\ff{9}{8}\nu_o(m_i/m_*)\sigma^{-4/3}e_o^{-1},
\ee
where $\sigma$ is the period ratio and and the approximations hold for small to moderate eccentricities.
In such cases, the rates of change of the two $2\!:\!1$ principal resonances are, from \rn{resang2},
\be
\dot\phi_{212}=\nu_o[(\sigma-2)-\ff{9}{4}(m_o/m_*)\sigma^{-1}e_i^{-1}]
\ee
and
\be
\dot\phi_{112}=\nu_o[(\sigma-2)-\ff{9}{8}(m_i/m_*)\sigma^{-4/3}e_o^{-1}].
\ee
For two Jupiter-mass planets orbiting a solar-mass star,
the positions of exact resonance (that is, the value of $\sigma$
for which $\dot\phi_{212}=0$ or $\dot\phi_{112}=0$) are therefore approximately a distance
\be
\delta\sigma_{212}=0.0011\,e_i^{-1}
\hand
\delta\sigma_{112}=0.0004\,e_o^{-1}
\label{split}
\ee
away from exact commensurability. These can be signficant for small eccentricities,
and this should be remembered when deciding whether or not an oberved system is
likely to be in resonance (subject to the caveat discussed
in footnote 2 on page~\pageref{dellask}).

It is interesting to note here that for non-coplanar systems, there are five independent labels including
$n$, $n'$ and $m$, and an additional two spherical harmonic $m$'s which we denote by $m_i$ and $m_o$
(non-coplanar systems will be studied in Paper III in this series).
The harmonic angle becomes
\bea
\phi_{m_i m_o m\, n\, n'}&=&
n M_i-n'M_o+m_i\,\omega_i-m_o\,\omega_o+m(\Omega_i-\Omega_o)\next
&=&
n\lambda_i-n'\lambda_o+(m_i-n)\varpi_i-(m_o-n')\varpi_o+(m-m_i)\Omega_i-(m-m_o)\Omega_o,
\eea
with respectively $\omega_i$ and $\omega_o$, and $\Omega_i$ and $\Omega_o$, 
the arguments of periastron and
the longitudes of the ascending nodes of the inner and
outer orbits respectively. Note that for coplanar systems, $m_i=m_o=m$.
The additional labels reflect 
the extra frequencies introduced when the problem becomes three dimensional.
The three frequencies 
associated with $m$, $m_i$ and $m_o$ are, respectively, the {\it difference}
in the rates of precession of the orbital planes about the total angular momentum vector,
and the rates of change of the inner and outer {\it arguments} of periastron.
We note also that the additional fine structure introduced when the orbits are not coplanar
has its own analogy with Zeeman splitting.
Before the latter phenomenon was understood in the context of quantum mechanics, physicists referred
to energy level splittings which were accurately predicted by the classical theory
of Lorentz as ``normal'' and those which were not as ``anomalous.''
Once the quantum mechanical concept of electron spin was introduced, it became clear that
the additional source of angular momentum was responsible for the ``anomalous'' fine structure,
with states not involving electron spin remaining ``normal.'' 
Suffice to say here that orbital precession introduces ``anomalous'' fine structure, with
``normal'' fine structure associated with principal resonances for which $m_i=m_o=m$.

As well as illustrating the phenomenon of resonance splitting, the fine structure calculation \rn{90} to \rn{split}
serves to further demonstrate
the ease with which the spherical harmonic expansion can be applied.
We now consider some specific applications which make more explicit the power
of the literal expansion.

\subsection{A second-order resonance}\label{53}
Written in the form \rn{R} with \rn{Rmnn}, it is easy to include terms to any order in the eccentricities and mass ratios.
For example, say one wanted to study the $5\!:\!3$ resonance for which $n=3$ and $n'=5$. 
The principal resonance angles are
(from \rn{resang2}), $\phi_{335}=3\lambda_i-5\lambda_o+2\varpi_o$, 
$\phi_{435}=3\lambda_i-5\lambda_o+\varpi_i+\varpi_o$ and $\phi_{535}=3\lambda_i-5\lambda_o+2\varpi_i$, with
${\cal R}_{335}= {\cal O}(e_o^2)$, ${\cal R}_{435}= {\cal O}(e_ie_o)$ and ${\cal R}_{535}={\cal O}(e_i^2)$.
If we choose to include terms, say, up to fourth order in the eccentricities, we would
include in the summation in \rn{Rmnn} terms up to $j=4$. Note that for this resonance, all terms
will be even-ordered in eccentricity so that, for example, $j=3$ terms will actually be fourth-order.
For instance, $F_{535}^{(3)}=\ff{9}{2}e_i^2e_o^2$.
Thus, to {\it third}-order in eccentricity, the terms in the disturbing function associated with the $5\!:\!3$ resonance are
\bea
{\cal R}&=&\frac{G\mu_i m_3}{a_o}\left\{\ldots+
\left(\ff{67}{8}{\cal A}_{03}+\ff{9}{4}{\cal A}_{13}+\ff{1}{4}{\cal A}_{23}\right)\,e_o^2\,
\cos(3\lambda_i-5\lambda_o+2\varpi_o)\right.\next
&&\hspace{2cm}-\left(18{\cal A}_{04}+\ff{9}{2}{\cal A}_{14}+\ff{1}{2}{\cal A}_{24}\right)\,e_ie_o\,
\cos(3\lambda_i-5\lambda_o+\varpi_i+\varpi_o)\next
&&\left.\hspace{4cm}+\left(\ff{75}{8}{\cal A}_{05}+\ff{9}{4}{\cal A}_{15}+\ff{1}{4}{\cal A}_{25}\right)\,e_i^2\,
\cos(3\lambda_i-5\lambda_o+2\varpi_i)+\ldots
\right\}.
\label{53gen}
\eea
Note for this example, however, that systems which exist stably in the $5\!:\!3$ resonance tend to have very small
values of the mass ratios $m_2/m_1$ and $m_3/m_1$ (generally of order $10^{-4}$),
in which case the approximations in Section~\ref{4.2.2} are reasonable.
One then obtains
\bea
{\cal R}&=&\frac{G m_2 m_3}{a_o}\left\{\ldots+
\left(\frac{67}{8}\,b_{1/2}^{(3)}(\alpha)
+\frac{9}{4}\,\alpha\,\frac{db_{1/2}^{(3)}}{d\alpha}
+\frac{1}{4}\alpha^2\,\frac{d^2b_{1/2}^{(3)}}{d\alpha^2}
\right)\,e_o^2\,
\cos(3\lambda_i-5\lambda_o+2\varpi_o)\right.\next
&&\hspace{2cm}
-\left(18\,b_{1/2}^{(4)}(\alpha)
+\frac{9}{2}\,\alpha\,\frac{db_{1/2}^{(4)}}{d\alpha}
+\frac{1}{2}\alpha^2\,\frac{d^2b_{1/2}^{(4)}}{d\alpha^2}
\right)\,e_ie_o\,
\cos(3\lambda_i-5\lambda_o+\varpi_i+\varpi_o)\next
&&\left.\hspace{4cm}+\left(
\frac{75}{8}\,b_{1/2}^{(5)}(\alpha)
+\frac{9}{4}\,\alpha\,\frac{db_{1/2}^{(5)}}{d\alpha}
+\frac{1}{4}\alpha^2\,\frac{d^2b_{1/2}^{(5)}}{d\alpha^2}
\right)\,e_i^2\,
\cos(3\lambda_i-5\lambda_o+2\varpi_i)+\ldots
\right\}.
\eea
On the other hand, one may wish, for example, to estimate the width of any of the $5\!:\!3$ resonances
for some arbitrary configuration (which may or may not be stable) in which case the expression \rn{53gen},
valid for arbitrary mass ratios, should be used.

As discussed in the previous section, 
in general there are $n'-n+1$ distinct principal resonances associated with the $n'\!:\!n$ resonance,
and each of these has $n'-n+1$ terms contributing to the lowest order in eccentricity.

\subsection{First-order resonances}\label{4.5}

Now consider those harmonic terms in the expansion which are first-order in eccentricity,
that is, those terms for which $j=0$ or 1 and for which 
\be
|m-n|+|m-n'|=n'-n=1
\ee
so that $m=n$ or $m=n+1$.
The relevant terms in the disturbing function are then
\bea
{\cal R}&=&\frac{G\mu_i m_3}{a_o}\left\{\ldots+
\ff{1}{2}\left[(2n+1){\cal A}_{0n}+{\cal A}_{1n}\right]\,e_o\,
\cos(n\lambda_i-(n+1)\lambda_o+\varpi_o)\right.\next
&&\left.\hspace{4cm}-\left[(n+1){\cal A}_{0n+1}+\ff{1}{2}{\cal A}_{1n+1}\right]\,e_i\,
\cos(n\lambda_i-(n+1)\lambda_o+\varpi_i)+\ldots
\right\},
\label{64}
\eea
and when $m_2\ll m_1$, this reduces to
\bea
{\cal R}&=&\frac{Gm_2 m_3}{a_o}\left\{\ldots+
\frac{1}{2}\left[(2n+1)b_{1/2}^{(n)}(\alpha)+\alpha\frac{db_{1/2}^{(n)}}{d\alpha}-4\alpha\,\delta_{n1}\right]\,e_o\,
\cos(n\lambda_i-(n+1)\lambda_o+\varpi_o)\right.\next
&&\left.\hspace{4cm}-\left[(n+1)b_{1/2}^{(n+1)}(\alpha)+\frac{\alpha}{2}\frac{db_{1/2}^{(n+1)}}{d\alpha}\right]\,e_i\,
\cos(n\lambda_i-(n+1)\lambda_o+\varpi_i)+\ldots
\right\}
\label{56}
\eea
which is consistent with \citet{papres}.
Here the term involving $\delta_{n1}$ comes from \rn{01} and \rn{11}.

\subsection{Resonance widths using the literal expansion}\label{4.6}
The literal expansion is especially suited to the study of systems with period ratios close to unity.
Stable systems in this category tend to have small mass ratios and at most modest eccentricities,
with the {\it induced} (forced) contributions to the latter being of order
$m_3/m_1$ and $m_2/m_1$ for the inner and outer eccentricities respectively (Paper II).
In this Section we derive an expression for the width of a general $[n'\!:\!n](m)$ resonance,
however, we will assume that the eccentricities are not so small that the apsidal advance
of one or both orbits contributes significantly to the resonance width; this case will be considered elsewhere.
Having said this, one should keep in mind the convergence issues discussed in the previous section.

Following the analysis in Section~\ref{secres} for the resonance width in
the case of the spherical harmonic expansion,
including the assumptions that the dynamics is dominated by a single harmonic
(not always true when the period ratio is close to 1)
and that $\dot\varpi_i$ and $\dot\varpi_o$ are negligible compared to $\nu_o$,
the librating angle $\phi_{mnn'}$ is governed by
\bea
\ddot\phi_{mnn'}&=&n\dot\nu_i-n'\dot\nu_o\next
&=&
-\frac{3}{2}\nu_o\left(n\sigma\frac{\dot a_i}{a_i}-n'\frac{\dot a_o}{a_o}\right)\next
&=&
\nu_o^2\left\{3\,n^2\left[\alpha\,\sigma^2\left(\frac{m_3}{m_{12}}\right)+\left(\frac{n'}{n}\right)^2
\left(\frac{m_1m_2}{m_{12}^2}\right)\right]
\sum_{j=0}^{j_{max}}{\cal A}_{jm}\,F_{mnn'}^{(j)}\right\}\sin\phi_{mnn'}\next
&\equiv& -\omega_{mnn'}^2\sin\phi_{mnn'},
\label{phiddl}
\eea
where $j_{max}\ge|m-n|+|m-n'|$ and is chosen to equal the highest order in eccentricity required.
Libration is around $\phi_{mnn'}=0$ if $\omega_{mnn'}^2>0$ and about
$\phi_{mnn'}=\pi$ if $\omega_{mnn'}^2<0$.
The angle $\phi_{mnn'}$ will librate when
\be
\dot\phi_{mnn'}=n\nu_o(\sigma-n'/n)<2\,\omega_{mnn'},
\ee
so that the width of the $[n'\!:\!n](m)$ resonance, that is, the maximum excursion of $\sigma$
away from $n'/n$ is given by \label{p25}
\be
\Delta\sigma_{mnn'}=2\sqrt{3}\left(\frac{n'}{n}\right)\left|\left[\alpha\,\left(\frac{m_3}{m_{12}}\right)+
\left(\frac{m_1m_2}{m_{12}^2}\right)\right]
\sum_{j=0}^{j_{max}}{\cal A}_{jm}\,F_{mnn'}^{(j)}\right|^{1/2},
\label{reswidth}
\ee
where we have put $\sigma=n'/n$ and $\alpha$ should be replaced by its value at exact
commensurability.

\subsubsection{Libration frequency}\label{libfreq}

As discussed in Section~\ref{3.1.2}, the libration frequency of a resonant harmonic is given by

\be
\omega_{mnn'}=\nu_o\Delta\sigma_{mnn'}/2.
\label{lfl}
\ee

\subsubsection{Widths of first-order resonances}\label{4.6.1}
The widths of the two principal first-order resonances are then
\be
\Delta\sigma_{n\,n\,n+1}=2\sqrt{3}\left(\frac{n+1}{n}\right)\left|\frac{1}{2}\left[\alpha\,\left(\frac{m_3}{m_{12}}\right)+
\left(\frac{m_1m_2}{m_{12}^2}\right)\right]
\left[(2n+1){\cal A}_{0n}+{\cal A}_{1n}\right]e_o\,\right|^{1/2}
\label{nnn}
\ee
and
\be
\Delta\sigma_{n+1\,n\,n+1}=2\sqrt{3}\left(\frac{n+1}{n}\right)\left|\left[\alpha\,\left(\frac{m_3}{m_{12}}\right)+
\left(\frac{m_1m_2}{m_{12}^2}\right)\right]
\left[(n+1){\cal A}_{0n+1}+\ff{1}{2}{\cal A}_{1n+1}\right]e_i\,\right|^{1/2}.
\label{nnnn}
\ee

\subsection{The secular disturbing function in the literal expansion}\label{sec2}

As for the spherical harmonic expansion (see Section~\ref{sec1}), 
the secular disturbing function is obtained by retaining the $n=n'=0$ terms only in \rn{R}. 
Again using the
notation $\tilde{\cal R}$ for the averaged disturbing function, we obtain
\be
\tilde{\cal R}=
\sum_{m=0}^\infty
\tilde{\cal R}_m\,\cos\left[m(\varpi_i-\varpi_o)\right],
\label{Rsec}
\ee
where now
\be
\tilde{\cal R}_m\equiv{\cal R}_{m00}=\frac{G\mu_i m_3}{a_o}\sum_{j=0}^\infty
{\cal A}_{jm}(\alpha;\beta_2)\,F_{m00}^{(j)}(e_i,e_o).
\ee
As before, ${\cal A}_{lm}$ and $F_{m00}^{(j)}$
are given by \rn{Alm} and \rn{Fmnn} respectively, but note that unlike for general $n,n'$,
closed-form expressions exist
for the Hansen coefficients $X_0^{k,m}(e_i)$ and $X_0^{-(k+1),m}(e_o)$;
these are given in Appendix~\ref{hansen}.
However, since the literal formulation for the disturbing function involves an expansion
in the eccentricities, it is still only correct to order $j_{max}$ in the eccentricities, where $j_{max}$
is the highest value of $j$ included in the expansion (see discussion in Section~\ref{4.1}).
Note that since $|m-n|+|m-n'|=2m$ is even, all terms in the secular expansion are even order
in eccentricity (including products of odd powers).

Recall from Section~\ref{sec1} on the spherical harmonic secular disturbing 
function that to octopole order (ie., including $l=2$ with $m=0,2$ and $l=3$ with $m=1,3$), only
terms with $m=0$ and $m=1$ are non-zero because
$X_0^{2,2}(e_o)=0$ and $X_0^{3,3}(e_o)=0$. Thus the only secular harmonic angle
appearing in the spherical harmonic development up to octopole level 
is $\phi_{100}=\varpi_i-\varpi_o$. 
From the point of view of the literal expansion, the coefficients of the 
``quadrupole'' and ``octopole'' harmonic angles
$\phi_{200}=2(\varpi_i-\varpi_o)$ and $\phi_{300}=3(\varpi_i-\varpi_o)$ are 
${\cal O}(e_i^2e_o^2)$ and ${\cal O}(e_i^3e_o^3)$ respectively, and are non-zero
because values of $j$ in addition to $j=2$ and $j=3$ contribute to them.
The literal planar secular disturbing function to second-order in the eccentricities 
and correct for any mass ratios is then
\be
\tilde{\cal R}=\frac{G\mu_i m_3}{a_o}
\left[{\cal A}_{00}+\ff{1}{2}(e_i^2+e_o^2)({\cal A}_{10}+{\cal A}_{20})
+\ff{1}{2}e_ie_o({\cal A}_{01}-{\cal A}_{11}-{\cal A}_{21})\cos(\varpi_i-\varpi_o)\right]+{\cal O}(e_i^2e_o^2)
\label{seclit}
\ee
so that to {\it first}-order in the eccentricities, the rates of change of the eccentricities and longitudes
of the periastra are, from Lagrange's planetary equations \rn{D1} and \rn{D2},
\bea
\frac{de_i}{dt}&=&\ff{1}{2}\nu_i\left(\frac{m_3}{m_{12}}\right)
e_o\,\alpha({\cal A}_{01}-{\cal A}_{11}-{\cal A}_{21})\sin(\varpi_i-\varpi_o),\label{121}\\
\frac{d\varpi_i}{dt}&=&\ff{1}{2}\nu_i\left(\frac{m_3}{m_{12}}\right)
\left[2\alpha({\cal A}_{10}+{\cal A}_{20})
+\left(\frac{e_o}{e_i}\right)\alpha({\cal A}_{01}-{\cal A}_{11}-{\cal A}_{21})\cos(\varpi_i-\varpi_o)\right],\\
\frac{de_o}{dt}&=&-\ff{1}{2}\nu_o\left(\frac{m_1m_2}{m_{12}^2}\right)
e_i\,({\cal A}_{01}-{\cal A}_{11}-{\cal A}_{21})\sin(\varpi_i-\varpi_o),\\
\frac{d\varpi_o}{dt}&=&\ff{1}{2}\nu_o\left(\frac{m_1m_2}{m_{12}^2}\right)
\left[2({\cal A}_{10}+{\cal A}_{20})
+\left(\frac{e_i}{e_o}\right)({\cal A}_{01}-{\cal A}_{11}-{\cal A}_{21})\cos(\varpi_i-\varpi_o)\right].
\label{124}
\eea
In the limiting case that $m_2/m_1\ll 1$, the disturbing function to second-order in the eccentricities 
becomes
\be
\tilde{\cal R}=\frac{Gm_2 m_3}{a_o}
\left[\ff{1}{2}b_{1/2}^{(0)}(\alpha)-1+\ff{1}{8}(e_i^2+e_o^2)
\left(2\alpha{\cal D}+\alpha^2{\cal D}^2\right)b_{1/2}^{(0)}(\alpha)
+\ff{1}{4}e_ie_o
\left(2-2\alpha{\cal D}-\alpha^2{\cal D}^2\right)b_{1/2}^{(1)}(\alpha)
\cos(\varpi_i-\varpi_o)\right],
\label{125}
\ee
where ${\cal D}\equiv d/d\alpha$.
This is consistent with equations (6.164--6.168) in \citet{murray} except for the additional
term $-Gm_2m_3/a_o$ which corresponds to their indirect term. Since in the secular case
this term is constant, it contributes nothing to the secular dynamics.

\section{Comparison of formulations to leading order in eccentricities}\label{acc}

In this Section we compare the two formulations in terms of the mass parameter
$|\beta_2|=m_2/m_{12}$ and the ratio of semimajor axes $\alpha$. 
The parameter $\beta_2$ is chosen because
it is taken to be zero in the classic literal expansion and is introduced here without restriction.
For each harmonic coefficient considered, the eccentricity dependence is factored out to leading order
and the resulting functional dependence on $|\beta_2|$ is compared. The dependence on $\alpha$ of the
resulting expression is therefore
exact in the literal case (to leading order in eccentricity), 
while in the spherical harmonic case it will depend on the number
of terms included in the summation over $l$. 

We start by defining the function ${\cal S}_{mnn'}(\alpha,\beta_2)$ such that the harmonic coefficients are given 
to leading order in the eccentricities by
\be
{\cal R}_{mnn'}\simeq \frac{G\mu_im_3}{a_o} {\cal S}_{mnn'}(\alpha,\beta_2)e_i^{|m-n|}e_o^{|m-n'|},
\ee
where for the spherical harmonic expansion,
\bea
{\cal S}_{mnn'}(\alpha,\beta_2)
&=&\sum_{l=l_{min}}^{l_{max}}\zeta_m c_{lm}^2\,{\cal M}_l\,
\alpha^l\, x_n^{l,m}x_{n'}^{-(l+1),m}\next
&=&\sum_{l=l_{min}}^{l_{max}}\zeta_m c_{lm}^2\,[(1-\beta_2)^{l-1}-\beta_2^{l-1}]\,
\alpha^l\, x_n^{l,m}x_{n'}^{-(l+1),m}
\label{Sspherical}
\eea
with $l_{min}$ given by \rn{s7b} and $l_{max}=l_{min}$ or $l_{min}+2$ (cases I and II),
while for the literal expansion,
\be
{\cal S}_{mnn'}(\alpha,\beta_2)=\sum_{j=0}^{|m-n|+|m-n'|}{\cal A}_{jm}(\alpha;\beta_2)\,f_{mnn'}^{(j)}
\label{Sliteral}
\ee
(case III).
Here
\be
x_n^{l,m}=\lim_{e_i\rightarrow 0}X_n^{l,m}(e_i)\,e_i^{-|m-n|},
\hspace{0.5cm}
x_{n'}^{-(l+1),m}=\lim_{e_o\rightarrow 0}X_{n'}^{-(l+1),m}(e_o)\,e_o^{-|m-n'|}
\ee
and
\be
f_{mnn'}^{(j)}=\lim_{e_i\rightarrow 0} \lim_{e_o\rightarrow 0}F_{mnn'}^{(j)}(e_i,e_o)e_i^{-|m-n|}e_o^{-|m-n|}
\ee
are the coefficients of $e_i^{|m-n|}$, $e_o^{|m-n'|}$ and $e_i^{|m-n|}e_o^{|m-n'|}$
in expansions of $X_n^{l,m}(e_i)$, $X_{n'}^{-{(l+1)},m}(e_o)$ and $F_{mnn'}^{(j)}(e_i,e_o)$ respectively.
Figure~\rn{PR21} 
\begin{figure}
\centering
\includegraphics[width=140mm]{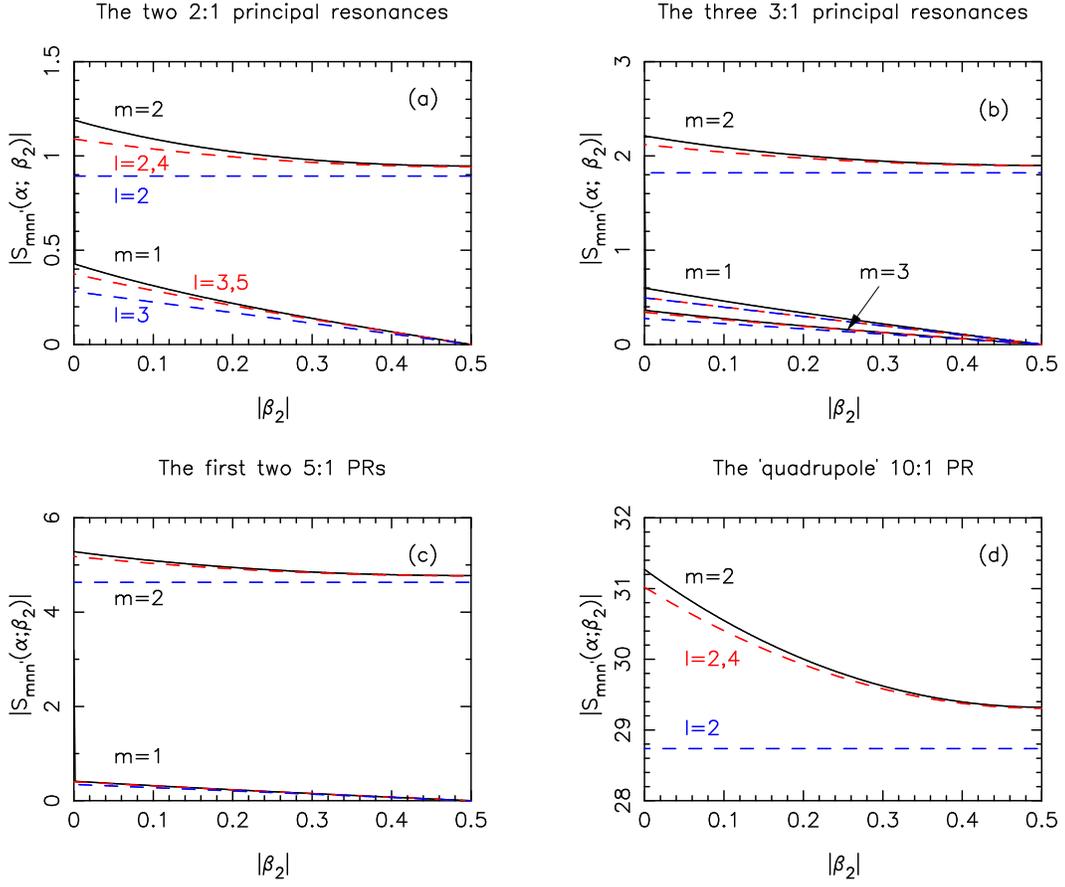}
\caption{Comparison of the dependence of harmonic coefficients on the inner mass ratio for the two expansions for various
harmonics. The function $S_{mnn'}(\alpha;\beta_2)$ is given by \rn{Sspherical} for the spherical harmonic expansion
and by \rn{Sliteral} for the literal expansion. One (blue dashed curves) and two (red dashed curves) values of $l$ are included
in the spherical harmonic expansion, while the literal expansion (black curves) is exact to the leading order in eccentricity.
For an $N\!:\!1$ harmonic, $\alpha=N^{-2/3}$ so that (a): $\alpha=0.63$,
(b): $\alpha=0.48$, (c): $\alpha=0.40$ and (a): $\alpha=0.22$.
}
\label{PR21}
\end{figure} 
compares $|{\cal S}_{mnn'}(\alpha,\beta_2)|$ for these three cases
for various $2\!:\!1$, $3\!:\!1$, $5\!:\!1$ and $10\!:\!1$ principal resonances,
while Figure~\rn{PR32} 
does the same for the two $3\!:\!2$ and $4\!:\!3$ principal resonances (see Section~\ref{4.3}
for their definition).
The biggest errors incurred are associated with $|\beta_2|=0$, and these decrease to zero for 
$|\beta_2|=0.5$. The main conclusion one draws from these comparisons is that even for
systems with period ratios as low as 1.5, including only the first two values of $l$ in the spherical
harmonic expansion produces quite accurate estimates of the harmonic coefficients. Since
the spherical harmonic expansion is the simplest to use of the two expansions, it is recommended for use
in preference to the literal expansion except when the period ratio is less than, say, 2.

Finally, while one might hope that the error incurred truncating the spherical harmonic expansion
at $\alpha^{l_{max}}$ is ${\cal O}(\alpha^{l_{max}+2})$, this is by no means guaranteed.
In fact, for the harmonics plotted in Figures~\rn{PR21} and \rn{PR32}, the error appears to be more
like ${\cal O}(\alpha^{l_{max}})$
for $\beta_2=0$, decreasing with increasing $\beta_2$.
For example, for the $[5\!:\!1](2)$ harmonic shown in Figure~\ref{PR21}(c),
$\alpha=5^{-2/3}=0.34$ so that including only the $l=2$ term in the spherical harmonic 
expansion (blue dashed line associated with $m=2$) incurs an error at $\beta_2=0$ of
$|\Delta{\cal S}_{215}/{\cal S}_{215}|={\cal O}(\alpha^2)\simeq 0.012$. Thus $|\Delta{\cal S}_{215}|\simeq 0.6$,
consistent with the figure.

\section{Equivalence of formulations}\label{equiv}

We now demonstrate the equivalence of the spherical harmonic and literal expansions,
which amounts to demonstrating the equivalence of the individual coefficients ${\cal R}_{mnn'}$
given by \rn{s7} and \rn{Rmnn}. Doing this will involve changing
summation orders as well as a change of variable. Throughout, one should keep in mind that
the indices $m,n,n'$ are fixed.

\begin{figure}
\centering
\includegraphics[width=150mm]{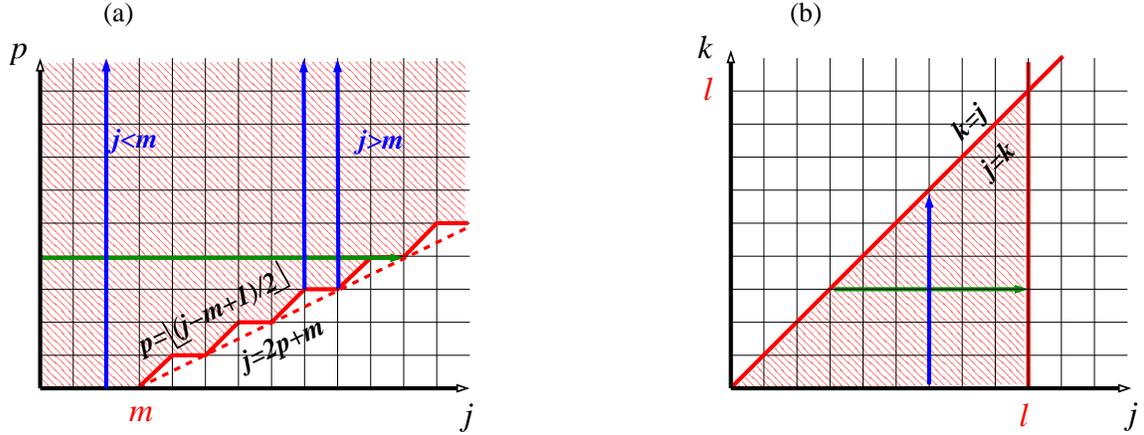}
\caption{Changing the order of summation. (a): Equations \rn{RRa} and \rn{RRaa}.
Notice how the summation boundary is stepped in the original order (solid red line), and smooth 
when the order is changed (dashed red line).
(b): Equations \rn{xi1} and \rn{xi2}.}
\label{grid1}
\end{figure}

Our aim is to show that
\be
{\cal R}_{mnn'}=\frac{G\mu_i m_3}{a_o}\sum_{l=l_{min},2}^\infty c_{lm}^2 {\cal M}_l\, \alpha^l\,
X_n^{l,m}(e_i)X_{n'}^{-(l+1),m}(e_o)
=\frac{G\mu_i m_3}{a_o}\sum_{j=0}^\infty {\cal A}_{jm}(\alpha;\beta_2)\,F_{mnn'}^{(j)}(e_i,e_o),
\label{Requiv}
\ee
where again,
\be
l_{min}=\left\{\begin{array}{ll}
2, & m=0\\
3, & m=1\\
m, &m\ge 2.
\end{array}
\right.
\label{s7bb}
\ee
Both ${\cal A}_{jm}$ and $F_{mnn'}^{(j)}$ can be expressed as series
given respectively by \rn{Alm} and \rn{A00} with \rn{e11} and \rn{Eplm}, and \rn{Fmnn}.
Noting the form of \rn{s7bb}, consider first $m\ge 2$.
Distinguishing the left and right sides of \rn{Requiv} by ${\cal R}_{mnn'}^{(L)}$ and ${\cal R}_{mnn'}^{(R)}$,
we have 
\be
{\cal R}_{mnn'}^{(L)}=\frac{G\mu_i m_3}{a_o}
\sum_{l=m,2}^\infty \frac{(-1)^{l+m}(l-m)!(l+m)!}{2^{2l-1}\left[((l+m)/2)!((l-m)/2)!\right]^2}
\left(\beta_1^{l-1}-\beta_2^{l-1}\right)\alpha^l
X_n^{l,m}(e_i)X_{n'}^{-(l+1),m}(e_o),
\label{RLa}
\ee
where we have used respectively \rn{clm2} and \rn{Ml} to replace $c_{lm}^2$ and ${\cal M}_l$,
and
\bea
{\cal R}_{mnn'}^{(R)}&=&\frac{G\mu_i m_3}{a_o}\sum_{j=0}^\infty 
\sum_{p=p_{min}}^\infty E_p^{(j,m)} \left[\beta_1^{m+2p-1}-\beta_2^{m+2p-1}\right]\,\alpha^{m+2p}\,
F_{mnn'}^{(j)}(e_i,e_o)\next
&=&
\frac{G\mu_i m_3}{a_o}\sum_{j=0}^\infty 
\sum_{p=p_{min}}^\infty 
\frac{2(2m+2p)!(m+2p)!(2p)!}{4^{2p+m}  j! (m+2p-j)! [p!(m+p)!]^2} 
\left[\beta_1^{m+2p-1}-\beta_2^{m+2p-1}\right]\,\alpha^{m+2p}\,
F_{mnn'}^{(j)}(e_i,e_o),
\label{RRa}
\eea
with
$p_{min}={\rm max}\left(0,\lfloor\ff{1}{2}(j-m+1)\rfloor\right)$ 
and $\lfloor\,\,\rfloor$ denoting the nearest lowest integer.
Referring to Figure~\ref{grid1}(a),
in the next step we change the order of summation of $j$ and $p$ in \rn{RRa}, then
make a change of variable for $p$ putting $l=m+2p$ 
so that
\bea
{\cal R}_{mnn'}^{(R)}&=&\frac{G\mu_i m_3}{a_o}\sum_{p=0}^\infty \sum_{j=0}^{2p+m}
\frac{2(2m+2p)!(m+2p)!(2p)!}{4^{2p+m}  j! (m+2p-j)! [p!(m+p)!]^2} 
\left[\beta_1^{m+2p-1}-\beta_2^{m+2p-1}\right]\,\alpha^{m+2p}\,
F_{mnn'}^{(j)}(e_i,e_o)
\label{RRaa}\\
&=&
\frac{G\mu_i m_3}{a_o}\sum_{l=m,2}^\infty\sum_{j=0}^l
\frac{(l+m)! \,l! \, (l-m)!}{2^{2l-1} j! (l-j)! \left[((l-m)/2)! ((l+m)/2)!\right]^2}
\left[\beta_1^{l-1}-\beta_2^{l-1}\right]\,\alpha^l\,
F_{mnn'}^{(j)}(e_i,e_o)\next
&=&
\frac{G\mu_i m_3}{a_o}\sum_{l=m,2}^\infty
c_{lm}^2\,
\left[\beta_1^{l-1}-\beta_2^{l-1}\right]\,\alpha^l\,
\left[\sum_{j=0}^l
\frac{l!}{j! (l-j)!}\,F_{mnn'}^{(j)}(e_i,e_o)\right]
\label{RRb}\\
&\equiv& 
\frac{G\mu_i m_3}{a_o}\sum_{l=m,2}^\infty
c_{lm}^2\,
\left[\beta_1^{l-1}-\beta_2^{l-1}\right]\,\alpha^l\,
\chi_{nn'}^{lm}.
\label{RRbb}
\eea
Note that $(-1)^{l+m}=1$ because $l+m$ is always even in the coplanar case.
Comparing \rn{RRb} with \rn{Requiv}, it remains to show that the expression in the large
square brackets in \rn{RRb} 
which we have defined as $\chi_{nn'}^{lm}$ in \rn{RRbb}
is equal to $X_n^{l,m}(e_i)X_{n'}^{-(l+1),m}(e_o)$.
Using the definition of $F_{mnn'}^{(j)}(e_i,e_o)$
from \rn{Fmnn}, we have
\be
\chi_{nn'}^{lm}=\sum_{j=0}^l\sum_{k=0}^j
(-1)^{j-k}
\frac{l!}{j! (l-j)!}
\frac{j!}{(j-k)!k!}
X_n^{k,m}(e_i)X_{n'}^{-(k+1),m}(e_o).
\label{xi1}
\ee
We therefore need to reduce the double summation over $j$ and $k$ to a single term.
With that aim in mind,
the next step involves gathering together the coefficients of each individual 
product $X_n^{k,m}(e_i)X_{n'}^{-(k+1),m}(e_o)$
by changing the order of summation of $j$ and $k$. Referring to Figure~\ref{grid1}(b), we then have
\bea
\chi_{nn'}^{lm}&=&\sum_{k=0}^l
\left[\sum_{j=k}^l
\frac{(-1)^{j-k}}{(l-j)!(j-k)!}\right]\frac{l!}{k!}
X_n^{k,m}(e_i)X_{n'}^{-(k+1),m}(e_o)\label{xi2}\\
&=&\sum_{k=0}^l
\delta_{kl}\,(l!/k!)\,
X_n^{k,m}(e_i)X_{n'}^{-(k+1),m}(e_o)\next
&=&
X_n^{l,m}(e_i)X_{n'}^{-(l+1),m}(e_o),
\eea
thereby verifying the equivalence of the formulations for $m\ge 2$.
When $m=1$, $\beta_1^{l-1}-\beta_2^{l-1}=0$ so that the summation over $l$ starts at $l=3$, consistent
with \rn{RLa}. When $m=0$, there is no contribution from $l=0$ because of the additional term
in the definition of ${\cal A}_{00}$ (see \rn{A00} and \rn{A00s}).
Thus the formulations are equivalent for all values of $m$.

\section{Comparisons with classic expansions}\label{compclass}

\subsubsection*{Kaula expansion}

Using our notation and setting the inclinations equal to zero, the \citet{kaula} expression for the disturbing function with $m_2=0$
{\it in units of energy per unit mass}, is
\be
{\cal R}_K=\sum_{l=2}^\infty {\cal R}_l,
\ee
where $l$ is the spherical harmonic degree, ${\cal R}_l$ is of the form (see his equation (10))
\be
{\cal R}_l=\frac{G m_3}{a_o}\left(\frac{a_i}{a_o}\right)^l\sum_{m=0}^l \sum_{\overline{n}=-\infty}^\infty\sum_{\overline{n}'=-\infty}^\infty
C_{lm}\,X_{l-m+\overline{n}}^{l,l-m}(e_i)\,X_{l-m+\overline{n}'}^{-(l+1),l-m}(e_o)\,
\cos \phi_{lm\overline{n}\overline{n}'},
\ee
$C_{lm}$ (which is different to our $c_{lm}$) is a constant involving lengthy expressions from \citet{kaula61},
and the harmonic angle is 
\be
\phi_{lm\overline{n}\overline{n}'}=(l-m+\overline{n})M_i-(l-m+\overline{n}')M_o+(l-m)(\varpi_i-\varpi_o).
\ee
Note that in this form, $l$ appears in the harmonic angle which is not the case in our spherical harmonic expansion
(see equation \rn{phi0}). In particular, the harmonic angle has four indices while ours has three, so
that all four cannot be independent.

Of special importance in our formulation is the simple form of the general harmonic angle and the clear relationship between the indices
and the natural frequencies in the problem (see Section~\ref{4.3} for a discussion of this point).
Moreover, by swapping the order of summation over the spherical harmonic indices $l$ and $m$ (equation~\rn{32}), 
they become effectively decoupled with
$m$ taking on the role of independent harmonic label (together with $n$ and $n'$), 
while $l$ retains the role of expansion index.

The new spherical harmonic expansion benefits especially from its general dependence on the mass ratios,
as well as its simple and evincing dependence on the eccentricities via power series and asymptotic approximations.
For the inner eccentricity dependence, power series expansions of the Hansen
coefficients associated with the dominant terms are accurate for most eccentricities less than unity 
(see Table~\ref{hansentab} and Figure~\ref{slmfig}(a)), while
for the outer eccentricity dependence, asymptotic expressions
demonstrate explicitly the exponential falloff of the harmonic coefficients with period ratio and eccentricity
(equations \rn{asymX} and \rn{Easym}).

\subsubsection*{Literal expansion}
 
Basing their analysis on the work of \citet{leverrier}, a lengthy derivation of 
the literal expansion of the direct and indirect parts of the disturbing function for the restricted problem
is given in  \citet{murray} to second order in the eccentricities and inclinations, and for either $m_2=0$ or $m_3=0$.
No general expression for the harmonic coefficients is given,\footnote{A general expression for the
coefficients of a hybrid Kaula-literal expansion \citep{ellis-murray} is given, with many
conditions on the maximum and minimum values of the summation indices. 
The derivation of this expression is not given, and a promise of such a derivation does not appear to have
been met. In addition, there is no discussion of the errors associated with the expansion. 
Note that there appears to be a typographical
error in equation (52) of \citet{ellis-murray}
which has carried over to equation (6.113) of \citet{murray}: the index $j$ on the Laplace coefficient should be $k$.
}
but rather several tables of the harmonic angles and their coeffiencients up to fourth order
in those elements are provided. 

One of the features of our formulation which simplifies the analysis is the fact that 
the expression for the general harmonic coefficient \rn{Rmnn} involves derivatives of
$b_{1/2}^{(m)}$ only, rather than $b_s^{(m)}$ for many values of the half-integer index $s$ (although the derivatives themselves involve
values of $s$ other than $s=1/2$ via \rn{C8} and \rn{C9}). 
This together with simple expressions for the
eccentricity functions makes it straightforward to write down any harmonic coefficient
(see Appendix~\ref{hansen} which gives series approximations for Hansen coefficients
as well as a short {\it Mathematica} program which generates a power series for $F_{mnn'}^{(j)}(e_i,e_o)$ to the order of the expansion).
We have demonstrated in Sections~\ref{53} and \ref{4.5} the ease with which this can be done.

It has previously been assumed that it is not possible to use Jacobi coordinates for a literal expansion 
of the disturbing function without performing an expansion in the mass ratios as well (see, for example,
the discussion on page 195 of \citet{laskar95}). Here
we avoid such an expansion by taking advantage of the symmetry in the mass ratios $m_1/m_{12}$ and $m_2/m_{12}$
in our form of the disturbing function \rn{distlit}. The novelty is in the use of two mass-weighted ratios of the semimajor axes
as the arguments of Laplace coefficients.
Note that the usual form for the disturbing function (in units of energy per unit mass) is 
\be
{\cal R}_i=Gm_i\left[\frac{1}{|{\bf R}-{\bf r}|}-\frac{{\bf r}\cdot{\bf R}}{R^3}\right],
\ee
where $m_i$ is $m_2\ll m_1$ or $m_3\ll m_1$, that is, there are two distinct disturbing functions, each with a
``direct'' and ``indirect'' term (the first and second terms respectively).

\section{Conclusion and highlights of new results}\label{6}

The aim of this paper has been to provide new general expansions of the disturbing function 
which are clear and accessible to anyone contributing to the rapidly expanding field of exoplanets,
as well as to many other fields of astrophysics.
The expansions are applicable to systems with arbitrary mass ratios, eccentricities and period ratios,
making them suitable for the study of any of the diverse
stellar and planetary configurations now being discovered in great numbers by surveys such as HARPS and Kepler.

The many applications include determining the rates of change of the orbital elements of both secular and
resonant systems; calculating the widths of resonances for the purpose of studying libration cycles 
and their effect on TTVs, or for the purpose of studying the stability characteristics of arbitrary
configurations (not just those with small eccentricities and masses); deriving analytical constraints
for use in orbit fitting procedures; and calculating the dynamical characteristics of circumbinary planetary systems.

Several new results and concepts have been introduced here including
\ben
\item
Arbitrary dependence of the disturbing function on the mass ratios for both the spherical harmonic 
and literal expansions;
\item
Simple general expressions for all harmonics to arbitrary order in the ratio of semimajor axes (spherical
harmonic expansion) and the eccentricities (literal expansion);
\item
Accurate and simple approximations for Hansen coefficients for $0\le e_i\le 1$ and $0\le e_o\le 1$
for the dependence of both expansions on eccentricities, 
including asymptotic expressions (Paper II) for the outer eccentricity which reveal the exponential dependence
of the disturbing function on $n'$ and $e_o$;
\item
The fact that
for a given level of accuracy, the order in eccentricity at which one truncates the series
depends on the configuration being studied.
For example, given the eccentricities, one requires fewer terms when studying the $2\!:\!1$ resonance than one does for
the $7\!:\!6$ resonance, even though they are both first order resonances;
\item
The equivalence of the spherical harmonic and literal expansions revealing the role of the spherical harmonic order $m$
in the literal expansion;
\item
The concept of ``principal resonances'' and the physical importance of the spherical harmonic order $m$
including ``Zeeman splitting'' of resonances;
\item
Comparison of the two expansions showing that the simpler spherical harmonic expansion can be used for 
problems with period ratios as low as 2.
\een

This work has revealed that
the link between the three-body problem and spherical harmonics is more than just a convenient
way to label Fourier terms. Via analogy with other physical problems involving spherical harmonics, the analysis presented here
has the potential to expose deep symmetries in this rich problem.

\section{Quick reference}\label{quick}

This section provides a quick reference to the main results for readers mainly interested 
in their application. Equation numbers corresponding to the main text are provided.

The paper derives two expansions, one in the ratio of semimajor axes
(the {\it spherical harmonic} expansion),
and the other in the eccentricities (the {\it literal} expansion). Both are valid for any masses.
The choice of which to use
depends on the configuration being studied as well as the application, and there is no clear boundary between them. 
In general one uses the expansion in eccentricity for systems with period ratios less than, say, two or three and for which 
the eccentricities are small,
while the expansion in semimajor axis ratio is best for wider eccentric systems. 

For both expansions 
the disturbing function is expressed as a triple Fourier series over the indices $m$, $n$ and $n'$,
where the frequencies associated with $n$ and $n'$ are those of the inner and outer orbits, and the 
frequency associated with $m$ is the difference in the apsidal motion rates.
We write this as
\[
{\cal R}=\sum_{m=0}^\infty \sum_{n=-\infty}^\infty \sum_{n'=-\infty}^\infty\,
{\cal R}_{mnn'}\left(e_i,e_o,a_i,a_o;m_1,m_2,m_3\right)
\,\cos\phi_{mnn'},
\hspace{71mm}
\rn{s6}, \rn{R}
\]
where the {\it harmonic angle} $\phi_{mnn'}$ 
can be written in terms of the mean anomalies of the inner and outer orbits,
$M_i$ and $M_o$, or the corresponding mean longitudes $\lambda_i$ and $\lambda_o$, as well as the 
longitudes of periastron $\varpi_i$ and $\varpi_o$, so that
\bean
\phi_{mnn'}&=&n M_i-n'M_o+m(\varpi_i-\varpi_o)
\\
&=&n\lambda_i-n'\lambda_o+(m-n)\varpi_i-(m-n')\varpi_o.
\hspace{90mm}
\rn{phi}, \rn{resang2}
\eean
The {\it harmonic coefficients} ${\cal R}_{mnn'}$ depend on the other parameters in the problem, namely
the inner and outer eccentricities $e_i$ and $e_o$, the semimajor axes $a_i$ and $a_o$,
and the masses $m_1$, $m_2$ and $m_3$, and are given below for the
semimajor axis and eccentricity expansions respectively.

\subsection{${\cal R}_{mnn'}$ for the semimajor axis expansion}\label{2.1}
In this case, the harmonic coefficients are given by
\bean
{\cal R}_{mnn'}&=&\frac{G\mu_i m_3}{a_o}\sum_{l=l_{min},2}^{l_{max}}
\zeta_m c_{lm}^2\,
{\cal M}_l\,
\alpha^l\,
X_{n}^{l,m}(e_i)X_{n'}^{-(l+1),m}(e_o)
\hspace{79mm}
\rn{s7}\\
&=&\frac{G\mu_i m_3}{R_p}\sum_{l=l_{min},2}^{l_{max}}
\zeta_m c_{lm}^2\,
{\cal M}_l\,
\rho^l\,
X_{n}^{l,m}(e_i)Z_{n'}^{-(l+1),m}(e_o).
\hspace{79mm}
\rn{rhoform}
\eean
where $\alpha=a_i/a_o$, $\rho=a_i/R_p$ with $R_p$ the outer periastron distance,
$l_{min}=m$ for $m\ge 2$ and 2 or 3 if $m=0$ or 1 respectively, $l_{max}$ is chosen
according to the accuracy required, the notation $\sum_{l=l_{min},2}$ means the summation is in steps of 2,
$\zeta_m$ takes on the values 1/2 or 1 according to whether $m$ is zero or not zero respectively,
$c_{20}^2=1/2$, $c_{22}^2=3/4$, $c_{31}^2=3/8$, $c_{33}^2=5/8$ with a general expression given by \rn{clm},
and 
\[
{\cal M}_l=\frac{m_1^{l-1}+(-1)^l m_2^{l-1}}{(m_1+m_2)^{l-1}}.
\hspace{135mm}
\rn{Ml}
\]
The eccentricity functions
$X_n^{l,m}(e_i)$
and $X_{n'}^{-(l+1),m}(e_o)$ are Hansen coefficients and $Z_{n'}^{-(l+1),m}(e_o)=(1-e_o)^{l+1}X_{n'}^{-(l+1),m}(e_o)$ is a modified Hansen coefficient.
These are defined and discussed in Appendix~\ref{hansen}.
The leading terms in their Taylor expansions are
${\cal O}(e_i^{|m-n|})$ and ${\cal O}(e_o^{|m-n'|})$ respectively.
They may be evaluated numerically
with as much precision as one requires, by closed-form expression when $n=n'=0$,
or approximately by series expansion (Sections~\ref{expan} and \ref{math}) or 
closed-form asymptotic approximation (equation~\rn{asymX}).

\subsubsection{Secular disturbing function to octopole order}\label{2.1.2}
The secular disturbing function $\tilde{\cal R}$ is given by \rn{s7} with $n=n'=0$.
To octopole order this is
\[
\tilde{\cal R}=\frac{G\mu_i m_3}{a_o}
\left[\frac{1}{4}\,\left(\frac{a_i}{a_o}\right)^2\,\frac{1+\ff{3}{2}e_i^2}{(1-e_o^2)^{3/2}}\,
-\frac{15}{16}\,\left(\frac{a_i}{a_o}\right)^3\,\left(\frac{m_1-m_2}{m_{12}}\right)\,
\frac{e_i e_o(1+\ff{3}{4}e_i^2)}{(1-e_o^2)^{5/2}}\,\cos(\varpi_i-\varpi_o)
\right].
\hspace{38mm}
\rn{Rsecoct}
\]
The secular rates of change of the eccentricities and apsidal longitudes are given by \rn{ei} to \rn{wo}.
Note that care should be taken when using secular expansions; see the discussion in Section~\ref{df}.

\subsubsection{Dominant non-secular terms}\label{domnon}
The dominant non-secular harmonics for systems well represented by the semimajor axis expansion tend to 
be those with $m=2$ and $n=1$. Including only $l=2$  
in \rn{s7} and using Table~\ref{hansentab} for $X_1^{2,2}(e_i)$
and the asymptotic approximation \rn{asymX} for  $X_{n'}^{-3,2}(e_o)$ (with accuracies indicated
in Table~\ref{hansentab} and Figure~\ref{asym} respectively), their coefficients are approximately
\be
{\cal R}_{21n'}=-\frac{G\mu_im_3}{a_o}\left(\frac{{\cal H}_{22}}{\sqrt{2\pi}}\right)
\,\alpha^2(3e_i-\ff{13}{8}e_i^3)(1-e_o^2)^{3/4}n'^{3/2}e^{-n'\xi(e_o)}e_o^{-2},
\label{explicit}
\ee
where $\xi(e_o)={\rm Cosh}^{-1}(1/e_o)-\sqrt{1-e_o^2}$ and ${\cal H}_{22}=0.71$.\footnote{Note the additional scaling
factors indicated in Figure~\ref{asym} for $n'\le 10$.}
Note the steep dependence on $n'\xi(e_o)$.
Note also that at this order of the expansion (quadrupole) there is no dependence on the inner mass ratio 
(apart from the factor $\mu_i$) because ${\cal M}_2=1$.

Most systems down to a period ratio of around 2 are well approximated by the spherical harmonic expansion
with only one or two values of $l$ included (see Section~\ref{acc}).

\subsubsection{Widths and libration frequencies of $[N\!:\!1](2)$ resonances}\label{2.1.1}

The spherical harmonic expansion is especially useful for studying the stability properties
of eccentric systems with moderate mass ratios (Paper II). Such systems tend to have
significant period ratios and hence can be quite accurately truncated at the quadrupole level.
For a system with period ratio close to the integer value $N$,
the harmonic angle of interest is the one corresponding to $n=1$, $n'=N$ and $m=2$, that is,
\be
\phi_{21N}=\lambda_i-N\lambda_o+\varpi_i+(N-2)\varpi_o.
\ee
Using the general notation $[n'\!:\!n](m)$ for an $n'\!:\!n$ resonance of spherical harmonic order $m$,
the width of the $[N\!:\!1](2)$ resonance is approximately
\[
\Delta\sigma_N=\frac{6{\cal H}_{22}^{1/2}}{(2\pi)^{1/4}}\left[\left(\frac{m_3}{m_{123}}\right)+N^{2/3}
\left(\frac{m_{12}}{m_{123}}\right)^{2/3}\left(\frac{m_1m_2}{m_{12}^2}\right)\right]^{1/2}
\left(\frac{e_i^{1/2}}{e_o}\right)(1-\ff{13}{24}e_i^2)^{1/2}(1-e_o^2)^{3/8}{N}^{3/4}\,{\rm e}^{-N\xi(e_o)/2}.
\hspace{16mm}
\rn{res}
\]
Note that $\lim_{e_o\rightarrow 0}\Delta\sigma_N$ is infinite for $N=1$, finite for $N=2$ and zero for $N\ge 3$.
The corresponding libration frequency $\omega_N$ is
\[
\omega_N=\nu_o\Delta\sigma_N/2,
\hspace{148mm}
\rn{libw}
\]
where $\nu_o$ is the outer orbital frequency.

\subsection{${\cal R}_{mnn'}$ for the eccentricity expansion}\label{2.2}

In this case, the harmonic coefficients are given by
\[
{\cal R}_{mnn'}=\frac{G\mu_i m_3}{a_o}\sum_{j=0}^{j_{max}}
{\cal A}_{jm}(\alpha;\beta_2)\,F_{mnn'}^{(j)}(e_i,e_o),
\hspace{103mm}
\rn{Rmnn}
\]
where $j_{max}$ is the order in eccentricity of the expansion,
the eccentricity functions $F_{mnn'}^{(j)}(e_i,e_o)$ are finite sums of products of Hansen coefficients 
given by
\[
F_{mnn'}^{(j)}(e_i,e_o)=\sum_{k=0}^j 
(-1)^{j-k}{j\choose k}
X_{n}^{k,m}(e_i)\,X_{n'}^{-(k+1),m}(e_o),
\hspace{90mm}
\rn{Fmnn}
\]
with the latter defined and discussed in Appendix~\ref{hansen},
and the ${\cal A}_{jm}$ depend on the semimajor axis ratio $\alpha$ and the mass ratios
$\beta_1=m_1/m_{12}$, $\beta_2=-m_2/m_{12}$ through
\[
{\cal A}_{jm}(\alpha;\beta_2)
=\zeta_m\,\left[\beta_1^{-1}{\cal B}_{1/2}^{(j,m)}(\alpha_1)-\beta_2^{-1}{\cal B}_{1/2}^{(j,m)}(\alpha_2)\right]
\hspace{98mm}
\rn{Alm}
\]
for all $j$, $m$ except when $j=m=0$ in which case
\[
{\cal A}_{00}(\alpha;\beta_2)
=\ff{1}{2}\,\left[\beta_1^{-1}b_{1/2}^{(0)}(\alpha_1)-\beta_2^{-1}b_{1/2}^{(0)}(\alpha_2)\right]
+\beta_1^{-1}\beta_2^{-1}.
\hspace{92mm}
\rn{A00}
\]
Here $\alpha_s=\beta_s\,\alpha$, $s=1,2$,
$b_{1/2}^{(m)}(\alpha_s)$ is a Laplace coefficient defined in \rn{e1b}
and
${\cal B}_{1/2}^{(j,m)}(\alpha_s)=(\alpha_s^j/j!)(d^j/d\alpha_s^j)b^{(m)}_{1/2}(\alpha_s)$.
The evaluation of these is discussed in Appendix~\ref{formulae}, while a {\it Mathematica} program for 
series expansions of $F_{mnn'}^{(j)}(e_i,e_o)$ is given
in Appendix~\ref{math}.

Examples are presented in Sections~\ref{53} and \ref{4.5} in which terms contributing to the $5\!:\!3$ resonance
and general first-order resonances are calculated.

\subsubsection{The secular disturbing function to second order in the eccentricities}\label{sdf}

To second-order in the eccentricities the secular disturbing function is
\[
\tilde{\cal R}=\frac{G\mu_i m_3}{a_o}
\left[{\cal A}_{00}+\ff{1}{2}(e_i^2+e_o^2)({\cal A}_{10}+{\cal A}_{20})
+\ff{1}{2}e_ie_o({\cal A}_{01}-{\cal A}_{11}-{\cal A}_{21})\cos(\varpi_i-\varpi_o)\right]+{\cal O}(e_i^2e_o^2)
\hspace{35mm}
\rn{seclit}
\]
which reduces to \rn{125} in the limit that $m_2/m_{12}\rightarrow 0$.
Equations governing the rates of change of the elements are given in \rn{121} to \rn{124}.
Note that care should be taken when using secular expansions; see the discussion in Section~\ref{df}.

\subsubsection{Widths and libration frequencies of $[n'\!:\!n](m)$ resonances}\label{rw}

The widths of the principal harmonics of the $n'\!:\!n$ resonance (those which are lowest order in the eccentricities) are given by
\[
\Delta\sigma_{mnn'}=2\sqrt{3}\left(\frac{n'}{n}\right)\left|\left[\alpha\,\left(\frac{m_3}{m_{12}}\right)+
\left(\frac{m_1m_2}{m_{12}^2}\right)\right]
\sum_{j=0}^{j_{max}}{\cal A}_{jm}\,F_{mnn'}^{(j)}\right|^{1/2},
\hspace{71mm}
\rn{reswidth}
\]
where $n\le m\le n'$. The corresponding libration frequencies are
\[
\omega_{mnn'}=\nu_o\Delta\sigma_{mnn'}/2.
\hspace{138mm}
\rn{lfl}
\]
To first-order in the eccentricities, the widths of the two principal first-order resonances are
\[
\Delta\sigma_{n\,n\,n+1}=2\sqrt{3}\left(\frac{n+1}{n}\right)\left|\left[\alpha\,\left(\frac{m_3}{m_{12}}\right)+
\left(\frac{m_1m_2}{m_{12}^2}\right)\right]
\left[(n+1){\cal A}_{0n}+\ff{1}{2}{\cal A}_{1n}\right]e_o\,\right|^{1/2}
\hspace{54mm}
\rn{nnn}
\]
and
\[
\Delta\sigma_{n+1\,n\,n+1}=2\sqrt{3}\left(\frac{n+1}{n}\right)\left|\frac{1}{2}\left[\alpha\,\left(\frac{m_3}{m_{12}}\right)+
\left(\frac{m_1m_2}{m_{12}^2}\right)\right]
\left[(2n+1){\cal A}_{0n+1}+{\cal A}_{1n+1}\right]e_i\,\right|^{1/2}.
\hspace{42mm}
\rn{nnnn}
\]
Note that expressions \rn{reswidth} to \rn{nnnn} do not include contributions from $\ddot\varpi_i$ and/or $\ddot\varpi_o$
which can be significant for first-order resonances when the eccentricities are very small.

\section*{Acknowledgments}\label{ack}
The author wishes to thank Jaques Laskar for suggesting the nomenclature 
``harmonic angle'' for the general (not necessarily resonant) angle occurring
in Fourier expansions of the disturbing function,
Stephane Udry for a thorough critical reading of the manuscript 
which improved its clarity for the non-celestial mechanician,
and the anonymous referee whose suggestions resulted in significant improvements to the paper.

\section*{Dedication}
For SDU.

\appendix

\section{Spherical harmonics}\label{shapp}

Using the definition of $Y_{lm}(\theta,\varphi)$ \citep{jackson}
\be
Y_{lm}(\theta,\varphi)=\sqrt{\frac{2l+1}{4\pi}\frac{(l-m)!}{(l+m)!}}
P_l^m(\cos\theta)\,e^{im\varphi}
\ee
with
\bea
P_l^m(x)&=&\frac{(-1)^m}{2^l l!}(1-x^2)^{m/2}\frac{d^{l+m}}{dx^{l+m}}(x^2-1)^l
=\frac{(-1)^m}{2^l l!}(1-x^2)^{m/2}\frac{d^{l+m}}{dx^{l+m}}\sum_{j=0}^l
{l\choose j}(-1)^{l-j}x^{2j}\next
&=&\frac{(-1)^m}{2^l l!}(1-x^2)^{m/2}
\sum_{j=\lfloor(l+m+1)/2\rfloor}^l (-1)^{l-j}{l\choose j}\frac{(2j)!}{(2j-l-m)!}x^{2j-l-m},
\label{Plm}
\eea
where $\lfloor\,\,\,\rfloor$ denotes the nearest lowest integer,
we have
\be
Y_{lm}(\pi/2,f)=\sqrt{\frac{2l+1}{4\pi}\frac{(l-m)!}{(l+m)!}}\,
P_l^{m}(0)\,e^{imf}\equiv \sqrt{\frac{2l+1}{4\pi}}\,c_{lm}\,e^{imf}
\label{Ylmclm}
\ee
where from equation~\rn{Plm},
\be
P_l^{m}(0)=
(-1)^{(l+m)/2}{l\choose (l+m)/2}\frac{(l+m)!}{2^l l!},\,\,\,\,  l+m\,\,{\rm even}
\label{Plm0}
\ee
and zero otherwise, so that
\be
c_{lm}^2=\frac{(l-m)!(l+m)!}{2^{2l-1}\left[\left((l+m)/2\right)! \left((l-m)/2\right)!\right]^2}.
\label{clm2}
\ee
Note that the association of non-zero values of $P_l^m(0)$ with even $l+m$ is consistent
with the sum in \rn{calR2} being in steps of 2.
Some values of $c_{lm}^2$ are listed in Table~\ref{eccfuns}.

\section{Hansen coefficients}\label{hansen}

The two formulations presented in this paper are distinguished by the expansion parameter;
for the spherical harmonic expansion the parameter is the ratio of semimajor axes and it places
no restrictions on the two eccentricities, while for the literal expansion the parameters are
the eccentricities, with no restriction on the ratio of semimajor axes (except that the orbits should not cross).
For the spherical harmonic expansion, we therefore require expressions for the Hansen coefficients
which are accurate for all eccentricities, while for the literal expansion, power series
representations are appropriate because the expansion is valid only to order $j_{max}$ in the 
combined powers of the eccentricities.

\subsection{Hansen coefficients relevant for the spherical harmonic expansion}\label{B1}

Hansen coefficients are defined such that
\be
X_n^{l,m}(e_i)=\frac{1}{2\pi}\int_0^{2\pi}(r/a_i)^l e^{imf_i}\,e^{-inM_i}dM_i
\ee
and
\be
X_{n'}^{-(l+1),m}(e_o)
=\frac{1}{2\pi}\int_0^{2\pi}\frac{e^{-imf_o}}{(R/a_o)^{l+1}}e^{in'M_o}dM_o.
\ee
Figures~\rn{slmfig} and \rn{flmfig} %
\begin{figure}
\centering
\includegraphics[width=140mm]{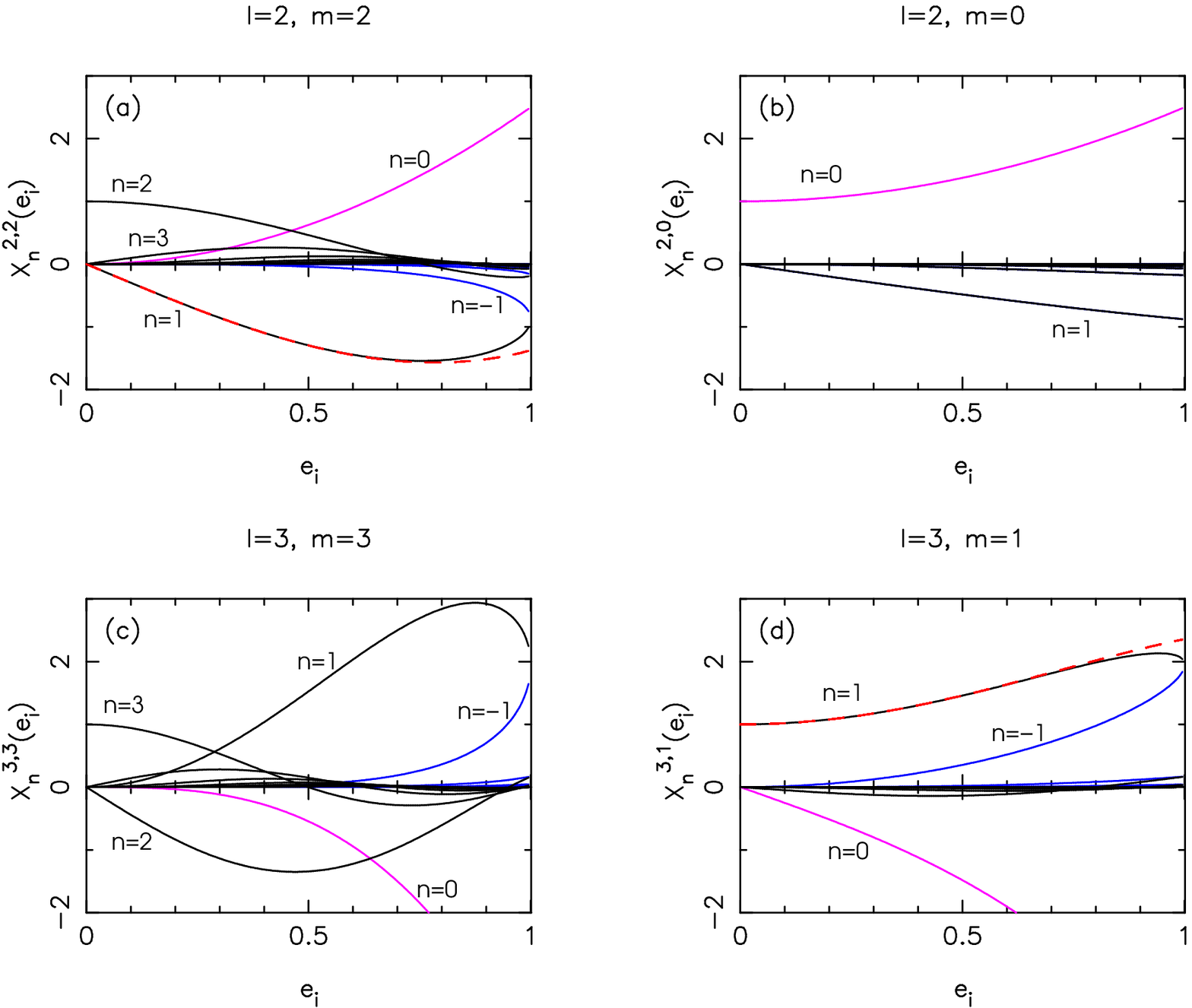}
\caption{Hansen coefficients $X_{n}^{l,m}(e_i)$ for quadrupole ($l=2$, $m=0,2$) and
octopole ($l=3$, $m=1,3$) values of $l$ and $m$,
and for various values of $n$, with dominant values labeled. Notice the ${\cal O}(e_i^{|m-n|})$ behaviour for small $e_i$.
The black curves are for $n=1,2,\ldots,10$, the blue curves are for $n=-10,-9,\ldots,-1$
and the pink curves are for $n=0$ for which closed-form expressions are given in Table~\ref{eccfuns}.
The red dashed curves are for the polynomial approximations $X_1^{2,2}(e_i)\simeq -3e_i+\ff{13}{8}e_i^3$.
and $X_1^{3,1}(e_i)\simeq 1+2e_i^2-\ff{41}{64}e_i^4$ (see Table~\ref{hansentab}).
}
\label{slmfig}
\end{figure}
\begin{figure}
\centering
\includegraphics[width=140mm]{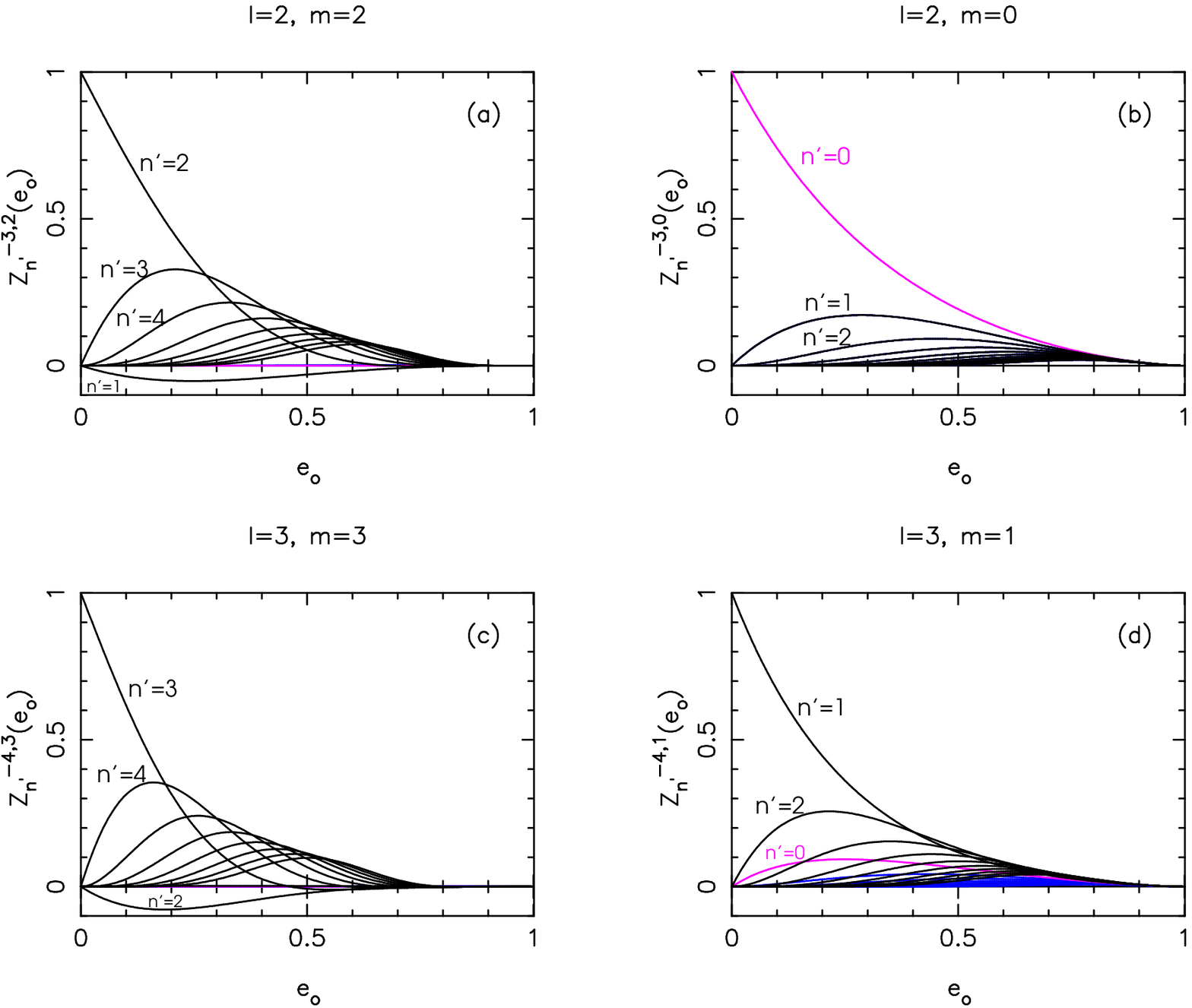}
\caption{The modified Hansen coefficients $Z_{n'}^{-(l+1),m}(e_o)=(1-e_o)^{l+1}X_{n'}^{-(l+1),m}(e_o)$ 
for quadrupole ($l=2$, $m=0,2$) and
octopole ($l=3$, $m=1,3$) values of $l$ and $m$,
and for various values of $n$, with dominant values labeled. 
The black curves are for $n=1,2,\ldots,10$, the blue curves are for $n=-10,-9,\ldots,-1$ (only visible in panel (d)),
and the pink curves are for $n=0$ for which closed-form expressions are given in Table~\ref{eccfuns}.
}
\label{flmfig}
\end{figure}
show the numerically integrated functions $X_{n}^{l,m}(e_i)$ and
$Z_{n'}^{-(l+1),m}(e_o)=(1-e_o)^{l+1}X_{n'}^{-(l+1),m}(e_o)$ 
for quadrupole and octopole values of $l$ and $m$, and for the first 10 positive values of $n'$ and $n$
(see equations~\rn{s4} and \rn{s5}).
The scaling factor $(1-e_o)^{l+1}$ replaces $a_o$ with the outer periastron separation
$R_p=a_o(1-e_o)$ in the definition \rn{s5} of $X_{n'}^{-(l+1),m}(e_o)$, factoring out the singularity
at $e_o=1$. We refer to the $Z_{n'}^{-(l+1),m}(e_o)$ as {\it modified} Hansen coefficients.
While no closed form expressions for these integrals exist (except for $n'=n=0$; see 
Section~\ref{hansen0}),
for many applications it is reasonable (CPU-wise) to integrate them numerically.
However, simple approximations exist as outlined below, the analytic form of which provides insight into 
the behaviour of the physical variables which depend on them.

In Section~\ref{expan} we give general power series expansions which are correct to fourth order
in the eccentricity. Amongst other things, these expressions demonstrate that the leading terms
are such that
\be
X_{n}^{l,m}(e_i)={\cal O}(e_i^{|m-n|})
\hand
X_{n'}^{-(l+1),m}(e_o)={\cal O}(e_i^{|m-n'|}),
\label{sF0}
\ee
consistent with Figures~\ref{slmfig} and \ref{flmfig}.

For most applications for which a spherical harmonic expansion of the disturbing function 
is appropriate, it suffices to know expressions for the Hansen coefficients associated with
the $[n'\!:\!1](2)$ harmonics for $l=2,4$ (see, for example, Sections~\ref{dom} and \ref{secres}),
that is, $X_1^{2,2}(e_i)$, $X_1^{4,2}(e_i)$, $X_{n'}^{-3,2}(e_o)$ and $X_{n'}^{-5,2}(e_o)$,
and perhaps those associated with 
the $[2n'+1\!:\!2](2)$ harmonics (those half-way between the $[n'\!:\!1](2)$ harmonics),
that is, $X_2^{2,2}(e_i)$, $X_2^{4,2}(e_i)$, $X_{2n'+1}^{-3,2}(e_o)$ and $X_{2n'+1}^{-5,2}(e_o)$.
Table~\ref{hansentab}
\begin{table*}
 \centering
 \begin{minipage}{140mm}
  \caption{Hansen coefficients for $n=1$ and $n=2$, errors ${\cal E}_{e_i}\equiv |\delta X_n^{l,m}/{\rm max}_{e_i}(X_n^{l,m})|$
  and scale factors
  }
  \label{hansentab}
   \begin{tabular}{@{}lllllll@{}}
  \hline
$l$ & $m$ & $X_1^{l,m}(e_i)$ & ${\cal E}_{0.7},\,{\cal E}_{0.9}$& $X_2^{l,m}(e_i)$  & ${\cal E}_{0.7},\,{\cal E}_{0.9}$  & ${\cal H}_{lm}$\\
   \hline
2 & 2 & $-3e_i+\ff{13}{8}e_i^3+\ff{5}{192}e_i^5$ & 0.006, 0.05 & $1-\ff{5}{2}e_i^2+\ff{23}{16}e_i^4-\ff{65}{288}e_i^6$ & 0.002, 0.025 & 0.71\\
&&&&\\
4 & 2 & $-4e_i-3e_i^3+\ff{79}{48}e_i^5$ &0.006, 0.025 & $1+e_i^2-\ff{43}{16}e_i^4+\ff{35}{36}e_i^6$ &0.007, 0.06 & 1.44\\
&&&&\\
3 & 1 & $1+2e_i^2-\ff{41}{64}e_i^4-\ff{37}{576}e_i^6$ & 0.0007, 0.014& $-\ff{1}{2}e_i+e_i^3-\ff{35}{96}e_i^5+\ff{23}{576}e_i^7$ 
& 0.002, 0.006 & 1.91\\
\hline
\end{tabular}
\end{minipage}
\end{table*}
gives the first few terms of the series expansions of
these Hansen coefficients for which $e_i$ is the argument, 
as well as the error
at $e_i=0.7$ and 0.9, defined such that
\be
{\cal E}_{e_i}\equiv \left|\frac{\delta X_n^{l,m}(e_i)}{{\rm max}_{e_i}[X_n^{l,m}]}\right|
\ee
where $\delta X_n^{l,m}(e_i)$ is the difference between the numerically integrated expression and the series approximation
at the given value of $e_i$,
and ${\rm max}_{e_i}[X_n^{l,m}]$ is the maximum value of $X_n^{l,m}$ over the interval $0\le e_i\le 1$.
The expansions are correct to
${\cal O}(e_i^6)$, except for 
$X_2^{3,1}(e_i)$ which is correct to ${\cal O}(e_i^8)$ because the errors are an order of magnitude smaller with the extra term.
Note that the error decreases monotonically with decreasing $e_i$ for each approximation.
The functions $X_1^{2,2}(e_i)$ and $X_1^{3,1}(e_i)$ are compared to the numerically calculated integrals in Figure~\ref{slmfig},
panels (a) and (d) respectively.

\begin{figure}
\centering
\includegraphics[width=120mm]{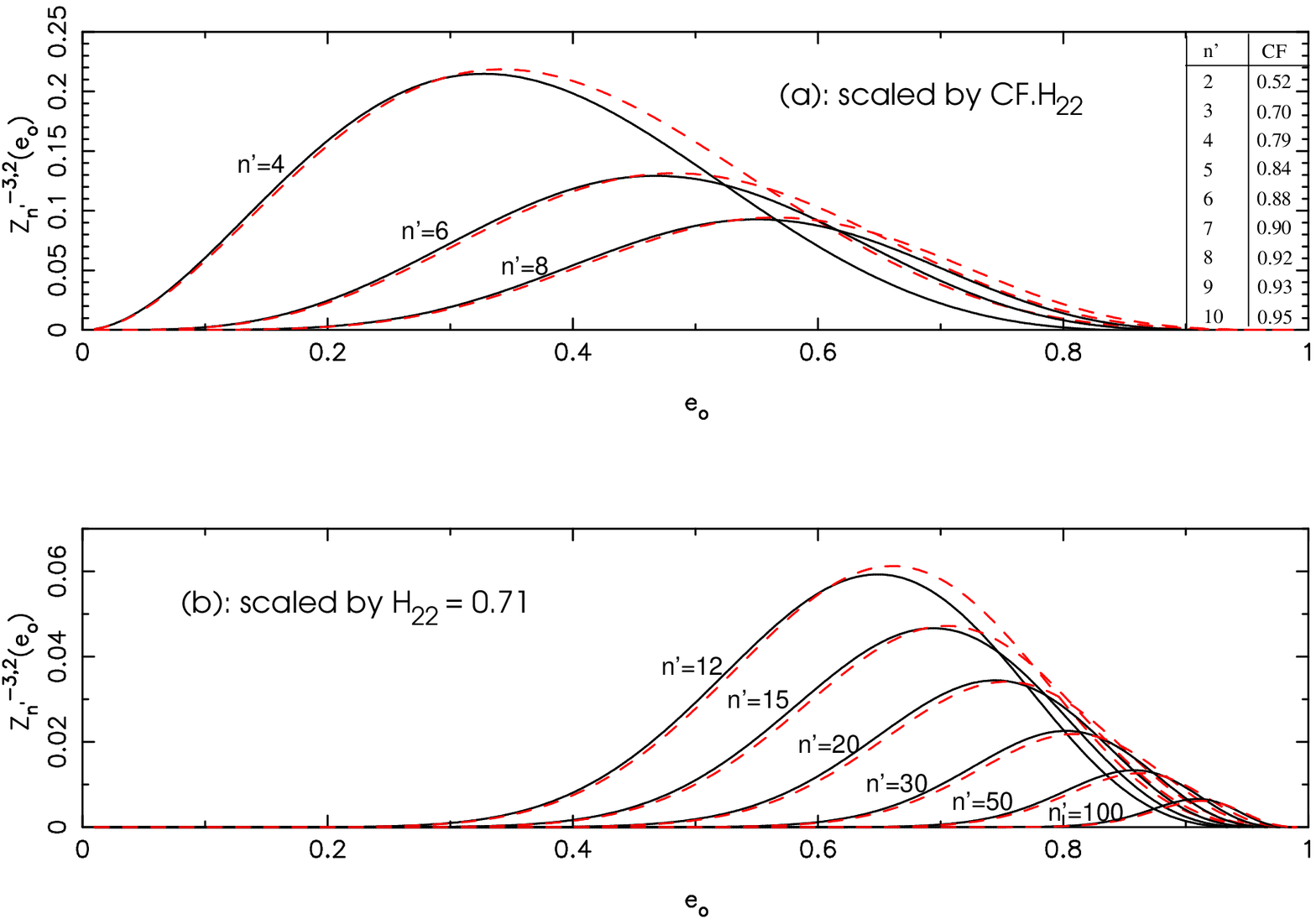}
\caption{Modified Hansen coefficients $Z_{n'}^{-3,2}(e_o)$ (black curves) and their
asymptotic approximations $\tilde Z_{n'}^{-3,2}(e_o)$ (equation \rn{asymX}, red dashed curves).
In panel (b) the approximations are scaled by correction factor ${\cal H}_{22}$ (already included in \rn{asymX}),
while in panel (a) they are scaled by the additional correction factors `CF', listed at the right of the panel.
}
\label{asym}
\end{figure}

A good approximation for modified Hansen coefficients governing the dependence of the
disturbing function on the {\it outer} eccentricity is given by the expression (Paper II)
\be
Z_{n'}^{-3,2}(e_o)\simeq (1-e_o)^3\cdot
\frac{4{\cal H}_{22}}{3\sqrt{2\pi}}
\frac{(1-e_o^2)^{3/4}}{e_o^2}\,{n'}^{3/2} e^{-n'\xi(e_o)}\equiv \tilde Z_{n'}^{-3,2}(e_o),
\label{asymX}
\ee
where
\be
\xi(e_o)={\rm Cosh}^{-1}(1/e_o)-\sqrt{1-e_o^2},
\label{xi}
\ee
and the constant ${\cal H}_{22}$ is an empirical scaling factor given in Table~\ref{hansentab}, determined
by comparing the maxima of $Z_{20}^{-3,2}(e_o)$ and $\tilde Z_{20}^{-3,2}(e_o)$ and scaling the latter
so that the values of their maxima are the same. All other $Z_{n'}^{-3,2}(e_o)$ are then scaled by the same factor,
except for $n\le 10$ which involve additional scale factors listed in panel (a) of 
Figure~\ref{asym}.
Note that $\lim_{e_o\rightarrow 0} e^{-n'\xi(e_o)}e_o^{-2}=0$, $n'\ge 3$, while for $n'=2$ the limit is ${\rm e}^2/2$.
However, for small values of $e_o$ it may be preferable to use a power series approximation for $X_{n'}^{-3,2}(e_o)$.

The derivation of \rn{asymX} uses the method of steepest decents to evaluate the integrals
and is therefore referred to as an asymptotic approximation. They are closely related to overlap integrals
which quantify the strength of the interaction between the orbits (Paper II).
Scale factors are necessary because approximations made in the analysis rely on the values of $e_o$ and 
$\sigma$ (and hence $n'$) being high, and accuracy is lost when they are not (although the shape of the curves
is preserved).
Figure~\ref{asym}
compares numerically evaluated integrals with their asymptotic approximations for selected values of $n'$ between 4 and 100.

The asymptotic approximation for general $l$,
$m\ge 0$ and $n'\ge 2$ is 
\be
Z_{n'}^{-(l+1),m}\simeq (1-e_o)^{l+1}\cdot
\frac{{\cal H}_{lm}}{\sqrt{2\pi}}
\frac{2^m}{(l+m-1)!!}
\frac{(1-e_o^2)^{(3m-l-1)/4}}{e_o^m}n'^{(l+m-1)/2}
e^{-n'\xi(e_o)}\equiv \tilde Z_{n'}^{-(l+1),m}(e_o).
\label{Easym}
\ee
Note that $\tilde Z_{n'}^{-(l+1),m}(e_o)={\cal O}(e_o^{|n'-m|})$, $n'\ge 2$, consistent with \rn{sF0}.

\subsection{Expansions up to fourth order in eccentricity for use in the literal expansion}\label{expan}
\begin{figure}
\centering
\includegraphics[width=140mm]{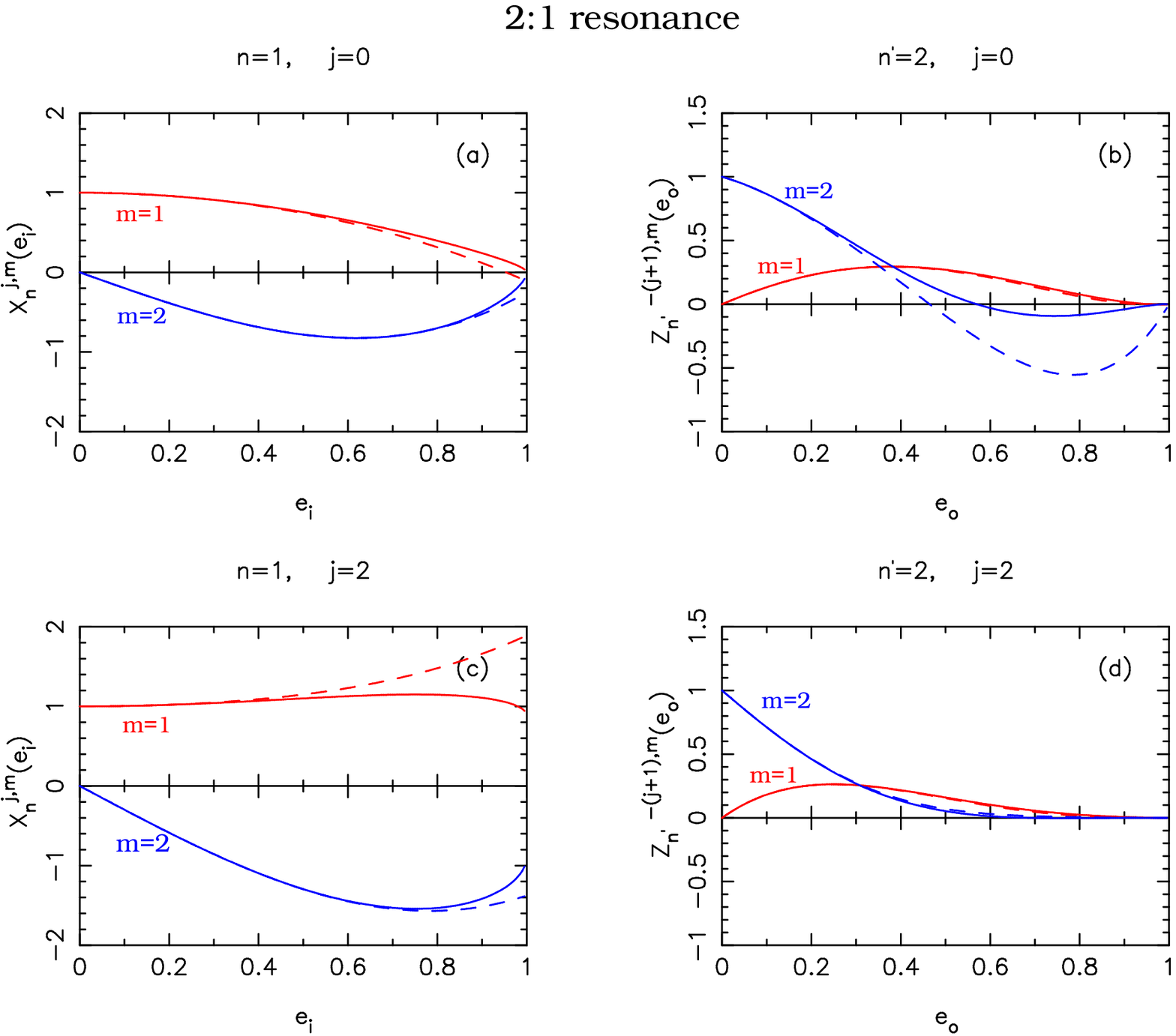}
\caption{Some Hansen coefficients associated with the first-order $2\!:\!1$ resonance. Plotted are
$X_1^{j,m}(e_i)$ and $Z_2^{-(j+1),m}(e_o)$ for $m=1$ (red) and
$m=2$ (blue), and for
$j=0$ (panels (a) and (b)) and
$j=2$ (panels (c) and (d)).
Solid curves: ``exact'' integration, dashed curves: fourth-order correct series expansions.
Accuracy is good up to at least $e_{i,o}=0.4$ for all 
Hansen coefficients shown here, this value increasing up to 1 in come cases (for example,
$Z_2^{-1,1}(e_o)$). These should be compared with series expansions for the 
first-order $7\!:\!6$ resonance which are less accurate given the higher values of $n$ and $n'$
(see Figure~\ref{hansencomp76}).
}
\label{hansencomp21}
\end{figure}

\begin{figure}
\centering
\includegraphics[width=140mm]{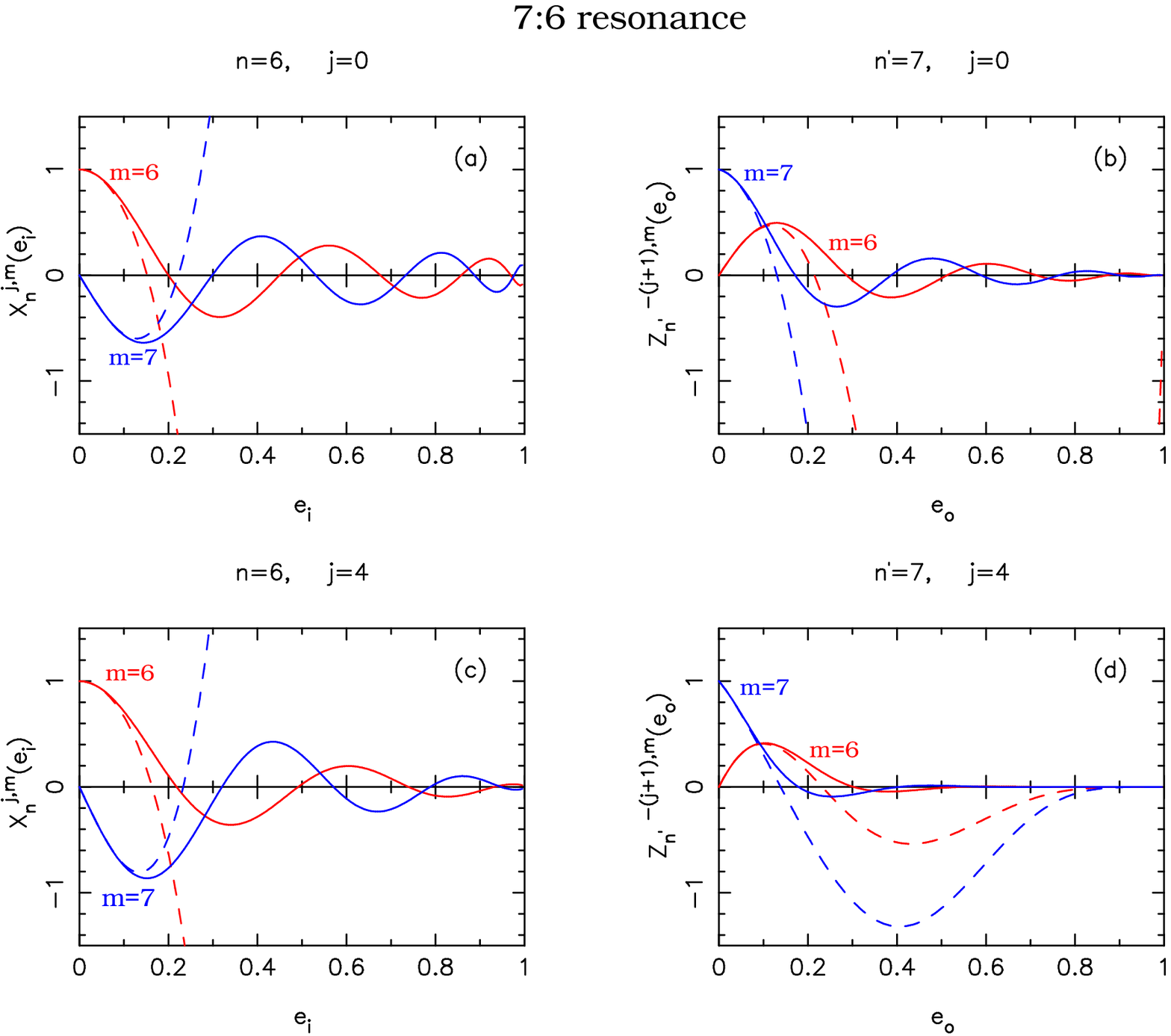}
\caption{Some Hansen coefficients associated with the first-order $7\!:\!6$ resonance. Plotted are
$X_6^{j,m}(e_i)$ and $Z_7^{-(j+1),m}(e_o)$ for $m=6$ (red) and
$m=7$ (blue), and for
$j=0$ (panels (a) and (b)) and
$j=4$ (panels (c) and (d)).
Solid curves: ``exact" integration, dashed curves: fourth-order correct series expansions.
While errors in the series expansions \rn{Xa} to \rn{Xb}
are formally of order of the fifth power of the eccentricity, inspection 
shows that errors are in fact ${\cal O}(n^5e^5)$. 
Putting $6^5e^5=0.1$,
the expansions for this case are accurate only for $e_{i,o}\lapp 0.1$ for a $10\%$ error.
}
\label{hansencomp76}
\end{figure}

\begin{figure}
\centering
\includegraphics[width=140mm]{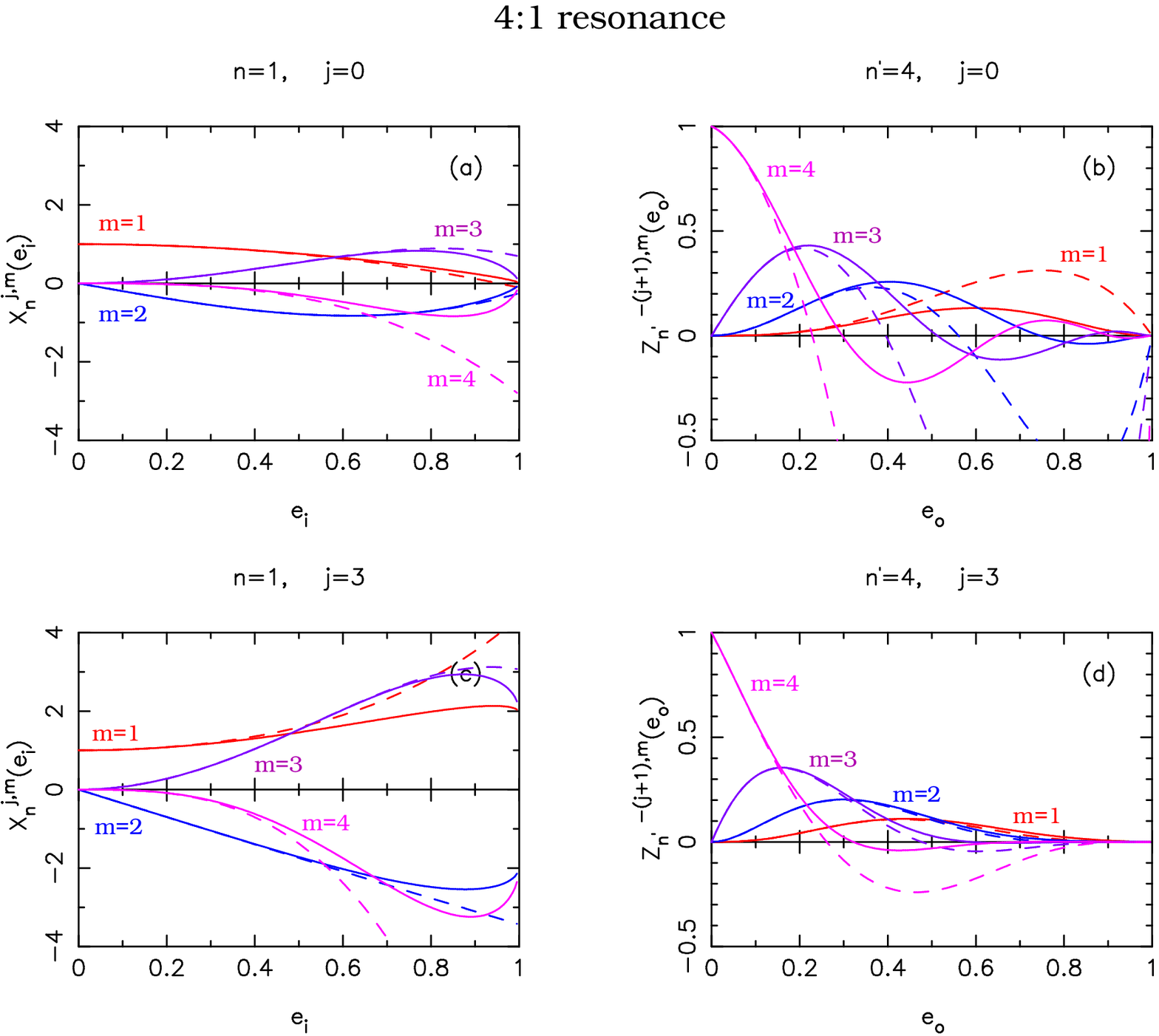}
\caption{Some Hansen coefficients associated with the third-order $4\!:\!1$ resonance.
Plotted are
$X_1^{j,m}(e_i)$ and $Z_4^{-(j+1),m}(e_o)$ for $m=1$ (red),
$m=2$ (blue), $m=3$ (purple) and $m=4$ (magenta),
and for
$j=0$ (panels (a) and (b)) and
$j=3$ (panels (c) and (d)).
Solid curves: `exact' integration, dashed curves: fourth-order correct series expansions.
While approximations are accurate for $e_i\lapp 0.5$ for $n=1$ (panels (a) and (c)),
they are accurate only for $e_o\lapp 0.15$ for $n=4$. As in the case of the $7\!:\!6$ resonance
in Figure~\ref{hansencomp76}, this is because the error in the expansions is ${\cal O}[n^5e^5]$. 
}
\label{hansencomp41}
\end{figure}

Using {\it Mathematica} or similar, it is easy to derive general 
power series expansions for $X_n^{j,m}(e)$ valid for any integers $j$ and $n$ and for
specific values of $m$. These series contain either even or odd powers of $e$ only.
Writing $\nu=n\,{\rm sgn}(m-n)$, we have for
$m=n$, $m=n\pm 1$, $m=n\pm 2$, $m=n\pm 3$ and $m=n\pm 4$ correct to ${\cal O}(e^4)$,
\bea
X_n^{j,n}(e)&=&1+\ff{1}{4}\left[j(j+1)-4\nu^2\right]\,e^2
+\ff{1}{64}\left[(j+1)j(j-1)(j-2)+\nu^2(16\nu^2-(8j^2+9))\right]\,e^4+\ldots,
\label{Xa}
\eea
\bea
X_n^{j,n\pm 1}(e)&=&
-\ff{1}{2}\left[(j+2)+2\nu\right]e
+\ff{1}{16}\left[-j(j+2)(j-1)+\nu(8\nu^2+2\nu(2j+7)-j(2j-3)+6)\right]e^3+\ldots,
\label{Xa2}
\eea
\bea
X_n^{j,n\pm 2}(e)&=&
\ff{1}{8}\left[(j+2)(j+3)+\nu(4\nu+4j+11)\right]e^2\next
&&\phantom{.}\hspace{-0.75cm}
+\ff{1}{96}\left[(j+3)(j+2)(j-1)(j-2)-2\nu(8\nu^3+34\nu^2+44\nu+13)
+j\nu(4j^2-3j-47-16\nu(\nu+3))\right]e^4+\ldots,
\eea
\bea
X_n^{j,n\pm 3}(e)&=&
-\ff{1}{48}\left[(j+4)(j+3)(j+2)+2\nu(4\nu^2+21\nu+31)
+3j\nu(4\nu+2j+13)\right]e^3+\ldots,
\eea
\bea
X_n^{j,n\pm 4}(e)&=&
\ff{1}{384}\left[(j+5)(j+4)(j+3)(j+2)+\nu(16\nu^3+136\nu^2+379\nu+394)\right.\next
&&\left. \hspace{4cm}+2j\nu(4j^2+45j+165+\nu(16\nu+12j+96))\right]e^4+\ldots
\label{Xb}
\eea
Notice that these are consistent with $X_n^{j,m}(e)={\cal O}(e^{|m-n|})$.
Comparison of these approximations with numerically evaluations of \rn{s4} and \rn{s5}
are shown in Figures~\ref{hansencomp21} to \ref{hansencomp41}. 
Note that for the outer eccentricity functions, modified Hansen coefficients are calculated.

Comparison of Figure~\ref{hansencomp21} with Figure~\ref{hansencomp76} 
for the $2\!:\!1$ and $7\!:\!6$ resonances respectively suggests that 
convergence of the series \rn{Xa} to \rn{Xb} is slower for the $7\!:\!6$ resonance. 
Inspection of the dependence of these series on $n$ and $j$ shows that the magnitude of the term 
proportional to $e^q$ is ${\cal O}(n\,e)^q$ or ${\cal O}(j\,e)^q$, whichever is the greatest,
although in some cases the errors may cancel to come extent (see, for example, the $m=1$ curve
in panel (d) of Figure~\ref{hansencomp21} 
which plots $Z_2^{-1,1}(e_o)=(1-e_o)X_2^{-1,1}(e_o)$, with the series expansion of $X_2^{-1,1}(e_o)$ given by \rn{Xa2}).

\subsection{Closed-form expressions}\label{hansen0}
\citet{hughes} has provided closed form expressions for Hansen coefficients with $n=0$ and $n'=0$.
Those associated with the inner orbit are
\bea
X_0^{l,m}(e_i)&=&\frac{1}{2\pi}\int_0^{2\pi}(r/a_i)^l e^{im f_i}dM_i\\
&=&\left(-\frac{e_i}{2}\right)^{m} {l+m+1\choose m}
F\left(\frac{m-l-1}{2},\frac{m-l}{2};m+1;e_i^2\right),
\eea
where $F(\,\,)$ is a hypergeometric function given here by (see
\citet{gradstein} for a general definition)
\be
F\left(\frac{m-l-1}{2},\frac{m-l}{2};m+1;e_i^2\right)
=1+\sum_{j=1}^{(l-m)/2}
\prod_{k=0}^{j-1}\frac{\left[(l-m+1)/2-k\right]\left[(l-m)/2-k\right]}
{(m+k+1)(k+1)}e^{2j},
\ee
with $F(-1/2,0;m+1;e_i^2)=1$ when $m=l$.
Hansen coefficients associated with the outer orbit are
\bea
X_0^{-(l+1),m}(e_o)&=&\frac{1}{2\pi}\int_0^{2\pi}
\frac{e^{-imf_o}}{(R/a_o)^{l+1}}dM_o\\
&=&\left(\frac{e_o}{2}\right)^{m}
(1-e_o^2)^{-(2l-1)/2}\sum_{j=0}^{\lfloor(l-m-1)/2\rfloor}
{l-1\choose 2j+m}{2j+m\choose j}\left(\frac{e_o^2}{2}\right)^j,
\eea
where $\lfloor\,\,\,\rfloor$ denotes the nearest lowest integer.

Quadrupole and octopole eccentricity functions are listed in Table~\ref{eccfuns}.
\begin{table*}
 \centering
 \begin{minipage}{140mm}
  \caption{Secular Hansen coefficients and $c_{lm}^2$}
  \label{eccfuns}
   \begin{tabular}{@{}ccccc@{}}
  \hline
   $l$     &     $m$       & $X_0^{l,m}(e_i)$ & $X_0^{-(l+1),m}(e_o)$ & $c_{lm}^2$\\
   \hline
2 & 2 & $\ff{5}{2}e_i^2$ & 0 & $\ff{3}{4}$\\
&&&\\
  &  0 & $1+\ff{3}{2}e_i^2$ & $(1-e_o^2)^{-3/2}$ & $\ff{1}{2}$ \\
  &&&\\
3 & 3 & $-\ff{35}{8}e_i^3$ & 0 & $\ff{5}{8}$ \\
&&&\\
  &  1 & $-\ff{5}{8}e_i(4+3e_i^2)$ & $e_o(1-e_o^2)^{-5/2}$ & $\ff{3}{8}$ \\
\hline
\end{tabular}
\end{minipage}
\end{table*}

\clearpage

\subsection{Mathematica programs for Hansen coefficients and $F_{mnn'}^{(j)}(e_i,e_o)$}\label{math}

It is the experience of the author that most colleagues in Astronomy 
have access to the software package {\it Mathematica} \citep{wolfram}, 
but many are not familiar with the syntax. For this reason the following short programs are included here. The first two
calculate series approximations for the Hansen coefficients
\be
X_n^{l,m}(e_i)=\frac{1}{2\pi}\int_0^{2\pi}r^l e^{imf_i}\,e^{-inM_i}dM_i
=\frac{1}{2\pi}\int_0^{2\pi}r^{l+1} \left[e^{if_i}\right]^m\,e^{-inM_i}dE_i
\ee
and
\be
X_n^{-(l+1),m}(e_o)=\frac{1}{2\pi}\int_0^{2\pi}\frac{e^{-imf_o}}{R^{l+1}}e^{in'M_o}dM_o
=\frac{1}{2\pi}\int_0^{2\pi}\frac{\left[e^{-if_o}\right]^m}{R^l}e^{in'M_o}dE_o,
\ee
where $E_i$ is the eccentric anomaly with
$r=1-e_i\cos E_i$, $M_i=E_i-e_i\sin E_i$, $\cos f_i=(\cos E_i-e_i)/(1-e_i\cos E_i)$ and
$\sin f_i=\sqrt{1-e_i^2}\sin E_i/(1-e_i\cos E_i)$, and similarly for $R$, $M_o$, $\cos f_o$ and $\sin f_o$.
The third program calculates $F_{mnn'}^{(j)}(e_i,e_o)$ according to \rn{Fmnn}. 
Recall that if $j_{max}$ is the order of the literal expansion
(ie., the highest combined powers of the eccentricities), then 
according to Section~\ref{4.1} $F_{mnn'}^{(j)}(e_i,e_o)$ should be expanded to $j_{max}$. 

\vspace{0.5cm}

\hrule

\vspace{2mm}

{\it Mathematica program to calculate} $X_1^{2,2}(e_i)=-3e_i+\ff{13}{8}e_i^3+\ff{5}{192}e_i^5+{\cal O}(e_o^7)$

\vspace{0.3cm}

{\tt
l = 2;

m = 2;

n = 1;

ne = 5;

r = 1 - e Cos[EA];

M = EA - e Sin[EA];

cosf = (Cos[EA] - e)/(1 - e Cos[EA]);

sinf = Sqrt[1 - e**2] Sin[EA]/(1 - e Cos[EA]);

expnM = Cos[n M] - I Sin[n M];

Xlmn = Normal[Series[r**(l + 1) (cosf + I sinf)**m expnM, \{e, 0, ne\}]];

Xlmnav = Simplify[Integrate[Xlmn/(2 Pi), \{EA, 0, 2 Pi\}]]
}

\vspace{0.5cm}

\hrule

\vspace{2mm}

{\it Mathematica program to calculate}  $X_2^{-3,2}(e_o)=1-\ff{5}{2}e_o^2+\ff{13}{16}e_o^4+{\cal O}(e_o^6)$

\vspace{0.3cm}

{\tt
l = 2;

m = 2;

n = 2;

ne = 4;

R = 1 - e Cos[EA];

M = EA - e Sin[EA];

cosf = (Cos[EA] - e)/(1 - e Cos[EA]);

sinf = Sqrt[1 - e**2] Sin[EA]/(1 - e Cos[EA]);

expnM = Cos[n M] + I Sin[n M];

Xlmn = Normal[Series[(cosf - I sinf)**m expnM/R**l, \{e, 0, ne\}]];

Xlmnav = Simplify[Integrate[Xlmn/(2 Pi), \{EA, 0, 2 Pi\}]]
}

\vspace{0.5cm}

\hrule

\vspace{2mm}

{\it Mathematica program to calculate $F_{435}^{(2)}(e_i,e_o)=-\ff{1}{2}e_i e_o-\ff{71}{16} e_i^3e_o
-\ff{97}{16}e_ie_o^3+{\cal O}(e^6)$ correct to 4th order}.
Notice that the 

leading term is of order $j=2$, and that all possible combinations of the powers of $e_i$ and $e_o$ are present 
given the 

constraints that only odd powers of $e_i\ge |m-n|=1$ and  $e_o\ge|m-n'|=1$ can appear. Terms 
of ${\cal O}(e^6)$ should be 

discarded.

\vspace{0.3cm}

{\tt

jmax=4;

j = 2;

m = 4;

n = 3;

no = 5;

nei = Max[Abs[m - n], jmax - Abs[m - no]];

neo = Max[Abs[m - no], jmax - Abs[m - n]];

\vspace{2mm}

r = (1 - ei Cos[EA]);

M = EA - ei Sin[EA];

cosf = (Cos[EA] - ei)/(1 - ei Cos[EA]);

sinf = Sqrt[1 - ei**2] Sin[EA]/(1 - ei Cos[EA]);

expnM = Cos[n M] - I Sin[n M];

\vspace{2mm}

R = (1 - eo Cos[EAo]);

Mo = EAo - eo Sin[EAo];

cosfo = (Cos[EAo] - eo)/(1 - eo Cos[EAo]);

sinfo = Sqrt[1 - eo**2] Sin[EAo]/(1 - eo Cos[EAo]);

expnMo = Cos[no Mo] + I Sin[no Mo];

\vspace{2mm}

Fjmnnp = Expand[Simplify[Sum[(-1)**(j - k) Binomial[j, k] 

\hspace{0.3cm}
Integrate[

\hspace{0.6cm}
Normal[Series[r**(k + 1) (cosf + I sinf)**m expnM, \{ei, 0, nei\}]]/(2 Pi), 

\hspace{0.3cm}
\{EA, 0, 2 Pi\}] 

\hspace{0.3cm}
Integrate[

\hspace{0.6cm}
Normal[Series[(cosfo - I sinfo)**m  expnMo/R**k , \{eo, 0, neo\}]]/(2 Pi), 

\hspace{0.3cm}
\{EAo, 0, 2 Pi\}], 

\{k, 0, j\}]]]
}

\vspace{0.5cm}

\hrule

\section{Laplace coefficients}\label{formulae}

\subsection{Series expansion}\label{C2}
%33/red/91a
The general Laplace coefficient is defined as
\be
b_s^{(m)}(x)=\frac{1}{\pi}\int_0^{2\pi}\frac{{\rm e}^{-im\psi}}{(1-2x\cos\psi+x^2)^s} d\psi,
\ee
where $s$ is a postive half integer.
A series expansion for this is
\be
b^{(m)}_{s}(x)
=x^m\sum_{p=0}^\infty C_p^{(m,s)}\,x^{2p}={\cal O}(x^m),
\label{e1bb}
\ee
where $s\ge 1/2$ is a half integer and
\be
C_p^{(m,s)}=\frac{2\Gamma(s+p)\Gamma(s+m+p)}{\left[\Gamma(s)\right]^2 p!(m+p)!}.
\label{e1bbb}
\ee
Recall that 
\be
\Gamma(n)=(n-1)!,
\ee
and
\be
\Gamma(n+\ff{1}{2})=\frac{(2n)!}{2^{2n}n!}\sqrt{\pi}.
\ee
%33/red/91
For $s=1/2$ we have
\be
C_p^{(m,1/2)}=\frac{2(2p)!(2m+2p)!}{2^{4p+2m}\left[p!(m+p)!\right]^2}.
\ee
By the ratio test, we have that the series \rn{e1bb} converges as long as $x>1$.

\subsection{Derivatives}\label{derivs}
The following formulae are taken from \citet{brouwer} p502 (and \citet{murray} who take their
formulae from \citet{brouwer}).
Defining the operator
\be
{\cal D}=\frac{d}{dx},
\ee
we have
\be
{\cal D}b_s^{(m)}(x)=s\left[b_{s+1}^{(m-1)}-2\,x\,b_{s+1}^{(m)}+b_{s+1}^{(m+1)}\right]
\label{C8}
\ee
and in general,
\be
{\cal D}^n b_s^{(m)}(x)=s\left[{\cal D}^{n-1}b_{s+1}^{(m-1)}
-2\,x\,{\cal D}^{n-1}b_{s+1}^{(m)}+{\cal D}^{n-1}b_{s+1}^{(m+1)}
-2(n-1){\cal D}^{n-2}b_{s+1}^{(m)}
\right], \hspace{0.5cm}n\geq 2.
\label{C9}
\ee

\subsubsection{Leading terms}\label{3.1}

Recalling the definition from \rn{bigB}
\be
{\cal B}_{1/2}^{(j,m)}(x)=\frac{x^j}{j!}\frac{d^j}{dx^j} \left[b_{1/2}^{(m)}(x)\right],
\ee
we have from \rn{e1bb} that
\bea
{\cal B}_{1/2}^{(j,m)}(x)
&=&\frac{1}{j!}\sum_{p=p_{min}}^\infty \frac{(m+2p)!}{(m+2p-j)!}\,C_p^{(m,1/2)}\,x^{m+2p}\\
&=&\sum_{p=p_{min}}^\infty E_p^{(j,m)}\,x^{m+2p},
\label{e11}
\eea
where 
$p_{min}={\rm max}\left(0,\lfloor\ff{1}{2}(j-m+1)\rfloor\right)$ 
with $\lfloor\,\,\rfloor$ denoting the nearest lowest integer,
and
\be
E_p^{(j,m)}={m+2p\choose j}\,C_p^{(m,1/2)}
=\frac{2(2m+2p)!(m+2p)!(2p)!}{4^{2p+m}  j! (m+2p-j)! [p!(m+p)!]^2}.
\label{Eplm}
\ee
The leading term of ${\cal B}_{1/2}^{(j,m)}(x)$ is therefore
\be
{\cal B}_{1/2}^{(j,m)}(x)=\left\{
\begin{array}{ll}
E_0^{(j,m)}\,x^m+\dots, &j\le m\\
E_{p_*}^{(j,m)}\,x^j+\ldots, & j\ge m,\,\,\, j-m\,\,\,{\rm even}\\
E_{p_*}^{(j+1,m)}\,x^{j+1}+\ldots, & j> m,\,\,\, j-m\,\,\,{\rm odd}\\
\end{array}
\right.
\label{lead}
\ee
where
\be
E_0^{(j,m)}=\frac{2(2m)!}{4^m  j! (m-j)! m!},
\label{E0}
\ee
and
\be
E_{p_*}^{(j,m)}=\frac{2(j+m)!((j-m))!}{4^j  [((j-m)/2)!((j+m)/2)!]^2}
\label{E0b}
\ee
with $p_*=\lfloor(j-m+1)/2\rfloor$.
Note that by the ratio test, the series \rn{e11} converges as long as $x<1$.

\subsection{Graphical representations}\label{C1}

\begin{figure}
\centering
\includegraphics[width=140mm]{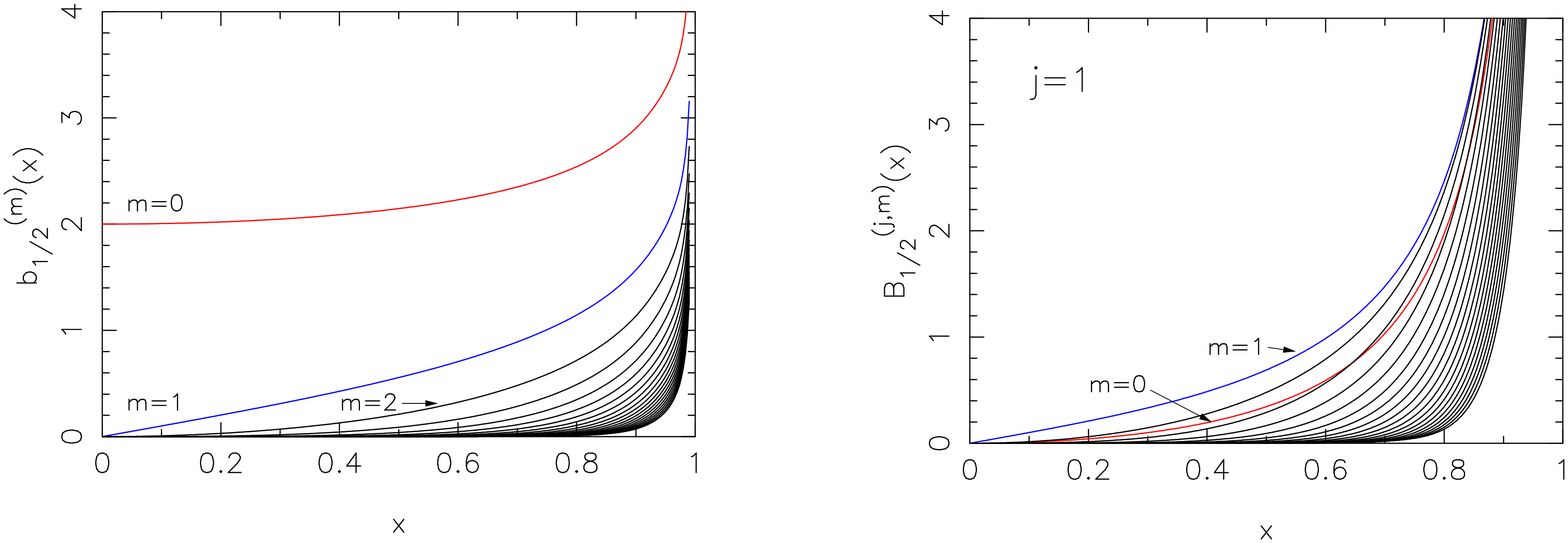}
\caption{$b_{1/2}^{(m)}(x)$ and ${\cal B}_{1/2}^{(j,m)}(x)$
for $j=1$, both
for $m=0,1,\ldots,20$. In particular, the red curves are for $m=0$ and
the blue for $m=1$.}
\label{laplace}
\end{figure}

Figure~\ref{laplace} 
shows $b_{1/2}^{(m)}(x)$ and ${\cal B}_{1/2}^{(j,m)}(x)$
for $j=1$, both
for $m=0,1,\ldots,20$. Notice how, for most values of $m$, ${\cal B}_{1/2}^{(1,m)}(x)>b_{1/2}^{(m)}(x)$.
In general, ${\cal B}_{1/2}^{(j_1,m)}(x)>{\cal B}_{1/2}^{(j_2,m)}(x)$ when $j_1>j_2$.
Note that $b_{1/2}^{(0)}(0)=2$, $b_{1/2}^{(m)}(0)=0$, $m\ge 1$ and ${\cal B}_{1/2}^{(j,m)}(0)=0$ for all $m$.

\section{Lagrange's planetary equations for the variation of the elements}\label{lagrange}

Because Lagrange's equations for the variation of the elements 
were developed for the restricted three-body problem (the mass of one of the three
bodies of interest is negligible compared to the other two
so that one orbit is fixed), they are normally
given in terms of a disturbing function which has the dimensions of 
energy per unit mass \citep{brouwer,murray}. Here the disturbing function 
has the dimensions of energy, and as a result the same function can be used
for the rates of change of the inner and outer orbital elements, that is,
there is no need to define separate inner and outer disturbing functions.
Note also that Lagrange's ``planetary'' equations hold for any mass ratios,
and in particular, there is no assumption about the smallness of ${\cal R}$.
In spite of the fact that in most applications the disturbing function acts to perturb the
Keplerian orbits from invariant elliptical motion,
the derivation of Lagrange's equations does not involve a perturbation technique. Rather it uses
the method of variation of parameters, those parameters being the orbital elements
which are constant when there is no interaction between the orbits (ie, when ${\cal R}=0$),
and vary once the orbits are allowed to interact via a non-zero ${\cal R}$.

For reference, the relevant Lagrange equations for the rates of change of the elements for coplanar systems are
\be
\frac{de}{dt}=-\frac{\varepsilon(1-\varepsilon)}{\mu\nu a^2 e}\pp{\cal R}{\lambda}
-\frac{\varepsilon}{\mu\nu a^2 e}\pp{\cal R}{\varpi},
\label{D1}
\ee
\be
\frac{d\varpi}{dt}=\frac{\varepsilon}{\mu\nu a^2 e}\pp{\cal R}{e},
\label{D2}
\ee
\be
\frac{d\epsilon}{dt}=-\frac{2}{\mu\nu a}\pp{\cal R}{a}
+\frac{\varepsilon(1-\varepsilon)}{\mu\nu a^2 e}\pp{\cal R}{e}
\label{epsilon}
\ee
and
\be
\frac{da}{dt}=\frac{2}{\mu\nu a}\pp{\cal R}{\lambda},
\label{dadt}
\ee
where $\varepsilon=\sqrt{1-e^2}$, and
with the set $\{a,e,\varpi,\epsilon;\mu,\nu\}$
representing either of the inner orbit $\{a_i,e_i,\varpi_i,\epsilon_i;\mu_i,\nu_i\}$
or the outer orbit $\{a_o,e_o,\varpi_o,\epsilon_o;\mu_o,\nu_o\}$. Note that in this form, $\lambda$ and $\epsilon$ are related
through $\lambda=\int_0^t \nu\, dt+\epsilon$ rather than the usual $\lambda=\nu t+\epsilon$
(see \citet{brouwer} page 285 for a discussion of this point, in particular that in using this definition one need not consider $\nu$ to be a function
of $a$ when evaluating $\partial{\cal R}/\partial a$). The former definition should always be used
when resonance plays a role, however, note that
the two definitions are equivalent when it doesn't (because in that case the semimajor axes are constant).

\section{The mean longitude at epoch}\label{ml}
Finally, a note on the mean longitude at epoch, $\epsilon\equiv M(T_0)+\varpi$,
where $T_0$ is the time (``epoch'') at which the osculating elements are determined
and $M(T_0)$ is the mean anomaly at that time.
For some applications $T_0$ is taken to be the time at periastron passage, while for others
it corresponds to some given time, for example, the mid-time of a particular set
of observations. While it is clear what it means for $\varpi$ to vary, one might reasonably
ask what it means for $M(T_0)$ to vary.

\begin{figure}
\centering
\includegraphics[width=60mm]{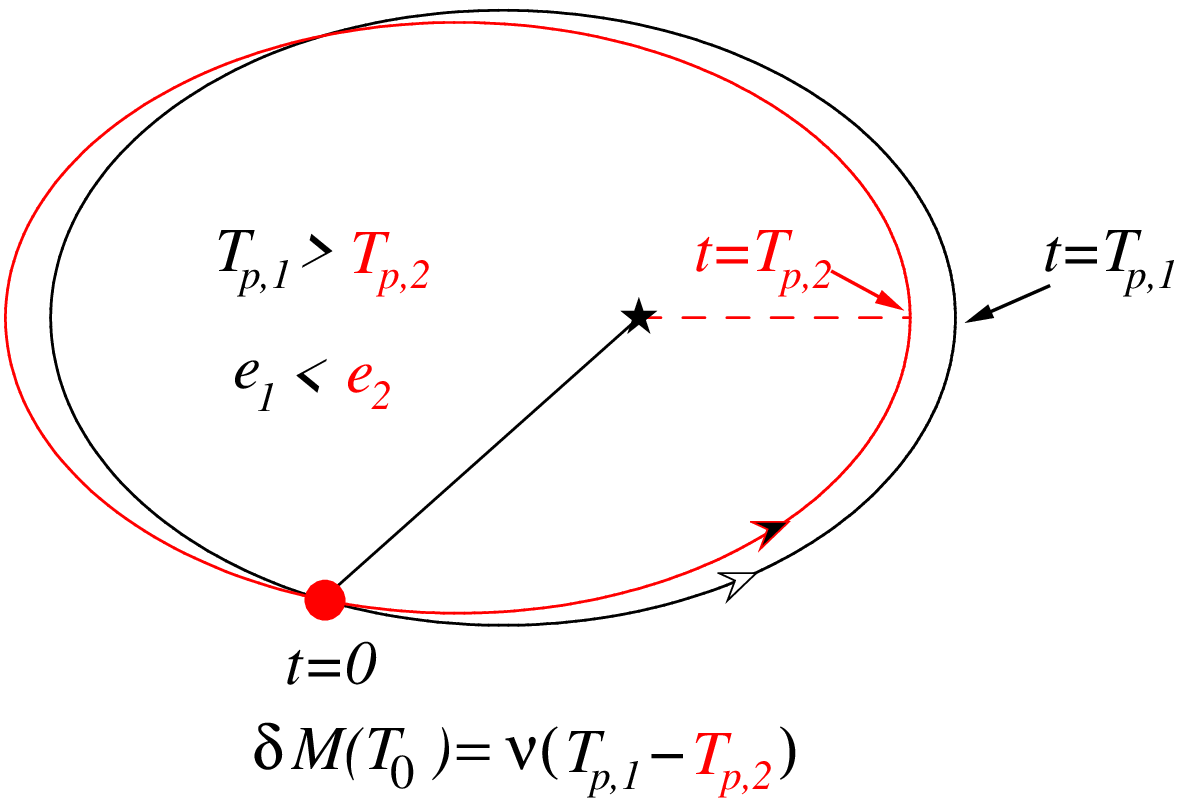}
\caption{Illustration of the effect on the mean anomaly at epoch, $M(T_0)=\nu(T_0-T_p)$, of increasing the eccentricity
while holding the semimajor axis constant.
If the perturbation is applied at $t=0$ when 
$M=-\nu \,T_{p,1}$, then although at the instant the force is applied there is no change in the {\it true} anomaly $f$
(the true position in the orbit), there is a change in the {\it mean} anomaly and hence the mean anomaly at epoch
(since $T_0$ is a fixed time). This is
given by $\delta M(T_0)=\nu(T_{p,1}-T_{p,2})$, where $T_{p,1}$ and $T_{p,2}$ are the times to periastron
before and after the perturbation is applied.
}
\label{ellipse}
\end{figure}

To answer this question, consider the usual definition of the mean anomaly.
This is given in terms of the orbital frequency (mean motion) $\nu$ and
the time at periastron passage $T_p$ as
\be
M=\nu(t-T_p),
\ee
or in terms of a general epoch $T_0$,
\be
M=\nu(t-T_0)+\nu(T_0-T_p)=\nu(t-T_0)+M(T_0).
\ee
Either way, $M(T_p)=0$.
Referring to Figure~\ref{ellipse}, 
imagine at $t=0$ a force acts on the system 
in such a way that only the eccentricity is changed.
If the eccentricity increases, the time to periastron passage, $T_p$, decreases, and since $\nu$
remains unchanged and $T_0$ is fixed, the mean anomaly at epoch, $M(T_0)=\nu(T_0-T_p)$, must {\it increase}.
The change in the mean anomaly at epoch must therefore be proportional to the change
in the time at periastron, that is,
\be
\delta M(T_0)=\nu(T_{p,1}-T_{p,2}),
\ee
where $T_{p,1}$ and $T_{p,2}$ are the times to periastron
before and after the perturbation is applied.
In general, an arbitrary force acting on the system will cause
the mean anomaly at epoch to change by an amount \citep[][page 36]{pollard}
\be
\delta M(T_0)=\left[\ff{3}{2}M-r\sqrt{1-e^2}/(e\sin f)\right](\delta a/a)+
\sqrt{1-e^2}\cot f\, (\delta e/e),
\ee
that is, only changes to the osculating eccentricity and semimajor axis affect this quantity (as reflected
in Lagrange's planetary equation \rn{epsilon} for
the rate of change of $\epsilon$).

\section{Notation}\label{not}

{\small
%\begin{table*}
 \centering
 \begin{minipage}{140mm}
%  \caption{Notation.}
  \begin{tabular}{@{}llr@{}}
  \hline
$f_i$, $f_o$ & inner and outer true anomalies& \pageref{p7}\\
$M_i$, $M_o$ & inner and outer mean anomalies& \pageref{p8}\\
$\lambda_i$, $\lambda_o$ & inner and outer mean longitudes& \pageref{p9}\\
$\varpi_i$, $\varpi_o$ & inner and outer longitudes of periastron& \pageref{p10}\\
$\epsilon_i$, $\epsilon_o$ & inner and outer mean longitudes at epoch& \pageref{p11}\\
$e_i$, $e_o$ & inner and outer eccentricities& \pageref{p12}\\
\vspace{2mm}
$a_i$, $a_o$ & inner and outer semimajor axes&  \pageref{p12}\\
$\alpha=a_i/a_o$ &&  \pageref{p13}\\
$\alpha_s=\beta_s\alpha$, $s=1,2$ &&  \pageref{p14}\\
$P_i$, $P_o$ & inner and outer orbital periods& \pageref{p15}\\
$\nu_i$, $\nu_o$ & inner and outer orbital frequencies (mean motions)& \pageref{p16}\\
\vspace{2mm}
$\sigma=P_o/P_i=\nu_i/\nu_o$ & period ratio& \pageref{p17}\\
${\bf r}$ & position of body 2 relative to body 1 & \pageref{p18}\\
\vspace{2mm}
${\bf R}$ & position of body 3 relative to the centre of mass of bodies 1 and 2 & \pageref{p18}\\
$m_{12}=m_1+m_2$ & & \pageref{p1}\\
$m_{123}=m_1+m_2+m_3$ && \pageref{p2}\\ 
$\beta_1=m_1/m_{12}$ && \pageref{p3}\\
$\beta_2=-m_2/m_{12}$ && \pageref{p4}\\
${\cal M}_l=\beta_1^{l-1}-\beta_2^{l-1}$ &&  \pageref{p4b}\\
$\mu_i=m_1m_2/m_{12}$ & reduced mass of inner binary & \pageref{p5}\\  
\vspace{2mm}
$\mu_o=m_{12}m_3/m_{123}$ & reduced mass of outer binary & \pageref{p6}\\  
${\cal R}$ & the disturbing function (interaction energy) & \pageref{p19}\\
$\tilde{\cal R}$ & the orbit-averaged disturbing function&  \pageref{p20}\\
${\cal R}_{mnn'}$ & harmonic coefficient&  \pageref{p21},  \pageref{p22}\\
$\phi_{mnn'}$ & harmonic angle &  \pageref{p13}, \pageref{p13b}\\
$\phi_N\equiv\phi_{21N}$ &  &  \pageref{p24}\\
\vspace{2mm}
$[n'\!:\!n](m)$ & harmonic associated with the angle $\phi_{mnn'}$& \pageref{p17}\\
$\Delta\sigma_N$ & width of $[N\!:\!1](2)$ resonance&  \pageref{p23}\\
$\Delta\sigma_{mnn'}$ & width of $[n'\!:\!n](m)$ resonance& \pageref{p25}\\
\vspace{2mm}
$\omega_N$, $\omega_{mnn'}$ & libration frequencies & \pageref{p26}, \pageref{libfreq}\\
$Y_{lm}(\theta,\varphi)$ & a spherical harmonic of {\it degree} $l$ and {\it order} $m$& \pageref{p27}\\
$c_{lm}^2$ & coefficients in the semimajor axis expansion & \pageref{pclm}\\
$\zeta_m$ & $\ff{1}{2}$ if $m=0$, 1 if $m\ge 1$ & \pageref{p31}\\
$m_{min}$ & 0 if $l$ is even, 1 if $l$ is odd& \pageref{p31}\\
\vspace{2mm}
$l_{min}$ & 2 if $m=0$, 3 if $m=1$, $m$ if $m\ge 2$& \pageref{p32}\\
$X_n^{l,m}(e_i)$ & a Hansen coefficient& \pageref{p28}\\
$Z_{n'}^{-(l+1),m}(e_o)$ & a modified Hansen coefficient& \pageref{p29}\\
\vspace{2mm}
$F_{mnn'}^{(j)}(e_i,e_o)$ & linear combination of Hansen coefficients & \pageref{pFmnn}\\
$b_{1/2}^{(m)}(\alpha_s)$ & a Laplace coefficient & \pageref{p30}\\
${\cal A}_{jm}(\alpha;\beta_2)$ & coefficient in eccentricity expansion & \pageref{pA}\\
\hline
\multicolumn{3}{@{}r@{}}{  {\scriptsize Numbers in the right-hand column refer to the page where
the variable is defined}}\\

\end{tabular}
\end{minipage}
%\end{table*}
 
 }

\label{lastpage}

\end{document}